\RequirePackage{pdf14}

\documentclass[letterpaper]{article}
\usepackage[usenames,dvipsnames,table]{xcolor}
\usepackage{graphicx}
\usepackage{caption}
\usepackage{subcaption}
\usepackage{array,setspace,mathrsfs,amsthm,amsfonts,amsmath,colonequals,mathtools,bbm}
\usepackage{./auxi/jheppub}
\usepackage{afterpage}
\usepackage{xspace}
\usepackage[normalem]{ulem}
\usepackage{cancel}

\makeatletter
\def\maketag@@@#1{\hbox{\m@th\normalfont\normalsize#1}}
\makeatother

\newcommand{\mbar}[1]{{\overline{ #1}}} 
\def\be{\begin{equation}}
\def\ee{\end{equation}}
\def\OO{{\mathcal{O}}}
\def\CC{{\mathcal{C}}}
\def\EE{{\mathcal{E}}}
\def\AA{{\mathcal{A}}}

\def\BB{{\mathcal{B}}}
\def\NN{{\mathcal{N}}}
\def\Nm{{\mathcal{N}}}

\def\DD{{\mathcal{D}}}
\def\QQ{{\mathcal{Q}}}

\def\WW{{\mathcal{W}}}
\def\JJ{\mathcal{J}}
\def\WB{\bar{\mathcal{W}}}
\def\OOh2{\mathcal{U}}
\def\calN{{\mathcal{N}}}
\def\rN{r_{\mathcal{N}=2}}
\def\RN{R_{\mathcal{N}=2}}

\def\DDb{\bar{\DD}}

\def\jb{\bar{\jmath}}
\def\zb{\bar{z}}

\def\th{\theta}
\def\bth{\bar{\theta}}

\def\nn{\nonumber}

\def\g{g}
\def\sl{g^{2d}}
\def\Ssl{g^{2d\;\NN=2}}

\def\zb{\bar{z}}

\def\SV{V}
\def\x{x}

\newcommand\beq{\begin{equation}}
\newcommand\bea{\begin{eqnarray}}
\newcommand\eea{\end{eqnarray}}
\newcommand\eeq{\end{equation}}

\newcommand\eg{{\it e.g.}}
\newcommand\ie{{\it i.e.}}

\def\nn{\nonumber}

\newcommand\qq{\mathbbmtt{Q}}

\newcommand\ospf{{\mathfrak{osp}}}

\newcommand\slf{\mathfrak{sl}}

\newcommand\fc{f}
\newcommand\Hc{H}
\newcommand\chiral{chiral }
\newcommand\nonchiral{non-chiral }

\newcommand\U{\mathrm{U}}
\newcommand\SU{\mathrm{SU}}
\newcommand\SL{\mathrm{SL}}
\newcommand\GL{\mathrm{GL}}

\newtheorem{remark}{Remark}

\title{Bootstrapping $\mathcal{N}=3$ superconformal theories}
\author[1]{Madalena Lemos,}
\author[1]{Pedro Liendo,}
\author[2]{Carlo Meneghelli,}
\author[3]{Vladimir Mitev}

\affiliation[1]{DESY Hamburg, Theory Group, Notkestrasse 85, D–22607 Hamburg, Germany}
\affiliation[2]{Simons Center for Geometry and Physics, Stony Brook University, Stony Brook, NY 11794-3636, USA}
\affiliation[3]{PRISMA Cluster of Excellence, Institut f\"ur Physik,
JGU Mainz,
Staudingerweg 7, 55128 Mainz, Germany}

\emailAdd{madalena.lemos@desy.de}
\emailAdd{pedro.liendo@desy.de}
\emailAdd{cmeneghelli@scgp.stonybrook.edu}
\emailAdd{vmitev@uni-mainz.de}

\preprint{DESY 16-237, MITP/16-132}

\bigskip
\abstract{
We initiate the bootstrap program for $\mathcal{N}=3$ superconformal field theories (SCFTs) in four dimensions. The  problem is considered from two fronts: the protected subsector described by a $2d$ chiral algebra, and crossing symmetry for half-BPS operators whose superconformal primaries parametrize the Coulomb branch of $\mathcal{N}=3$ theories. With the goal of describing a protected subsector of a family of $\mathcal{N}=3$ SCFTs, we propose a new $2d$ chiral algebra with super Virasoro symmetry that depends on an arbitrary parameter, identified with the central charge of the theory.
Turning to the crossing equations, we work out the superconformal block expansion and apply standard numerical bootstrap techniques in order to constrain the CFT data. We obtain bounds valid for any theory but also, thanks to input from the chiral algebra results, we are able to exclude solutions with $\mathcal{N}=4$ supersymmetry, allowing us to zoom in on a specific $\mathcal{N}=3$ SCFT.
}


\begin{document}
\setcounter{tocdepth}{2}
\maketitle
\setcounter{page}{1}

\section{Introduction}
\label{sec:intro}

The study of superconformal symmetry has given invaluable insights into quantum field theory, and in particular
into the nature of strong-coupling dynamics. The presence of supersymmetry gives us additional analytical tools and allows for computations that are otherwise hard to perform. A cursory look at the superconformal literature in four dimensions shows a vast number of works on $\Nm=2$ and $\Nm=4$ superconformal field theories (SCFTs), with the intermediate case of $\Nm=3$ almost absent. The main reason for this is that, due to CPT invariance, the Lagrangian formulation of any $\Nm=3$ theory becomes automatically $\Nm=4$. By now, however, there is a significant amount of evidence that superconformal theories are not restricted to just Lagrangian examples, and this has inspired recent papers that revisit the status of $\Nm=3$ SCFTs.

Assuming these theories exist, the authors of \cite{Aharony:2015oyb} studied several of their properties. They found in particular that the $a$ and $c$ anomaly coefficients are always the same, that pure $\Nm=3$ theories (\ie, theories whose symmetry does not enhance to $\NN=4$) have no marginal deformations
and are therefore always isolated, and also, 
in stark contrast with the most familiar $\Nm=2$ theories,
that pure $\Nm=3$ SCFTs cannot have a flavor symmetry that is not an R-symmetry. Moreover, since the only possible free multiplet of an $\NN=3$ SCFT is a vector multiplet, the low energy theory at a generic point on the moduli space must involve vector multiplets, and the types of short multiplets whose expectations values can parametrize such branches were analyzed in \cite{Aharony:2015oyb}. When an $\NN=3$ vector multiplet is decomposed in $\NN=2$, it contains both an $\NN=2$ vector and hyper multiplet, which implies that the theories possess both $\NN=2$ Higgs and Coulomb branches that are rotated by $\NN=3$.

Shortly after \cite{Aharony:2015oyb}, the authors of \cite{Garcia-Etxebarria:2015wns} presented the first evidence for $\Nm=3$ theories by studying $N$ D3-branes in the presence of an S-fold plane, which is a generalization of the standard orientifold construction that also includes the S-duality group. The classification of different variants of $\NN=3$ preserving S-folds was done in \cite{Aharony:2016kai}, leading to additional $\NN=3$ SCFTs. In \cite{Garcia-Etxebarria:2016erx} yet another generalization was considered, in which in addition to including the S-duality group in the orientifold construction, one also considers T-duality. This background is known as a U-fold, and the study of M5-branes on this background leads to $\Nm=3$ theories associated with the exceptional $(2,0)$ theories.

The systematic study of rank one $\NN=2$ SCFTs (\ie, with a one complex dimensional Coulomb branch) through their Coulomb branch geometries  \cite{Argyres:2015ffa, Argyres:2015gha, Argyres:2016xua, Argyres:2016xmc} has recovered the known $\NN=3$ SCFTs, but also led to new ones \cite{Argyres:2016xua,Argyres:2016yzz}.
Some of these theories are obtained by starting from $\NN=4$ SYM with gauge group $\U(1)$ or $\SU(2)$ and gauging discrete symmetries, while others correspond to genuine $\NN=3$ SCFTs which are not obtained by discrete gauging. Note that, as emphasized in \cite{Aharony:2016kai,Argyres:2016yzz}, gauging by a discrete symmetry does not change the local  dynamics of the theory on $\mathbb{R}^4$, only the spectrum of local and non-local operators. In particular, the central charges and correlation functions remain the same.

Of the class of theories constructed in \cite{Aharony:2016kai}, labeled by the number $N$ of D3-branes and by integers $k,\ell$ associated to the S-fold, 
some have enhanced $\NN=4$ supersymmetry, 
or arise as discretely gauged versions of $\NN=4$.
The non-trivial $\NN=3$ SCFT with the smallest central charge 
corresponds to the theory labeled by $N=1$ and $\ell=k=3$ in \cite{Aharony:2016kai}, with central charge given by $\tfrac{15}{12}$. 
This corresponds to a rank one theory with Coulomb branch parameter of scaling dimension three.
Since the Coulomb branch operators of $\NN=3$ theories must have integer dimensions \cite{Aharony:2015oyb}, and since theories with a Coulomb branch generator of dimension one or two enhance to $\NN=4$, it follows that dimension three is the smallest a genuine  $\NN=3$ theory with a Coulomb branch can have, and that this theory could indeed correspond to the ``minimal'' $\NN=3$ SCFT. By increasing the number of D3-branes, higher rank versions of this minimal theory can be obtained . More generally, the rank $N$ theories with $k=\ell$, are not obtained from others by discrete gauging, and have an $N$ dimensional Coulomb branch.

Since pure $\NN=3$ SCFTs have no relevant or marginal deformations, they are hard to study by standard field theoretical approaches. Apart from the aforementioned papers, recent progress in understanding $\Nm=3$ theories includes \cite{Nishinaka:2016hbw,Imamura:2016abe,Imamura:2016udl,Agarwal:2016rvx}.
The classification of all $\NN=3$ SCFTs is not complete yet, and one can wonder if there are theories not arising from the S-fold (and generalizations thereof) constructions. On the other hand, one would like to obtain more information on the spectrum of the currently known theories.
In this paper we take the superconformal bootstrap approach to address these questions, and tackle $\NN=3$ SCFTs by studying the operators that parametrize the Coulomb branch. These operators sit in half-BPS multiplets of the $\Nm=3$ superconformal algebra, and when decomposed in $\NN=2$ language contain both Higgs and Coulomb branch operators. 
We will mostly focus on the simplest case of Coulomb branch operators of dimension three.

The bootstrap approach does not rely on any Lagrangian or perturbative description of the theory. It depends only on the existence of an associative local operator algebra and on the symmetries of the theory in question, and is therefore very well suited to the study of $\NN=3$ SCFTs.
Since the original work of \cite{Rattazzi:2008pe} there have been many papers that study SCFTs through the lens of the numerical bootstrap \cite{Poland:2010wg,Poland:2011ey,Berkooz:2014yda,Poland:2015mta,Beem:2014zpa,Beem:2013qxa,Alday:2013opa,Alday:2014qfa,Chester:2014fya,Chester:2014mea,Chester:2015qca,Bashkirov:2013vya,Bobev:2015vsa,Bobev:2015jxa,Beem:2015aoa}. 
A basic requirement in any superconformal bootstrap analysis is the computation of the superconformal blocks relevant for the theory in question, although correlation functions of half-BPS operators in various dimensions have been studied \cite{Dolan:2001tt,Dolan:2004mu,Nirschl:2004pa}, the case of $\NN=3$ has not yet been considered, and calculating the necessary blocks will be one of the goals of this paper. For literature on superconformal blocks see \cite{Dolan:2001tt,Dolan:2004mu,Nirschl:2004pa,Fortin:2011nq,Fitzpatrick:2014oza,Khandker:2014mpa,Bissi:2015qoa,Doobary:2015gia,Li:2016chh,Liendo:2016ymz}.

Also relevant for our work is the information encoded in the $2d$ chiral algebras associated to $4d$ SCFTs \cite{Beem:2013sza,Beem:2014rza, Lemos:2014lua,Chacaltana:2014nya,Buican:2015ina,Cordova:2015nma,Liendo:2015ofa,Buican:2015tda,Song:2015wta,Cecotti:2015lab,Lemos:2015orc,Arakawa:2016hkg,Buican:2016arp, Cordova:2016uwk,Xie:2016evu}. The original analysis of \cite{Beem:2013sza} implies
that any four-dimensional $\NN \geqslant 2$ SCFT contains a closed subsector of local operators isomorphic to a $2d$ chiral algebra. 
For $\Nm=3$ theories, part of the extra supercharges, with respect to a pure $\NN=2$  theory, make it to the chiral algebra and therefore its symmetry enhances to $\Nm=2$ super Virasoro \cite{Nishinaka:2016hbw}. The authors of \cite{Nishinaka:2016hbw} constructed a family of chiral algebras conjectured to  describe the rank one $\NN=3$ theories,
generalizing these chiral algebras in order to accommodate the higher-rank cases will be another subject of this work.

The paper is organized as follows. Section \ref{sec:chiral algebra} 
studies the two-dimensional chiral algebras associated with $\NN=3$ SCFTs, determining the $\NN=3$ superconformal multiplets they capture, and some of their general properties. 
We then construct a candidate subalgebra of the chiral algebras for higher rank $\ell=k=3$ theories.  
In section \ref{sec:superblocksmain} we use harmonic superspace techniques in order to obtain the superconformal blocks that will allow us to derive the crossing equations for half-BPS operators of section \ref{sec:crossing}. We focus mostly on a dimension three operator, but also present the dimension two case as a warm-up. Section \ref{sec:numerics} presents the results of the numerical bootstrap, both for generic $\NN=3$ SCFTs and also attempting to zoom in to the simplest known $\NN=3$ theory by inputting data from the chiral algebra analysis of section \ref{sec:chiral algebra}.
We conclude with an overview of the paper and directions for future research in section \ref{sec:conclusions}.

\section{\texorpdfstring{$\NN=3$}{N=3} chiral algebras}
\label{sec:chiral algebra}

Every $4d$ $\NN \geqslant 2$ SCFT contains a protected sector that is isomorphic to a $2d$ chiral algebra, obtained by passing to the cohomology of a nilpotent supercharge \cite{Beem:2013sza}. Because $\NN=3$ is a special case of $\Nm=2$, one can also study chiral algebras associated to $\Nm=3$ SCFTs.
This program was started for rank one theories in \cite{Nishinaka:2016hbw}, and here we explore possible modifications such that one can describe higher-rank cases as well. We will put particular emphasis on theories containing a Coulomb branch operator with scaling dimension three, since these are the correlators we will study numerically in section \ref{sec:numerics}. We propose a set of generators that, under certain assumptions, describes a closed subalgebra of theories with a dimension three Coulomb branch operator, and write down an associative chiral algebra for them. Associativity fixes all OPE coefficients in terms of a single parameter: the central charge of the theory.

In order to do this we will need extensive use of the representation theory of the $\NN=3$ superconformal algebra; this was studied in \cite{Dobrev:1985qv,Minwalla:1997ka,Kinney:2005ej,Aharony:2015oyb,Cordova:2016xhm,Cordova:2016emh} and is briefly reviewed in appendix~\ref{app:shortening}.
We will also leverage previous knowledge of chiral algebras for $\NN=2$ SCFTs, and so it will be useful to view $\NN=3$ theories from an $\NN=2$ perspective. Therefore, we will pick an 
$\NN=2$ subalgebra of $\NN=3$, with the $\SU(3)_R \times \U(1)_r$ R-symmetry of the latter decomposing in
$\SU(2)_{\RN} \times \U(1)_{\rN} \times \U(1)_f$. The first two factors make up the R-symmetry of the $\NN=2$ superconformal algebra and the last corresponds to a global symmetry.
Therefore, from the $\NN=2$ point of view, all $\NN=3$ theories necessarily have a $\U(1)_f$ flavor symmetry arising from the extra R-symmetry currents. 
The additional supercharges and the $\U(1)_f$ flavor symmetry imply that the Virasoro symmetry expected in chiral algebras of $\NN=2$ theories algebras will be enhanced to a super Virasoro symmetry in the $\Nm=3$ case \cite{Nishinaka:2016hbw}.

Let us start reviewing the essentials of the chiral algebra construction (we refer the reader to \cite{Beem:2013sza} for more details). The elements of the protected sector are given by the cohomology of a nilpotent supercharge $\qq$ that is a linear combination of a Poincar\'{e} and a conformal supercharge,
\be 
\label{eq:Q}
\qq = \QQ_{\phantom{1}-}^1 + \tilde{S}^{2 \, \dot{-}}\,.
\ee
In order to be in the cohomology operators have to lie on a fixed plane $\mathbb{R}^2 \subset \mathbb{R}^4$. The global conformal algebra on the plane $\slf(2) \times \bar{\slf}(2)$ is a subalgebra of the four-dimensional conformal algebra. While the generators of the $\slf(2)$ commute with \eqref{eq:Q}, those of $\bar{\slf}(2)$ do not, and an operator in the cohomology at the origin will not remain in the cohomology if translated by the latter. However, it is possible to introduce twisted translations obtained by the diagonal subalgebra of the $\bar{\slf}(2)$ and a complexification, $\slf(2)_R$, of the R-symmetry algebra $\mathfrak{su}(2)_R$, such that the supercharge satisfies 
\be 
[\qq,\slf(2)] = 0\,, \qquad [\qq, \text{something}] = \text{diag}(\bar{\slf}(2) \times \slf(2)_R)\,.
\label{eq:twistedtransl}
\ee
From these relations one can prove that $\qq$-closed operators restricted to the plane have meromorphic correlators.
We call the operators that belong to the cohomology of $\qq$ ``Schur'' operators.
The Schur operators in $\calN=2$ language are local conformal primary fields which obey the conditions
\beq
\label{eq:SchurOperatorCondition}
\Delta-(j+ \jb) -2\RN =0 \,,\qquad \jb - j - \rN =0\,.
\eeq
The  cohomology classes of the twisted translations of any such operator $\OO$ corresponds to a $2d$ local meromorphic operator
\be 
\OO(z) = \left[ \OO(z,\zb) \right]_\qq\,.
\ee
The two important Schur operators that we expect to have in any $\NN=2$ theory with a flavor symmetry are\footnote{We follow the conventions of \cite{Dolan:2002zh} for $\NN=2$ superconformal multiplets.}
\begin{itemize}
\item $\hat{\CC}_{0(0,0)}$: The highest-weight component of the $\SU(2)_{\RN}$ current (with charges $\Delta=3, j=\bar \jmath=\tfrac{1}{2}, \RN=1$, $\rN=0$) corresponding to the $2d$ stress tensor $T(z)$. 
\item $\hat{\BB}_{1}$: The highest-weight component $J^{11}$ of the moment map operator ($\Delta=2, j=\bar \jmath=0, \RN=1$ and $\rN=0$) that is mapped to the affine current $J(z)$ of the flavor group.  
\end{itemize}
These two Schur operators give rise to a Virasoro and an affine symmetry in the chiral algebra respectively, with the two-dimensional central charges obtained in terms of their four-dimensional counterparts by
\be\label{c2dk2d}
c_{2d}= - 12 c_{4d}\,, \qquad k_{2d} = - \frac{k_{4d}}{2}\,.
\ee
Note that, since we insist on having unitarity in the four-dimensional theory, the $2d$ chiral algebra will be necessarily non-unitary.

The chiral algebra description of a protected subsector is extremely powerful. By performing the twist of \cite{Beem:2013sza} on a four-dimensional correlation function of Schur operators, we are left with a meromorphic $2d$ correlator that is completely determined by knowledge of its poles and residues. The poles can be understood by taking various OPE limits, thus fixing the correlator in terms of a finite number of parameters corresponding to OPE coefficients. In the cases we will study in this paper (see for example subsection \ref{sec:determinationoff}), the meromorphic piece can be fixed using crossing symmetry in terms of a single parameter, which can be identified with the central charge of the theory. Let us emphasize that this can be done without knowledge of which particular chiral algebra is relevant for the SCFT at hand. 


\subsection{Generalities of \texorpdfstring{$\NN =3 $}{N=3} chiral algebras}

Let us now study the $\NN=3$ case in more detail.
Any local $\NN=3$ SCFT will necessarily contain a stress tensor multiplet, which in table~\ref{tab:RepresentationsN3} corresponds to $\hat{\BB}_{[1,1]}$. 
After an $\NN=2$ decomposition of this multiplet (shown in \eqref{eq:Neq2decST}) one finds four terms, each contributing with a single representative to the chiral algebra. These four multiplets are related by the action of the extra supercharges enhancing $\NN=2$ to $\NN=3$, and four of these ($\QQ_{\phantom{3}+}^{3}$ and $\tilde{\QQ}_{3\, \dot{+}}$ and their conjugates) commute with $\qq$ \cite{Nishinaka:2016hbw}. Therefore, acting on Schur operators with these supercharges produces new Schur operators, and the representatives of the four multiplets will be related by these two supercharges. The multiplets and their representatives are:
\begin{itemize}
\item A multiplet containing the $\U(1)_f$ flavor currents ($\hat{\BB}_1$), whose moment map $M^{IJ}$ gives rise to a two-dimensional current $J(z) = \left[ M(z,\zb) \right]_{\qq}$ of an  $\U(1)_f$ affine Kac-Moody (AKM) algebra,
\item Two ``extra'' supercurrents, responsible for the enhancement to $\NN=3$, contribute as operators of holomorphic dimension $\tfrac{3}{2}$. These are obtained from the moment map  by the action of the supercharges $G(z)= \left[ \QQ_{\phantom{3}+}^{3} M(z,\zb)\right]_{\qq}$ and $\tilde{G}(z)= \left[ \tilde{\QQ}_{3\, \dot{+}} M(z,\zb) \right]_{\qq}$ \cite{Nishinaka:2016hbw}.\footnote{These arise from $\NN=2$ multiplets $\mbar{\DD}_{\tfrac{1}{2}(0,0)}$ and $\DD_{\tfrac{1}{2}(0,0)}$ respectively in the notation of \cite{Dolan:2002zh}.}
\item The stress-tensor multiplet ($\hat{\CC}_{0,(0,0)}$) which gives rise to the stress tensor of the chiral algebra $T(z) = \tfrac{1}{2} \left[ \left[\QQ_{\phantom{3}+}^{3} ,\tilde{\QQ}_{3\, \dot{+}} \right]M(z,\zb) \right]_{\qq}$ \cite{Nishinaka:2016hbw}.
\end{itemize}
The supercharges $\QQ_{\phantom{3}+}^{3}$ and $\tilde{\QQ}_{3\, \dot{+}}$ have charges $\pm 1$ under the $\U(1)_f$ flavor symmetry, where we follow the $\U(1)_f$ charge normalizations of \cite{Nishinaka:2016hbw}. Therefore the operators $G(z)$ and $\tilde{G}(z)$ have a $J$ charge of $+1$ and $-1$ respectively.
This multiplet content is exactly the one we would expect from the considerations in the beginning of this section, with the extra supercharges, that commute with $\qq$, producing a global $d=2$, $\NN=2$ superconformal symmetry.\footnote{The holomorphic $\slf(2)$ that commutes with the supercharge $\qq$,
 more precisely the $\qq$-cohomology of the superconformal algebra, is enhanced to a $\slf(2|1)$.}
Moreover, the operator content we just described corresponds precisely to the content of an $\NN=2$ stress tensor superfield which we denote by $\JJ$, enhancing the Virasoro algebra to an $\NN=2$ super Virasoro algebra \cite{Nishinaka:2016hbw}.

\subsubsection{$\NN=3$ superconformal multiplets containing Schur operators}
\label{subsec:schurop}
Our next task is to understand which multiplets of the $\NN=3$ superconformal algebra contribute to the chiral algebra, aside from the already discussed case of the stress-tensor multiplet.

Instead of searching for superconformal multiplets that contain conformal primaries satisfying \eqref{eq:SchurOperatorCondition}, we will take advantage of the fact that this was already done in \cite{Beem:2013sza} for $\NN=2$ multiplets, and simply search for $\NN=3$ multiplets that contain $\NN=2$ Schur multiplets. To accomplish this, we decompose $\NN=3$ multiplets in $\NN=2$ ones by performing the decomposition of the corresponding characters. In appendix~\ref{app:shortening} we present a few examples of such decompositions.
Going systematically through the multiplets,\footnote{One can quickly see that in table~\ref{tab:RepresentationsN3} multiplets that obey no $\NN=3$ shortening conditions on the one of the sides also obey no $\NN=2$ shortening condition on one of the sides, and these are known \cite{Beem:2013sza} not to contain Schur  operators. Therefore we must go only through the multiplets that obey shortening conditions on both sides.} we find the following list of $\NN=3$ Schur multiplets:
\begin{small}
\begingroup
\allowdisplaybreaks[1]
\begin{align}
\hat{\CC}_{[R_1,R_2],(j,\jb)}\vert_{\mathrm{Schur}}=& u_f^{R_2-R_{1}+2(\jb-j)}\Bigg[\hat{\CC}_{\tfrac{R_1+R_2}{2},(j,\jb)} \oplus u_f^{-1} \hat{\CC}_{\tfrac{R_1+R_2}{2},(j,\jb+\tfrac{1}{2})} \oplus u_f \, \hat{\CC}_{\tfrac{R_1+R_2}{2},(j+\tfrac{1}{2},\jb)} \nn \\
\label{eq:CCSchur}
&\oplus \hat{\CC}_{\tfrac{R_1+R_2}{2},(j+\tfrac{1}{2},\jb+\tfrac{1}{2})}\Bigg]\,,\\
\hat{\BB}_{[R_1,R_2]}\vert_{\mathrm{Schur}}=& u_f^{R_2-R_1}\Bigg[\hat{\BB}_{\tfrac{R_1+R_2}{2}} \oplus u_f^{-1} \DD_{\tfrac{R_1+R_2-1}{2},(0,0)} \oplus u_f\, \mbar{\DD}_{\tfrac{R_1+R_2-1}{2},(0,0)}\nonumber \\
\label{eq:BBSchur}
&\oplus \hat{\CC}_{\tfrac{R_1+R_2-2}{2},(0,0)}\Bigg]\,,\qquad \text{ for } R_1 R_2 \neq 0\,,\\
\label{eq:Bschurantichiral}
\hat{\BB}_{[R_1,0]}\vert_{\mathrm{Schur}} =& u_f^{-R_1}\left[\hat{\BB}_{\tfrac{R_1}{2}} \oplus u_f \mbar{\DD}_{\tfrac{R_1-1}{2},(0,0)}\right] \,,\\
\label{eq:Bschurchiral}
\hat{\BB}_{[0,R_2]}\vert_{\mathrm{Schur}} =& u_f^{R_2}\left[\hat{\BB}_{\tfrac{R_2}{2}} \oplus u_f^{-1} \DD_{\tfrac{R_2-1}{2},(0,0)}\right] \,,\\
\DD_{[R_1,R_2],\jb}\vert_{\mathrm{Schur}}&=u_f^{R_2-R_1+2\jb+2}\Bigg[ \DD_{\tfrac{R_1+R_2}{2},(0,\jb)} \oplus u_f^{-1} \DD_{\tfrac{R_1+R_2}{2},(0,\jb+\tfrac{1}{2})} \oplus u_f \hat{\CC}_{\tfrac{R_1+R_2-1}{2},(0,\jb)} \nn \\
\label{eq:DSchur}
&\oplus \hat{\CC}_{\tfrac{R_1+R_2-1}{2},(0,\jb+\tfrac{1}{2})}\Bigg]\qquad \text{ for } R_1>0\,,
\\
 \mbar{\DD}_{[R_1,R_2],j}\vert_{\mathrm{Schur}}&=u_f^{R_2-R_1-2j-2}\Bigg[ \mbar{\DD}_{\tfrac{R_1+R_2}{2},(j,0)} \oplus u_f^{-1} \hat{\CC}_{\tfrac{R_1+R_2-1}{2},(j,0)} \oplus u_f \mbar{\DD}_{\tfrac{R_1+R_2}{2},(j+\tfrac{1}{2},0)} \nn \\
\label{eq:DbSchur}
&\oplus \hat{\CC}_{\tfrac{R_1+R_2-1}{2},(j+\tfrac{1}{2},0)}\Bigg]\qquad \text{ for } R_2>0\,,
\\
\label{eq:DSchurR10}
 \DD_{[0,R_2],\jb}\vert_{\mathrm{Schur}}&=u_f^{R_2+2\jb+2}\left[ \DD_{\tfrac{R_2}{2},(0,\jb)} \oplus u_f^{-1} \DD_{\tfrac{R_2}{2},(0,\jb+\tfrac{1}{2})}  \right] \,, \\
 \label{eq:DbSchurR20}
  \mbar{\DD}_{[R_1,0],j}\vert_{\mathrm{Schur}}&=u_f^{-R_1-2j-2}\left[ \mbar{\DD}_{\tfrac{R_1}{2},(j,0)} \oplus  u_f \mbar{\DD}_{\tfrac{R_1}{2},(j+\tfrac{1}{2},0)}\right] \,.
\end{align}
\endgroup
\end{small}
\noindent
Let us stress again that we are not showing the full decomposition in $\NN=2$ multiplets, but only the Schur multiplets. 
In performing the decompositions we kept the grading of the $\NN=2$ multiplets with respect to the $U(1)_f$ flavor symmetry, denoting the corresponding fugacity by $u_f$.

Some noteworthy multiplets in this list are the stress-tensor multiplet $\hat{\BB}_{[1,1]}$, already discussed in the beginning of this subsection, as well as the half-BPS operators $\hat{\BB}_{[R_1,0]}$ (and their conjugates $\hat{\BB}_{[0,R_1]}$) which are connected to the Coulomb branch, as discussed in section \ref{sec:intro}. Due to their physical significance we present their full decomposition in $\NN=2$ multiplets in \ref{eq:Neq2decST} and \ref{eq:BR0fulldec}. 
As described in \cite{Cordova:2016xhm}, there are no \textit{relevant} Lorentz invariant supersymmetric deformations of $\NN=3$ theories, while the only such deformations that are \textit{exactly marginal} are contained in the multiplet $\hat{\BB}_{[2,0]}$  (and conjugate $\hat{\BB}_{[0,2]}$). However, these multiplets also contain additional supersymmetry currents, as can be seen from their $\NN=2$ decomposition, that allow for the enhancement of $\NN=3$ to $\NN=4$, and thus pure $\NN=3$ theories are not expected to have exactly marginal operators.
Let us also recall that the multiplets $\hat{\CC}_{[0,0],(j,\bar\jmath)}$ contain conserved currents of spin larger than two, and therefore are expected to be absent in interacting theories \cite{Maldacena:2011jn,Alba:2013yda}.

\subsubsection*{Quasi-primaries and Virasoro primaries}

Each of the $\NN=2$ multiplets listed above will contribute to the chiral algebra with exactly one \emph{global conformal primary} (also called quasi-primary), with holomorphic dimension as given in table~1 of \cite{Beem:2013sza} and with $\U(1)_f$ charge $f$, under the $J(z)$ current, as can be read off from the $u_f$ fugacity in the above decompositions. These multiplets generically will not be Virasoro primaries.
Only the so-called Hall-Littlewood (HL) operators\footnote{Following \cite{Beem:2013sza} we refer to operators which are $\NN=1$ chiral and satisfy the Schur condition as Hall-Littlewood operators.} ($\hat{\BB}_R$, $\mbar{\DD}_{R,(j,0)}$ and $\DD_{R,(0,\bar{\jmath})}$) are actually guaranteed to be Virasoro primaries. The remaining multiplets will appear in the chiral algebras sometimes as Virasoro primaries, sometimes only as quasi-primaries.

\subsubsection*{Super Virasoro primaries}

Similarly, each $\NN=3$ multiplet gives rise in the chiral algebra to a global supermultiplet consisting of a global superprimary and its three global superdescendants obtained by the action of $\QQ_{\phantom{3}+}^{3}$ and $\tilde{\QQ}_{3\, \dot{+}}$.\footnote{Recall that the global superprimary is annihilated only by the $G_{\tfrac{1}{2}}, \tilde{G}_{\tfrac{1}{2}}, L_{1}$ modes of $G(z), \tilde{G}(z), T(z)$, and global super descendants are obtained by the action of $G_{-\tfrac{1}{2}}$ and  $\tilde{G}_{-\tfrac{1}{2}}$}
Generically however, these multiplets will not be super Virasoro primaries, even if the global superprimary corresponds to a Virasoro primary.
Recall that a super Virasoro primary must, in addition to being a Virasoro primary, have at most a pole of order one in its OPE with both $G(z)$ and $\tilde{G}(z)$, and have at most a singular term of order one in the OPE with $J(z)$.\footnote{These conditions translate into the following modes annihilating the superprimary state: $L_{n \geqslant 0}$, $G_{n\geqslant +\tfrac{1}{2}}, \tilde{G}_{n\geqslant +\tfrac{1}{2}}$ and $J_{n \geqslant 0}$.} This last condition corresponds to being an AKM primary.

Let us consider the operators which have as a global superprimary a Virasoro primary. 
For the case of $\hat{\BB}_{[R_1,R_2]}$ multiplets, we see that its two (or one in case $R_1 R_2=0$) level $\tfrac{1}{2}$ descendants are HL operators, and thus Virasoro primaries. The two-dimensional superconformal algebra then implies that the global superprimary is not only a Virasoro primary, but that it is also annihilated by all the modes $G_{n\geqslant +\tfrac{1}{2}}, \tilde{G}_{n\geqslant +\tfrac{1}{2}}$. However, this is not enough to make it a super Virasoro primary, as it is not guaranteed that these operators are AKM primaries. An obvious example is the stress tensor multiplet $\hat{\BB}_{[1,1]}$, where the AKM current is clearly not an AKM primary. Similar considerations apply to $\DD_{[R_1,R_2],\jb}$ multiplets, with the subtlety that even though one of its level $\tfrac{1}{2}$ descendants is not a HL operator, it is still a Virasoro primary \cite{Nishinaka:2016hbw}.

In certain cases it is possible to show that the operators in question are actually super Virasoro primaries, and concrete examples will be given below. For example, if one considers a $\hat{\BB}_{[R_1,R_2]}$ generator that is not the stress tensor multiplet, then the OPE selection rules for the $\NN=2$ $\hat{\BB}_{\tfrac{R_1+R_2}{2}}$ multiplet \cite{Arutyunov:2001qw} imply it is also an AKM primary \cite{Lemos:2014lua}.
\subsubsection*{Chiral and anti-chiral operators}

Finally we note that the multiplets in \eqref{eq:Bschurantichiral} and \eqref{eq:Bschurchiral} give rise, in two dimensions, to anti-chiral and chiral operators: they are killed by $\tilde{\QQ}_{3\, \dot{+}}$ and $\QQ_{\phantom{3}+}^{3}$ respectively. These two-dimensional superfields have holomorphic dimension satisfying $h=\tfrac{R_2}{2}=-\tfrac{f}{2}$ and  $h=\tfrac{R_1}{2}=\tfrac{f}{2}$ respectively.


\subsection{\texorpdfstring{$[3,0]$}{[3,0]} chiral algebras}
\label{subsec:keq3chiralalg}

We are now in a position to describe the general features of the chiral algebras associated to the known $\NN=3$ theories.
We will describe the chiral algebra in terms of its generators, by which we mean operators that cannot be expressed as normal-ordered products and/or (super)derivatives of other operators. In what follows we assume the chiral algebra to be finitely generated.
Although there is yet no complete characterization of what should be the generators of the chiral algebra of a given four-dimensional theory, it was shown in \cite{Beem:2013sza} that all generators of the HL chiral ring are generators of the chiral algebra.
Moreover, the stress tensor is always guaranteed to be present and, with the exception of cases where a null relation identifies it with a composite operator, it must always be a generator.
However this is not necessarily  the complete list, and indeed examples with more generators than just the above have been given in \cite{Beem:2013sza,Lemos:2014lua}.\footnote{
One possible way to determine which generators a given chiral algebra should have is through a Schur index \cite{Kinney:2005ej,Gadde:2011ik,Gadde:2011uv,Rastelli:2014jja} analysis, as done in \cite{Beem:2013sza,Lemos:2014lua}.
}
The chiral algebras associated to $4d$ SCFTs do not always correspond to known examples in the literature, and in such situations one must construct a new associative two-dimensional chiral algebra. This problem can be bootstrapped by writing down the most general OPEs for the expected set of generators and then imposing associativity by solving the Jacobi identities. Chiral algebras are very rigid structures and in the cases so far considered \cite{Lemos:2014lua,Nishinaka:2016hbw}, the Jacobi identities are powerful enough to completely fix all OPE coefficients, including the central charges.

\subsubsection*{Rank one chiral algebras}

In \cite{Nishinaka:2016hbw}, the authors assumed that the only generators of the chiral algebras corresponding to the rank one $\NN=3$ SCFTs described in section \ref{sec:intro} (with $k=\ell,$ $N=1$) were the stress tensor and the generators of its Higgs branch:
\be 
\hat{\BB}_{[1,1]}\,, \qquad \hat{\BB}_{[\ell,0]}\,, \qquad   \hat{\mathcal{B}}_{[0,\ell]}\,, \qquad \ell =3,4\,.
\ee
Recall the first multiplet gives rise, in two dimensions, to the stress tensor multiplet, and the last two to anti-chiral and chiral operators respectively.
With these assumptions they were able to write an associative chiral algebra for the cases $\ell=3,4$ only for a single central charge for the first case and a finite set of values for the second. 
This set was further restricted to the correct value expected for the known $\NN=3$ theories 
\be
c_{4d}=a_{4d} = \frac{2 \ell -1}{4}\,,
\label{eq:rankone}
\ee
by imposing the expected Higgs branch chiral ring relation 
\be 
\hat{\BB}_{[\ell,0]} \hat{\BB}_{[0,\ell]} \sim  \left(\hat{\BB}_{[1,1]}\right)^\ell\,,
\ee
which appears as a null state in the chiral algebra.
Note that above, by abuse of notation, we denoted the Higgs branch chiral ring operator by the superconformal multiplet it belongs to.
Associativity then fixes all other OPE coefficients of the chiral algebra.
The authors of \cite{Nishinaka:2016hbw} were also able to construct an associative chiral algebra for $\ell=5$ and $\ell=6$ satisfying the Higgs branch relation if the central charge is given by \eqref{eq:rankone}. However, as they point out, $\ell=5$ does not correspond to an allowed value for an $\NN=3$ SCFTs, as five is not an allowed scaling dimension for the Coulomb branch of a rank one theory, following from Kodaria's classification of elliptic surfaces (see, \eg, \cite{Argyres:2015ffa,Nishinaka:2016hbw}).
The case $\ell=6$ is in principle allowed, however no such $\Nm=3$ theory was obtained in the S-fold constructions of \cite{Aharony:2016kai}.\footnote{We emphasize that the existence of a two-dimensional chiral algebra does not imply that there exists a four-dimensional theory that gives rise to it. In fact it is still not clear what are the sufficient conditions for a chiral algebra to correspond to a physical four-dimensional theory.}
\subsubsection*{Higher rank theories}

We now attempt to generalize the chiral algebras of \cite{Nishinaka:2016hbw} to the higher-rank case (with $k=\ell,$ $N>1$). In particular, we focus on the theories whose lowest dimensional generator corresponds to a  $\hat{\BB}_{[3,0]}$ and its conjugate, since these are the ones relevant for the following sections.
To compute OPEs and Jacobi identities we will make extensive use of the Mathematica package \cite{Krivonos:1995bk}. Following its conventions, we use the two-dimensional $\NN=2$ holomorphic superspace with bosonic coordinate $z$ and fermionic coordinates $\theta$ and $\bar{\theta}$, and define the superderivatives as
\be
\DD = \partial_{\theta} - \tfrac{1}{2} \bth \partial_z \,, \qquad \DDb = \partial_{\bth} - \tfrac{1}{2} \theta \partial_{z} \,.
\ee
We will denote the two-dimensional generators arising from the half-BPS Higgs branch generators $\hat{\BB}_{[0,3]}$ ($\hat{\mathcal{B}}_{[3,0]}$) by $\WW$ ($\WB$).\footnote{Note that in \cite{Krivonos:1995bk} what is called chiral primary is what we call anti-chiral primary, \eg, $\WB$ which obeys $\DD \WB =0$. } Furthermore,  we denote the two-dimensional superfield arising from the stress tensor ($\hat{\BB}_{[1,1]}$) by $\JJ$.
The OPE of $\JJ$ with itself is fixed by superconformal symmetry,
\be 
\JJ(Z_1) \JJ(Z_2) \sim \frac{c_{2d}/3 + \th_{12} \bth_{12} \JJ}{Z_{12}^2} + \frac{- \th_{12} \DD \JJ +\bth_{12} \DDb \JJ + \th_{12} \bth_{12} \partial \JJ}{Z_{12}}\,,
\label{eq:STOPE}
\ee
where we defined
\be 
Z_{ij}= z_1 - z_2 + \tfrac{1}{2} \left(\th_1 \bth_2 - \th_2 \bth_1 \right) \,, \qquad \th_{12} = \th_1 - \th_2 \,, \qquad \bth_{12}=\bth_1 - \bth_2\,.
\ee
The OPEs of $\JJ$ with $\WW$ and $\WB$, given in \eqref{eq:superPrimaryOPE},  are fixed by demanding that these two operators be super Virasoro primaries. As discussed in the previous subsection, $\WW$ and $\WB$ could fail to be super Virasoro primaries only if their global superprimary (arising from an $\NN=2$ $\hat{\BB}_{3/2}$) failed to be an AKM primary. However, since we are assuming the $\hat{\BB}_{3/2}$ multiplet to be a generator, and since the AKM current comes from a $\hat{\BB}_1$ $\NN=2$ multiplet, it is clear from the selection rules of $\NN=2$ $\hat{\BB}_R$ operators \cite{Arutyunov:2001qw} that these must be AKM primaries.

The self OPEs of the chiral (anti-chiral) $\WW$ ($\WB$) superfields are regular, which is consistent with the $\NN=3$ OPE selection rules shown in \ref{chiralchiralselectionsrules}.
For $\WW \WB$ the most general OPE in terms of all of the existing generators is \cite{Nishinaka:2016hbw}
\be 
\WW(Z_1)\WB(Z_2) \sim \sum_{h=0}^2 \frac{1}{Z_{12}^{3-h}}\left( \frac{3-h}{2} \frac{\th_{12} \bth_{12}}{Z_{12}} + 1 + \th_{12} \DD \right)\lambda_{\OO_h} \OO_h \,,
\label{eq:WWBOPE}
\ee
where the sum runs over all uncharged operators, including composites and (super)derivatives.

The authors of \cite{Nishinaka:2016hbw} showed that, considering just these three fields as generators, one finds an associative chiral algebra only if $c_{2d}=-15$, which indeed corresponds to the correct value for the simplest known non-trivial $\NN=3$ SCFT ($k=\ell=3$ and $N=1$ in the notation of \cite{Aharony:2016kai}). However, there are higher rank versions of this theory ($k=\ell=3$ and $N>1$), that contain these half-BPS operators plus higher-dimensional ones. The list of half-BPS operators is \cite{Aharony:2016kai}
\be 
\hat{\BB}_{[0,R]}\,, \quad \hat{\BB}_{[R,0]}\,, \quad \text{with} \; R=3,6,\ldots, 3N\,,
\label{eq:halfBPSgen}
\ee
giving rise in two dimensions to additional chiral and antichiral operators with charges $f=\pm 6, \ldots \pm 6N$, and holomorphic dimension $h=|f|/2$.
One can quickly see that the extra generators never appear in the OPEs of $\WW, \WB, \JJ$, as the only OPE not fixed by symmetry is the $\WW \WB$, and $\U(1)_f$ charge conservation forbids any of the  $\hat{\BB}_{[R,0]}$ with $R \geqslant 6$ to appear.
If the generators of the chiral algebras of higher rank theories corresponded \emph{only} to the half-BPS operators plus the stress tensor, then we would reach a contradiction: $\WW, \WB, \JJ$ would form a closed subalgebra of the full chiral algebra, but the central charge would be frozen at $c_{2d}=-15$, which is not the correct value for rank greater than one.

To resolve this contradiction we must allow for more generators in the higher-rank case, and at least one of these must be exchanged in the $\WW \WB$ OPE.
The \emph{only} freedom in this OPE is to add an uncharged dimension two generator. From the OPE selection rules shown in \ref{selectionchiantichi} one can see that this operator must correspond to a $\hat{\mathcal{B}}_{[2,2]}$. There is another possibility, namely a $\hat{\CC}_{[0,0],0}$ multiplet, but in four dimensions it contains conserved currents of spin greater than two, which should be absent \cite{Maldacena:2011jn,Alba:2013yda} in interacting theories such as the ones we are interested in.
The minimal resolution is to add the generator corresponding to $\hat{\mathcal{B}}_{[2,2]}$. We then \emph{assume} that the generators of the chiral algebra associated with the $\ell=k=3$ theories with $N > 1$ are
\begin{itemize}
\item The stress tensor $\JJ$,
\item (Anti-)chiral operators arising from the generators of the Coulomb branch operators $\hat{\BB}_{[0,R]}$ ($\hat{\BB}_{[R,0]}$) with $R=3,6,\ldots, 3N$,
\item A generator corresponding to $\hat{\BB}_{[2,2]}$ which we denote by $\OOh2$.
\end{itemize}
As before we denote by $\WW$ and $\WB$ the generators arising from $\hat{\BB}_{[0,3]}$ and its conjugate.\footnote{The fact that we do not allow for any other operator of dimension one (or smaller) prevents the symmetry of the chiral algebra from enhancing to the small $\NN=4$ superconformal algebra one gets from $4d$ $\NN=4$ theories, thereby excluding $\NN=4$ solutions from our analysis. And by not allowing for additional dimension $3/2$ generators we also exclude discretely gauged versions of $\NN=4$.}
Even though examples are known where the number of generators not arising from generators of the HL ring grows with the number of HL generators \cite{Lemos:2014lua}, the addition of a single operator $\OOh2$ is the minimal modification that unfreezes the value of the central charge.

We can now proceed to write down the most general OPEs, it is easy to check that in the ones involving
\be
\JJ\,, \quad \WW\,, \quad \WB \,,\;\; \text{and} \quad  \OOh2\,,
\label{eq:generatorschiralalg}
\ee
the operators in \eqref{eq:halfBPSgen} with $R\geqslant6$ cannot be exchanged.
Therefore, if our assumption above is correct, the generators in \eqref{eq:generatorschiralalg} form a closed subalgebra.

In what follows we write down the most general ansatz for the OPEs of these operators which, as explained above, are all super Virasoro primaries with the exception of $\JJ$.
The regularity of the self OPEs of $\WW$ and $\WB$ follows simply from OPE selection rules, while the OPE between $\WW$ and $\WB$ is given by \eqref{eq:WWBOPE}, allowing for the exchange of $\OOh2$ as well.
The OPEs involving $\OOh2$ are quite long and therefore we collect them in appendix \ref{app:chiral_algebra}.
Imposing Jacobi identities we were able to fix all the OPE coefficients in terms of a single coefficient: the central charge. In our construction we did not need to impose null states for closure of the algebra.


\subsection{Fixing OPE coefficients}
\label{sec:fixingB33}

In the next sections we will study numerically the \emph{complete} four-point function of two $\hat{\BB}_{[3,0]}$  and two $\hat{\BB}_{[0,3]}$ operators, thanks to the chiral algebra we can compute the OPE coefficients of all operators appearing in the right hand side of the $\WW \WB$ OPE.
However, we still need to identify the four-dimensional superconformal multiplet that each two-dimensional operator corresponds to.
Let us start by examining the low dimensional operators appearing in this OPE: we can write all possible operators with a given dimension that can be made out of the generators by normal ordered products and (super) derivatives. Furthermore, they must be uncharged, since the product $\WW\WB$ is. 
All in all we find the following list:
\begin{center}
\begin{tabular}{cc}
\hline\hline
dimension & operators\\ 
\hline 
0 & Identity \\
1 & $\JJ$  \\
2 &  $\OOh2$, $\JJ\JJ$, $\DD\DDb \JJ$, $\JJ'$ \\
3 &  $\WW \WB$, $\JJ \DD \DDb \JJ$, $\JJ''$, $\JJ' \JJ$, $\JJ \JJ\JJ$, $\DD \DDb \JJ'$, $\DD \JJ \DDb \JJ$, $\DD\DDb \OOh2$, $\JJ \OOh2$, $\OOh2'$ \\
\ldots & \ldots \\
\hline 
\end{tabular} 
\end{center}
From these operators we are only interested in the combinations that are global superprimary fields, as the contributions of descendants will be fixed from them.\footnote{Note that this is only true because $\WW$ and $\WB$ are chiral and anti-chiral, and therefore their three-point function with an arbitrary superfield has a unique structure, being determined by a single number.}
Note also that, if we are interested in the four point function of $\langle \WW \WB \WW \WB \rangle$, we only see, for the exchange of an operator of a given dimension, a sum of the contributions of all global primaries, and we cannot distinguish between individual fields. 

At dimension $h=1$ there is only one operator -- the superprimary of the stress-tensor multiplet -- and its OPE coefficient squared can be computed to be (after normalizing the identity operator to appear with coefficient one in the four-point function decomposition, and normalizing the $\JJ$ two-point function to match the normalization for the blocks ($\Ssl$, see \eqref{eq:OSP22superblocks}) that we use in the following sections)
\be\label{lambdaWWbJJsec2}
\big|\lambda_{\WW \WB \JJ}\big|^2 = - \frac{27}{c_{2d}}\,.
\ee
This does not depend on the particular chiral algebra at hand, as the OPE coefficient with which the the current $\JJ$ is exchanged, is totally fixed in terms of their charge $f$ and the central charge.
As we will show in \ref{sec:determinationoff}, the two-dimensional correlation function of the two $\WW$ and two $\WB$, is fixed in terms of one parameter which we take to be the OPE coefficient of $\JJ$, and thus related to $c_{2d}$. This implies that, for the exchange of operators of dimension larger than one, any sum of OPE coefficients corresponds to a universal function of $c_{2d}$.

At dimension $h=2$ we find two global superprimaries, one corresponding to $\OOh2$ itself, and the other containing $\JJ \JJ$. From the four-dimensional OPE selection rules, shown in \eqref{selectionchiantichi}, it follows that both superprimaries must correspond to $\hat{\BB}_{[2,2]}$ supermultiplets in four dimensions, as the only other option is $\hat{\CC}_{[0,0],0}$ which should be absent in interacting theories.
This means that, even from the point of view of the four-dimensional correlation function, these two operators are indistinguishable. Thus, all we can fix is the sum of two OPE coefficients squared:
\be 
\big|\lambda_{\WW \WB \OOh2}\big|^2 + \big|\lambda_{\WW \WB (\JJ\JJ)}\big|^2 = -\frac{18}{c_{2d}}\,,
\ee
where we used the same normalizations as before, and fixed an orthonormal basis for the operators.
This number is again independent of the particular details of the chiral algebra: it only requires the existence of $\WW$, $\WB$ and $\JJ$.

At dimension $h=3$,  we find four global superprimaries made out of the fields listed above, three of which are Virasoro primaries.
In this case, however, these  three different operators must belong to two different types of four-dimensional multiplets (once again we are excluding the multiplet containing higher-spin currents). Namely, they must correspond to $\hat{\BB}_{[3,3]}$ and $\hat{\CC}_{[1,1],0}$, and distinguishing them from the point of view of the chiral algebra is hard.
The two-dimensional operators arising from $\hat{\BB}_{[3,3]}$  are guaranteed to be Virasoro primaries, while those of $\hat{\CC}_{[1,1], 0}$ could be or not.
\emph{Assuming} that all Virasoro primaries come exclusively from $\hat{\BB}_{[3,3]}$ we can compute the OPE coefficient with which this multiplet is exchanged by summing the squared OPE coefficients of all Virasoro primaries
\be
\label{eq:B33inchiralalg} 
\sum_{i=1}^{3} \big|\lambda_{\JJ \JJ \mathrm{Vir}_{i,\, h=3}}\big|^2  = \frac{2 (c_{2d} (5 c_{2d}+127)+945)}{5 c_{2d} (2 c_{2d}+13)}\,.
\ee
We can take the large $c_{4d}=-\tfrac{c_{2d}}{12}$ limit, where the solution should correspond to generalized free field theory. In this case we can find from the four-point function given in appendix \ref{app: generalized free theory} that the OPE coefficient above should go to $1$, and indeed this is the case.
We could also have assumed that different subsets of the three Virasoro primaries correspond to  $\hat{\BB}_{[3,3]}$. Not counting the possibility used in \eqref{eq:B33inchiralalg}, there is one possibility which does not have the correct behavior as $c_{4d} \to \infty$, and two that have:
\begin{align}
\label{eq:B33inchiralalg2} 
\sum_{i=1}^{2} \big|\lambda_{\JJ \JJ \mathrm{Vir}_{i,\, h=3}}\big|^2  &= \frac{351378-10 c_{2d} (c_{2d} (c_{2d} (c_{2d}+22)-260)-8430)}{(c_{2d}-1) c_{2d} (2 c_{2d}+13) (12-5 c_{2d})}\,,\\
\label{eq:B33inchiralalg3} 
\sum_{i=2}^{3} \big|\lambda_{\JJ \JJ \mathrm{Vir}_{i,\, h=3}}\big|^2  &= \frac{2 (c_{2d}+15) (c_{2d} (c_{2d} (5 c_{2d}+37)+39)+4482)}{5 (c_{2d}-1) c_{2d} (c_{2d}+6) (2 c_{2d}-3)}\,.
\end{align}
We can now also compute for each of the above cases the OPE coefficient of the $\hat{\CC}_{[1,1],0}$ multiplet, and we find that only \eqref{eq:B33inchiralalg} and \eqref{eq:B33inchiralalg3} are compatible with $4d$ unitarity (the precise relation between $2d$ and $4d$ OPE coefficients is given by \eqref{lambdafromb}).

If we now go to higher dimension, the list of operators keeps on growing, and their four-dimensional interpretation is always ambiguous. A dimension $h$ global superprimary can either be a $\hat{\CC}_{[2,2], h-4}$ or a $\hat{\CC}_{[1,1], h-3}$ four-dimensional multiplet, and in this case there does not seem to be an easy way to resolve the ambiguity.\footnote{One possibility would be to find two sets of OPEs such that in each set, one of the above multiplets is forbidden to appear by selection rules.}

\subsubsection*{Rank one case}

Let us now comment on what happens for the case of the rank one theory, where $c_{2d}=-15$ and the extra generator $\OOh2$ is absent.
In this case we find a single (non-null) Virasoro primary at dimension three.\footnote{There is another Virasoro primary, which is a composite operator that is null for this central charge. This null corresponds precisely to the Higgs branch relation of the form $\WW \WB \sim \JJ \JJ^3$ described in \cite{Nishinaka:2016hbw}.}  This implies that either there is no  $\hat{\BB}_{[3,3]}$ multiplet and that the OPE coefficient is zero, or, which seems like a more natural option, that the Virasoro primary corresponds to this multiplet, with OPE coefficient
\be 
\big|\lambda_{\JJ \JJ \mathrm{Vir}_{h=3}}\big|^2 = \frac{22}{85}\,.
\ee
The above corresponds to setting $c_{2d}=-15$ in both \eqref{eq:B33inchiralalg} and \eqref{eq:B33inchiralalg2}, as expected since for this value the extra generator is not needed and decouples. The possibility that there is no $\hat{\BB}_{[3,3]}$ multiplet in the rank one theory and thus that the OPE coefficient is zero
corresponds to the $c_{2d}=-15$ case of \eqref{eq:B33inchiralalg3}.
If this last possibility were  true, then we would have that the operator $\WW \WB \sim \JJ^3$ is not in the Higgs branch, since Higgs branch operators correspond to $\hat{\BB}_R$ multiplets in $\NN=2$ language. Hence, there would be a relation setting $\JJ^3 =0$ in the Higgs branch, which does not seem plausible. 
In any case, we will allow for \eqref{eq:B33inchiralalg3} for generic values of the central charge. It might be possible to select among the two options (\eqref{eq:B33inchiralalg} and \eqref{eq:B33inchiralalg3}) by making use of the considerations in \cite{ChrisLeonardounpublished} about recovering the Higgs branch out of the chiral algebra, but we leave this for future work.

\section{Superblocks}
\label{sec:superblocksmain}

In this section we will use harmonic 
superspace techniques in order to study correlation functions of half-BPS operators. We will follow closely \cite{Liendo:2015cgi,Liendo:2016ymz}, where a similar approach was used to study correlation functions in several superconformal setups.

\subsection{Superspace}
We introduce the  superspace  $\mathscr{M}$ as a coset
$\mathscr{M}\,\simeq\, \SL(4|3)\,\big{/}\,G_{\leq 0}$.
Here, the factor $G_{\leq 0}$ corresponds to lower triangular block matrices with respect to the decomposition given in 
\eqref{eq:definitionofcosetrepresentative} below.
We take $E(\mathbf{p})\in G_{>0}$ as coset representative explicitly given by
\beq
\label{eq:definitionofcosetrepresentative}
E(\mathbf{p}):=
\exp\text{\footnotesize{$
\begin{pmatrix}
0_{(2|1)} & \SV & X \\
0 & 0 & \overline{\SV}\\
0 & 0 &  0_{(2|1)}
\end{pmatrix}$}}\,=\,
\text{\footnotesize{$\begin{pmatrix}
\mathbf{1}_{(2|1)} & \SV & X_+ \\
0 & 1 & \overline{\SV}\\
0 & 0  &  \mathbf{1}_{(2|1)} 
\end{pmatrix}$}}\,,
\eeq
where
\beq
\label{Xandstuff}
X\,=\,
\begin{pmatrix}
x^{\alpha \dot{\alpha}} & \lambda^{\alpha} \\
\pi^{\dot{\alpha}} & y
\end{pmatrix}\,,
\qquad
\SV\,=\,
\begin{pmatrix}
\theta^{\alpha} \\
v
\end{pmatrix}\,,
\qquad
\overline{\SV}\,=\,
\begin{pmatrix}
\bar{\theta}^{\dot \alpha} &
\bar{v}
\end{pmatrix}\,.
\eeq
In the above, $\alpha \in\{1,2\}$, $\dot{\alpha}\in\{1,2\}$ are the familiar Lorentz indices and the coordinates 
$\{\lambda^{\alpha},\pi^{\dot\alpha}, \theta^{\alpha}, \bar{\theta}^{\dot\alpha}\}$ are fermionic,
while the $y,v,\bar{v}$ are bosonic R-symmetry coordinates.
The action of $\SL(4|3)$ on this superspace follows from the coset construction and  is summarized in table \ref{tab:Xpmtransf}.
\begin{table}
\centering
\renewcommand{\arraystretch}{1.6}
\begin{tabular}{%
| l
                |>{\centering }m{2.2cm}
     |>{\centering }m{2.2cm}
        |>{\centering }m{2.2cm}
 |>{\centering }m{2.35cm}
             |>{\centering\arraybackslash}m{2.35cm}|
}
\hline
& 
$G_{(+,0)}$
&
$G_{(0,+)}$
&
$G_{(0,0)}$
&
$G_{(-,0)}$
&
$G_{(0,-)}$
\\\hline
\footnotesize{$\!g\circ (X_+,\overline{\SV})\!\!$}
&
\footnotesize{$ \!(X_+\!+\!b\overline{\SV},\overline{\SV})\!$}
&
\footnotesize{$\!(X_+,\overline{\SV}+\bar{b})\!$}
&
\footnotesize{$(AX_+,\overline{\SV})D^{-1}\!\!$}
&
\footnotesize{$(X_+, \overline{\SV}\!+\!\bar{c}X_+)$}
&
\footnotesize{$(X_+,\overline{\SV})\bar{h}\!\!\!$}
\\\hline
\footnotesize{$\!g\circ (X_-,\SV)\!\!$}
&
\footnotesize{$\!(X_-,\SV+b)\!$}
&
\footnotesize{$\!(X_-\!-\!\SV\bar{b},\SV)\!$}
&
\footnotesize{$A(X_-D^{-1},\SV)\!\!$}
&
\footnotesize{$h(X_-,\SV)\!\!\!$}
&
\footnotesize{$(X_-, \SV-X_-c)$}
\\\hline
\end{tabular}
\renewcommand{\arraystretch}{1.0}
\caption{
We used the definitions 
$h:=(\mathbf{1}_3+\SV \bar{c})^{-1}$ and
$\bar{h}:=(\mathbf{1}_3+c\overline{\SV})^{-1}$.
The transformations corresponding to $G_{(+,+)}$, $G_{(-,-)}$  are generated by the ones above.
For convenience we give the explicit form of special superconformal transformations $G_{(-,-)}$:
$(X_+,\overline{\SV})\mapsto (X_+,\overline{\SV})(1+CX_+)^{-1}$ and 
$(X_-,\SV)\mapsto (1+X_-C)^{-1}(X_-,\SV)$.}
\label{tab:Xpmtransf}
\end{table}
Notice that $\SL(4|3)$ acts invariantly within the superspaces $\mathscr{M}_+$, $\mathscr{M}_-$
with coordinates $\{X_+,\overline{\SV}\}$, $\{X_-,\SV\}$
 respectively, where we have defined $X_{\pm}=X\pm\tfrac{1}{2}\SV\overline{\SV}$.
The basic covariant objects extracted from the invariant product $E(\mathbf{p}_2)^{-1}E(\mathbf{p}_1)$ are
\beq\label{covdistanceX}
X_{\bar{1}2}:=X_{+,1}-X_{-,2}-\SV_2\overline{\SV}_1\,,
\qquad
\SV_{12}:=\SV_1-\SV_2\,,
\qquad
\overline{\SV}_{12}:=\overline{\SV}_1-\overline{\SV}_2\,.
\eeq
We  also define $X_{2\bar{1}}:=-X_{\bar{1}2}$.

\subsubsection*{Superfields for superconformal multiplets}
The supermultiplets $\hat{\mathcal{B}}_{[R_1,R_2]}$ correspond to ``scalar'' superfields on $\mathscr{M}$.
Among them, as discussed in the previous section, the ones with $R_1R_2=0$ are special
in the sense that they satisfy certain chirality conditions.
We call \textit{chiral (anti-chiral)} a superfield
 that depends only on the coordinates 
$\{X_-,\SV\}$ ($\{X_+,\overline{\SV}\}$).\footnote{
This is not the standard terminology for chiral superfields in $\mathcal{N}$-extended superspace.
We hope this  will not cause any confusion to the reader.} Within this terminology, the operators $\hat{\mathcal{B}}_{[0,R]}$ are chiral
while  the $\hat{\mathcal{B}}_{[R,0]}$ are antichiral. More general supermultiplets can be described as superfields  on $\mathscr{M}$
with $\SL(2|1)\times \SL(2|1)$ indices which extend the familiar Lorentz indices.
We will not need to develop the dictionary between  $\mathcal{N}=3$ superconformal representations and  
$\SL(2|1)\times \SL(2|1)\times \GL(1)\times \GL(1)$ induced representations  in this work and thus leave it for the future.

\begin{remark}
\label{remark:N2reduction}
\normalfont 
 The subspace $\mathscr{M}_{\mathcal{N}=2}$ corresponding to setting
 $\SV=\overline{\SV}=0$ is acted upon by the $\mathcal{N}=2$ superconformal group $\SL(4|2)$.
The corresponding superspace is well known, see e.g.~\cite{Doobary:2015gia}.
The superfields corresponding to the $\mathcal{N}=3$ supermultiplets $\hat{\mathcal{B}}_{[R_1,R_2]}$
reduce to the $\mathcal{N}=2$ supermultiplet 
 $\hat{\mathcal{B}}_{\frac{1}{2}(R_1+R_2)}$ when restricted to the superspace $\mathscr{M}_{\mathcal{N}=2}$.
 The other operators in  the decomposition of
 $\hat{\mathcal{B}}_{[R_1,R_2]}$  in $\mathcal{N}=2$ supermultiplets, 
 see \eqref{eq:Neq2decST}, \eqref{eq:BR0fulldec},
  roughly  corresponds
 to the expansion of the superfield in $\SV$ and $\overline{\SV}$.
 There is also an $\mathcal{N}=1$ subspace $\mathscr{M}_{\mathcal{N}=1}$ , which is not a subspace of 
 $\mathscr{M}_{\mathcal{N}=2}$, defined by setting $\lambda^{\alpha}, \pi^{\dot{\alpha}},  v,\bar{v}$ to zero.
An $\SL(4|1)\times \SL(2)$ subgroup of $\SL(4|3)$ acts on $\mathscr{M}_{\mathcal{N}=1}$.
 This observation will be useful in the derivation of the superconformal blocks in section \ref{sec:superblocks}.
 \end{remark}

\subsubsection*{Examples of two- and three-point functions}
We denote superfields and supermultiplets in the same way.
Let us list some relevant examples of two- and three-point  functions of $\hat{\mathcal{B}}$-operators of increasing complexity:
\begin{align}
\langle  \hat{\mathcal{B}}_{[R_1,R_2]}(1)\hat{\mathcal{B}}_{[R_3,R_4]}(2)\rangle \,&=\,\delta_{R_1,R_4}\delta_{R_2,R_3}
(\bar{2}1)^{R_1}\,(\bar{1}2)^{R_2}\,,\\
\label{threepoint1}
\langle  \hat{\mathcal{B}}_{[0,R]}(1)\hat{\mathcal{B}}_{[R,0]}(2)\hat{\mathcal{B}}_{[S,S]}(3)\rangle \,&=\,
(\bar{2}1)^{R-S} \left((\bar{2}3)(\bar{3}1)\right)^S\,,
\\
\label{threepoint1b}
\langle  \hat{\mathcal{B}}_{[0,R]}(1)\hat{\mathcal{B}}_{[0,R]}(2)\hat{\mathcal{B}}_{[R_1,R_2]}(3)\rangle \,&=\,
\delta_{R_1,2R}\,\delta_{R_2,0}\,
\left((\bar{3}1)(\bar{2}1)\right)^R\,,\\
\label{threepoint2}
\langle  \hat{\mathcal{B}}_{[R,R]}(1)\hat{\mathcal{B}}_{[R,R]}(2)\hat{\mathcal{B}}_{[R,R]}(3)\rangle \,&=\,
\left((\bar{2}1)(\bar{3}2)(\bar{1}3)\right)^R\,P_R(C)\,,
\end{align}
where we have defined
\beq
(\bar{1}2)\colonequals \frac{1}{\text{sdet}(X_{\bar{1}2})}\,,
\qquad C\colonequals\frac{(\bar{3}1)(\bar{1}2)(\bar{2}3)}{(\bar{2}1)(\bar{3}2)(\bar{1}3)}\,.
\eeq
In \eqref{threepoint1} superspace analyticity implies that $S\leq R$ and that the correlation function vanishes otherwise.
Similarly, in \eqref{threepoint2}, 
$C$ is a superconformal invariant and superspace analyticity implies that $P_R(C)$ is a polynomial of degree $R$ in $C$.
Since the three operators are identical, one further imposes Bose symmetry which translates to $P_R(x)=x^R P_{R}(x^{-1})$.
Equation \eqref{threepoint2} specialized to the case $R=1$ corresponds to the three-point function
 of the stress-tensor supermultiplet $\hat{\BB}_{[1,1]}$ and the argument above implies that $P_1(x)=\text{const}\times(1+x)$. 
This provides a quick proof of the fact that for $\mathcal{N}=3$ superconformal theories one has the relation  $a=c$
as first derived in \cite{Aharony:2015oyb}.

Let us consider the three-point functions relevant for the \nonchiral OPE
 $ \hat{\mathcal{B}}_{[R,0]}\times \hat{\mathcal{B}}_{[0,R]}$. 
A simple superspace analysis reveals that the three-point function of a chiral and an anti-chiral operator
 with a generic operator takes the form 
\beq\label{threepoint3}
\langle  \hat{\mathcal{B}}_{[0,R]}(X_{-,1},\SV_1)\hat{\mathcal{B}}_{[R,0]}(X_{+,2},\overline{\SV}_2)
\mathcal{O}(X_3,\SV_3,\overline{\SV}_3)\rangle\,=\,
(\bar{2}1)^{R} \rho_{\mathcal{O}}\left(X_{\bar{3}1}^{}X_{\bar{2}1}^{-1}X_{\bar{2}3}^{}\right)\,.
\eeq
The quantity $\rho_{\mathcal{O}}$ is determined uniquely up to a multiplicative constant
 by the requirement that \eqref{threepoint3} is superconformally covariant.
It is not hard to verify that one can set the coordinates $\SV_1$, $\overline{\SV}_2,\SV_3,\overline{\SV}_3$ 
to zero by an $\SL(4|3)$ transformation which is not part of the $\mathcal{N}=2$ superconformal group  $\SL(4|2)$ 
(with the embedding specified in the remark~\ref{remark:N2reduction} above).
This means that \eqref{threepoint3} is zero  if its $\mathcal{N}=2$ reduction
 (\ie, the result obtained after setting $\SV_i=\overline{\SV}_i=0$) is zero, 
as confirmed by the selection rules result  \eqref{selectionchiantichi} that we derive later in section~\ref{sec:selectionrules}.

Turning to the three-point functions relevant for the \chiral OPE $ \hat{\mathcal{B}}_{[R,0]}\times \hat{\mathcal{B}}_{[R,0]}$,
it is not hard to convince oneself that they take the form
\beq\label{threepoint3cc}
\langle  \hat{\mathcal{B}}_{[0,R]}(X_{-,1},\SV_1)\hat{\mathcal{B}}_{[0,R]}(X_{-,2},\SV_2)
\widetilde{\mathcal{O}}(X_3,\SV_3,\overline{\SV}_3)\rangle\,=\,
\left((\bar{3}1)(\bar{3}2)\right)^{R} \widetilde{\rho}_{\widetilde{\mathcal{O}}}
\big(\widehat{X},\widehat{V}\big)\,,
\eeq
where
\beq
\widehat{X}=\,\left(X_{2\bar{3}}^{-1}-X_{1\bar{3}}^{-1}\right)^{-1}\,,
\qquad
\widehat{V}=\,X_{2\bar{3}}^{-1}V_{23}^{}-X_{1\bar{3}}^{-1}V_{13}^{}\,,
\eeq
and $ \widetilde{\rho}_{\widetilde{\mathcal{O}}}$ is fixed by requiring superconformal covariance of \eqref{threepoint3cc}.
 It is important to remark that, as opposed to \eqref{threepoint3}, in this case one cannot set the coordinates 
 $\SV_1$, $\SV_2,\SV_3,\overline{\SV}_3$ to zero using superconformal transformations. However, they can be set to the values
 \beq
 \{(X_{-,1},\SV_1),(X_{-,2},\SV_2),(X_{-,3},\SV_3,\overline{\SV}_3)\}\,\mapsto\,
\{(\infty,0),\widehat{X}(1,\widehat{\SV}),(0,0,0)\}\,.
 \eeq
 The combinations $\widehat{X}$ and $\widehat{\SV}$ carry non trivial superconformal weights only with respect to the third point corresponding to the operator $\widetilde{\mathcal{O}}$.

\subsection{Superconformal Ward identities}
\label{sec:WI}
We will now derive, along the same lines as  \cite{Dolan:2004mu,Liendo:2015cgi,Liendo:2016ymz},  the superconformal Ward identities for the four-point correlation function 
$\langle \hat{\BB}_{[0,R]}\hat{\BB}_{[R,0]}\hat{\BB}_{[0,R]} \hat{\BB}_{[R,0]} \rangle$.
Let us first introduce super cross-ratios for this four point function.
The eigenvalues of the graded matrix
\beq\label{Zmatrix}
\mathcal{Z}\,:=\,
X_{1\bar{2}}^{}\,
X_{3\bar{2}}^{-1}\,
X_{3\bar{4}}^{}\,
X_{1\bar{4}}^{-1}
\,,
\eeq
are invariant and will be denoted by $\x_1,\x_2,y$.
This can be seen from the fact that all fermionic coordinates in this four-point function
can be set to zero by a superconformal transformation.
It follows that 
\beq
\label{eq:BBbBBbcorr}
\langle \hat{\BB}_{[0,R]}(1) \hat{\BB}_{[R,0]}(\bar 2) \hat{\BB}_{[0,R]}(3)  \hat{\BB}_{[R,0]}(\bar 4) \rangle 
 = (1\bar{2})^R\,(3\bar{4})^R\,G_R(\x_1,\x_2,y)\,,
\eeq
where $G_R(\x_1,\x_2,y)=G_R(\x_2,\x_1,y)$.
The form of $G_R(\x_1,\x_2,y)$ is strongly restricted by the requirement of superspace analyticity.
Firstly, after setting all fermionic variables to zero
\beq
\mathcal{Z}|_{\text{ferm}=0}\,=\,
\begin{pmatrix}
x_{12}^{}\,
x_{32}^{-1}\,
x_{34}^{}\,
x_{14}^{-1}
& 0\\
0 & \frac{y_{1\bar{2}}y_{3\bar{4}}}{y_{3\bar{2}}y_{1\bar{4}}}
\end{pmatrix}\,,
\qquad
 y_{\bar{\imath}j}:=y_i-y_j-v_j\bar{v}_i\,,
\eeq
 polynomiality in the R-symmetry variables 
implies that $G_R(\x_1,\x_2,y)$  is a polynomial of degree $R$ in $y^{-1}$.
Secondly, one has to make sure that the fermionic coordinates can be turned on without introducing extra singularities in the R-symmetry variables. By looking at the expansion of the eigenvalues of  \eqref{Zmatrix} in fermions, one concludes that the absence 
of spurious singularities is equivalent to the conditions
 \beq\label{eq:WardIdentities}
 \left(\partial_{\x_1}+\partial_y\right)G_R(\x_1,\x_2,y)\big{|}_{\x_1=y}\,=\,0\,,\qquad
 \left(\partial_{\x_2}+\partial_y\right)G_R(\x_1,\x_2,y)\big{|}_{\x_2=y}\,=\,0\,.
 \eeq
These equations imply in particular that  $G_R(x,\x_2,x)=f_R(\x_2)$ and $G_R(\x_1,x,x)=f_R(\x_1)$.
This is a consequence of the protected subsector discussed in section~\ref{sec:chiral algebra}, where setting $x_1=x$ (or $x_2=x$) follows from the twisted translations \eqref{eq:twistedtransl}, as originally discussed in \cite{Beem:2013sza}. The chiral algebra further tells us that $f_R(\x_1)$ is a meromorphic function of $x_1$, corresponding to a two-dimensional correlation function of the twisted-translated Schur operators, with each $\hat{\BB}_{[0,R]}$ ($\hat{\BB}_{[R,0]}$) multiplet giving rise to a two-dimensional $\NN=2$ chiral (anti-chiral) operator, as discussed in \ref{subsec:schurop}.

The general solution of the Ward identities can be parametrized as
\beq\label{solutionWI}
G_R(\x_1,\x_2,y) = \frac{(\x_1^{-1} - y^{-1} )f_R(\x_1)-  (\x_2^{-1} - y^{-1} )f_R(\x_2)}{\x_1^{-1}-\x_2^{-1}} 
+  \left(\x_1^{-1} - y^{-1}  \right)\left(\x_2^{-1} - y^{-1}  \right)\,H_R(\x_1,\x_2,y)\,,
\eeq
where $H_R(\x_1,\x_2,y)$ is a polynomial of degree $R-2$ in $y^{-1}$.
In particular, it is zero for the case $R=1$ corresponding to a free theory.
For the following analysis it is useful to introduce the variables $z,\bar{z},w$ as
\beq\label{z1z2yfromzzw}
\x_1\,=\,\frac{z}{z-1}\,,
\qquad
\x_2\,=\,\frac{\bar{z}}{\bar{z}-1}\,,
\qquad
y\,=\,\frac{w}{w-1}\,.
\eeq
This change of variable is an involution in the sense that $z=\frac{\x_1}{\x_1-1}$ and so on.
They are related to the more familiar cross-ratios as
\beq\label{uvzzbcrossratios}
u\,=\,\frac{x_{12}^2x_{34}^2}{x_{13}^2x_{24}^2}=z\zb |_{\text{ferm}=0}\,,
\qquad
v\,=\,\frac{x_{14}^2x_{23}^2}{x_{13}^2x_{24}^2}=(1-z)(1-\zb) |_{\text{ferm}=0}\,.
\eeq
Notice that the WI \eqref{eq:WardIdentities} take the same form in the new variables and that moreover
\begin{subequations}
\label{idforWI}
\begin{align}
(z^{-1}-w^{-1})(\bar{z}^{-1}-w^{-1})\,&=\,(\x_1^{-1}-y^{-1})(\x_2^{-1}-y^{-1})\,,\\
\frac{(\x_1^{-1} - y^{-1} )f(\x_1)-  (\x_2^{-1} - y^{-1} )f(\x_2)}{\x_1^{-1}-\x_2^{-1}} \,&=\,
\frac{(z^{-1} - w^{-1} )f(\x_1)-  (\bar{z}^{-1} - w^{-1} )f(\x_2)}{z^{-1}-\bar{z}^{-1}} \,,
\end{align}
\end{subequations}
for any function $f(x)$.

\subsection{Selection rules}
\label{sec:selectionrules}

In this subsection we analyze the possible multiplets allowed by superconformal symmetry in the \nonchiral and  \chiral OPEs. This is a crucial ingredient for the crossing equations and are usually called the OPE selection rules.

\subsubsection*{Non-chiral channel}
  The OPE in the \nonchiral channel $\hat{\mathcal{B}}_{[R,0]} \times \hat{\mathcal{B}}_{[0,R]}$ can be obtained by using the superconformal Ward identities just derived,
  together  with the fact that the three-point function 
$\langle \hat{\mathcal{B}}_{[R,0]} \hat{\mathcal{B}}_{[0,R]} \mathcal{O}\rangle$, 
where $\mathcal{O}$ is a generic operator, is non-zero only if the three-point function of the corresponding superprimary states is non-zero by conformal and R-symmetry.
The latter condition can be derived by recalling that the fermionic coordinates in this three-point function can be set 
to zero by a superconformal transformation. A simple analysis shows that 
\beq\label{selectionchiantichi}
\hat{\mathcal{B}}_{[R,0]}\times \hat{\mathcal{B}}_{[0,R]}\,=\,
\mathcal{I}+ \sum_{a=1}^R\,\hat{\mathcal{B}}_{[a,a]}
+\sum_{\ell=0}^\infty
\left[\sum_{a=0}^{R-1}\hat{\mathcal{C}}_{[a,a],\ell}+
\sum_{a=0}^{R-2}\mathcal{A}^{\Delta}_{[a,a],r=0,\ell}\right]\,.
\eeq
Notice that these relations are remarkably similar to the $\hat{\mathcal{B}}_{R/2}\times \hat{\mathcal{B}}_{R/2}$ 
OPE in  the $\mathcal{N}=2$ case, see  \cite{Arutyunov:2001qw}. 
The three upper bounds on the finite summations $R,R-1,R-2$ could be derived by imposing that the  three-point function
$\langle \hat{\mathcal{B}}_{[R,0]} \hat{\mathcal{B}}_{[0,R]} \mathcal{O}\rangle$ is free of  superspace singularities.
Equivalently, it can be derived by requiring that the associated superconformal block takes the form \eqref{solutionWI}.
We followed the latter strategy as it seemed more economical. 

\subsubsection*{Chiral channel}

The \chiral channel selection rules are obtained by requiring that a given multiplet can only contribute if it contains an operator annihilated by all the supercharges that annihilate the highest weight of $\hat{\BB}_{[R,0]}$, and if said operator transforms in one of the representations appearing in the tensor product of the R-symmetry representations $[R,0] \times [R,0]$, and with the appropriate spin to appear in the OPE of the external scalars. We have performed this calculation for $R=2,3$ and based on it we propose that the expression for general $R$ is
\begin{align}
\label{chiralchiralselectionsrules}
\hat{\mathcal{B}}_{[R,0]}\times \hat{\mathcal{B}}_{[R,0]} &= 
\hat{\mathcal{B}}_{[2R,0]}+\sum_{a=2}^R\,\overline{\mathcal{B}}_{[2(R-a),a],r=4R,0}\,+\,\nonumber\\
&
+\,\sum_{\ell=0}^{\infty}\left[
\hat{\mathcal{C}}_{[2R-2,0],(\frac{\ell+1}{2},\frac{\ell}{2})}+\sum_{a=2}^{R}
\left(\overline{\mathcal{C}}^{\,r=4R-1}_{[2(R-a),a-1],(\frac{\ell+1}{2},\frac{\ell}{2})}+
\mathcal{A}^{\Delta,r=4R-2}_{[2(R-a),a-2],(\frac{\ell}{2},\frac{\ell}{2})}\right)\right]
\,.
\end{align}
We have checked the above in several cases for $R>3$ and superspace arguments suggest it is indeed the correct selection rule.
Note that in \eqref{chiralchiralselectionsrules} the $\mathcal{B}$-type multiplets have $r=4R$,
the $\mathcal{C}$-type multiplets $r=4R-1$, and
the $\mathcal{A}$-type multiplets $r=4R-2$.
Moreover, if we are considering identical $\hat{\BB}_{[R,0]}$, then Bose symmetry further constraints the spin of the operators appearing on the right-hand-side according to their $\SU(3)_R$ representation.

\subsection{Superconformal blocks}
\label{sec:superblocks}
We will now derive the superconformal blocks relevant for the expansion of the  four-point function \eqref{eq:BBbBBbcorr}.
The  superconformal Ward identities alone turn out not to be sufficient to  uniquely determine all the superblocks.
We resolve the leftover ambiguity by requiring that they are linear combinations
 of $\SL(4|1)\times \SU(2)$ ($\mathcal{N}=1$)
superblocks. 
There are two types of blocks corresponding to the two channels: \nonchiral \eqref{selectionchiantichi} and \chiral \eqref{chiralchiralselectionsrules}.
The two kinds of blocks are closely connected to $\mathcal{N}=2$ superconformal blocks relevant for the four-point function of $\hat{\mathcal{B}}$-type operators and are collected in tables 
\ref{tab:blockssummary} and  \ref{tab:blockssummary2}.
When the kinematics is restricted to $(z,\zb,w)=(z,w,w)$, only superconformal blocks corresponding to the exchange of 
Schur operators, defined in section  \ref{subsec:schurop},
are non-vanishing. Moreover, they reduce to $2d$ (global) superblocks for the 
$\mathcal{N}=2$ superconformal algebra $\slf(2|1)\simeq \ospf(2|2)$ in the appropriate channel.

\subsubsection{Superconformal blocks for the non-chiral channel.}
\label{sec:Superconformal blocks for the non-chiral channel.}

On general grounds, the $\mathcal{N}=3$ superconformal blocks contributing to the four-point function \eqref{eq:BBbBBbcorr}
in the \nonchiral channel can be written as an expansion in terms of  conformal times $\SU(3)$ R-symmetry blocks:
\beq\label{blockassum}
\mathcal{G}_{\chi}(z,\bar{z},w)\,=\,\sum_{\alpha\,\in\,S_\chi}\,c_{\alpha}(\chi)\,
g_{\Delta_{\alpha},\ell_{\alpha}}(z,\bar{z})\,h_{[R_{\alpha},R_{\alpha}]}(w)\,.
\eeq 
The explicit form of the conformal blocks $g_{\Delta,\ell}$ is given in Appendix \ref{app:blockology}.
The $\SU(3)$ R-symmetry blocks take the form 
\beq\label{Rblocks}
h_{[a,a]}(w)\,=\,\binom{2a+1}{a+1}^{-1}\,_{2}F_1(-a,a+2,1,y^{-1})\,,\qquad
y\,=\,\frac{w}{w-1}\,.
\eeq
The normalization in \eqref{Rblocks} is chosen so that 
$h_{[a,a]}(w)=w^{-a}+\dots$ for $w\rightarrow 0$.
The set $S_\chi$ is determined by considering the decomposition of the $\mathcal{N}=3$ representation being exchanged 
into representations of the bosonic subalgebra (this can be done using superconformal characters).
The normalization can be fixed by taking for instance $c_{\alpha}(\chi)=1$ for the label $\alpha$ corresponding to the minimum value of
 $\Delta_{\alpha}$ in the supermultiplet.

Consider the superblocks corresponding to the \nonchiral OPE channel of \eqref{selectionchiantichi}.
Concerning the superblocks for the $\hat{\mathcal{B}}_{[a,a]}$ exchange, it turns out that they are uniquely fixed 
by imposing the superconformal WI on \eqref{blockassum}.
 The superblocks corresponding to the exchange of a $\hat{\mathcal{C}}_{[a,a],\ell}$
 on the other hand are not uniquely fixed by  the this procedure.
The remaining ambiguity can be resolved by requiring that they reduce to 
$\ospf(2|2)$ (this is the global part of the chiral half of the $d=2,\mathcal{N}=2$ super Virasoro algebra) 
superblocks
when restricted to $(z,\bar{z},w)=(z,w,w)$. Recall that this restriction reduces the correlator to that of $2d$ $\NN=2$ chiral and anti-chiral operators, and thus the exchange of an operator in the non-chiral channel is captured by the $\ospf(2|2)$ superblocks of \cite{Fitzpatrick:2014oza}. Specifically, this amounts to requiring
\beq
f_{\hat{\mathcal{C}}_{[a,a],\ell}}(z)=
\mathcal{G}_{\hat{\mathcal{C}}_{[a,a],\ell}}(z,w,w)=(-1)^{\ell+1}
\Ssl_{a+\ell+2}(\tfrac{z}{z-1})\,,
\eeq
where the $\ospf(2|2)$ superblock is \cite{Fitzpatrick:2014oza}
\beq
\label{eq:OSP22superblocks}
\Ssl_h(x)=x^h\,_{2}F_1(h,h,2h+1,x)\,,
\eeq
and $f(z)$ corresponds to the parametrization \eqref{solutionWI}.\footnote{To each superblock $\mathcal{G}_\chi$ corresponds a function $\fc_\chi$ and a function $\Hc_\chi$ by using the parametrization \eqref{solutionWI}.}
The superblocks for the exchange of long operators $\mathcal{A}^{\Delta}_{[a,a],r=0,\ell}$ are not uniquely determined
 by the two conditions given above.  The leftover ambiguity can be resolved by studying  the Casimir equations.
However, we will take a shortcut and use the knowledge of $\mathcal{N}=1$ superblocks.
 The relevant superblocks, which were derived in \cite{Poland:2010wg,Fitzpatrick:2014oza}, are given by 
\beq
\label{N=1blocks}
\mathcal{G}^{\mathcal{N}=1}_{\Delta,\ell}(z,\bar{z})\,=\,
(z\bar{z})^{-\frac{1}{2}}\,g^{\Delta_{12}=\Delta_{34}=1}_{\Delta+1,\ell}(z,\bar{z})\,.
\eeq
It follows from the remark~\ref{remark:N2reduction},
 that the $\mathcal{N}=3$  superblocks can be expanded in $\mathcal{N}=1$ 
times $SU(2)$ ``flavor symmetry''  blocks as
\beq\label{expansion_in_N=1}
\mathcal{G}^{\mathcal{N}=3}_{\mathcal{A}^{\Delta}_{[a,a],r=0,\ell}}(z,\bar{z},w)\,=\,
\tilde{d}_{\Delta,\ell}^{(0,0)}(w)\,
\mathcal{G}^{\mathcal{N}=1}_{\Delta,\ell}(z,\bar{z})+
\tilde{d}_{\Delta,\ell}^{(1,1)}(w)\, 
\mathcal{G}^{\mathcal{N}=1}_{\Delta+1,\ell+1}(z,\bar{z})+
\dots +\tilde{d}_{\Delta,\ell}^{(4,0)}(w)\, \mathcal{G}^{\mathcal{N}=1}_{\Delta+4,\ell}(z,\bar{z})\,.
\eeq
On the right hand side, the sum runs over the terms  
\beq
(\Delta,\ell)\,,
\quad
(\Delta+1,\ell\pm1)\,,
\quad
(\Delta+2,\ell\pm2)\,,
\quad
(\Delta+2,\ell)\,,
\quad
(\Delta+3,\ell\pm1)\,,
\quad
(\Delta+4,\ell)\,.
\eeq
Imposing that the form \eqref{blockassum}, subject to the WI, 
can be expanded as in \eqref{expansion_in_N=1}, 
fixes the leftover ambiguity in the $\mathcal{N}=3$ superblocks and the  coefficient functions
  $\tilde{d}_{\Delta,\ell}^{(a,b)}(w)$ up to an overall normalization.
The solution can then be rewritten in the compact form 
\beq
\label{eq:chiralantichirallongblocks}
\mathcal{G}^{\mathcal{N}=3}_{\mathcal{A}^{\Delta}_{[a,a],r=0,\ell}}(z,\bar{z},w)\,=\,
(-1)^a(z^{-1}-w^{-1})(\bar{z}^{-1}-w^{-1})\,\mathcal{G}^{\mathcal{N}=1}_{\Delta+2,\ell}(z,\bar{z}) \,h_{[a,a]}(w)\,.
\eeq
The simplicity of this expression will be justified in remark~\ref{remark: Hesloppaper} below.
This concludes the derivation of superconformal blocks relevant for the \nonchiral channel, 
the results are summarized in table  \ref{tab:blockssummary}.

\begin{table}[t]
\centering
\renewcommand{\arraystretch}{1.6}
\begin{tabular}{%
| l
                |>{\centering }m{3.2cm}
             |>{\centering\arraybackslash}m{6.35cm}|
}
\hline
$\chi$
& 
$\fc_{\chi}$
&
$\Hc_{\chi}$
\\\hline
identity
&
$1$
&
0
\\\hline
$\hat{\mathcal{B}}_{[a,a]}$
&
$
\Ssl_{a}$
&
$(-1)^a\sum_{k=0}^{a-2}\,\mathcal{G}^{\mathcal{N}=1}_{a+k+2,a-k-2}\,h_{[k,k]}$
\\\hline
$\hat{\mathcal{C}}_{[a,a],\ell}$
&
$(-1)^{\ell+1}\,\Ssl_{a+\ell+2}$
&
$(-1)^{a+1}\,\sum_{k=0}^{a-1}\,\mathcal{G}^{\mathcal{N}=1}_{a+\ell+k+4,a+\ell-k}\,h_{[k,k]}$ 
\\\hline
$\mathcal{A}^{\Delta}_{[a,a],\ell}$
&
$0$
&
$(-1)^a\mathcal{G}^{\mathcal{N}=1}_{\Delta+2,\ell}\,h_{[a,a]}$
\\\hline
\end{tabular}
\renewcommand{\arraystretch}{1.0}
\caption{Superconformal blocks contributing to \eqref{selectionchiantichi}
in the parametrization \eqref{solutionWI}. 
These expressions are consistent with the decompositions of superblocks at unitarity bounds,
see \eqref{unitarityonbloks}.
We recall that the explicit expressions for the blocks entering the table
are given in
\eqref{Rblocks}, \eqref{eq:OSP22superblocks}, and \eqref{N=1blocks}.
Notice that for the stress-tensor supermultiplets $\hat{\mathcal{B}}_{[1,1]}$, the function $\Hc_{\hat{\mathcal{B}}_{[1,1]}}$ is zero.
}
\label{tab:blockssummary}
\end{table}

%
Before turning to the discussion of the superblocks relevant for the \chiral channel,
 we perform a consistency check on the blocks just derived.
As can be seen in table~\ref{tab:blockssummary}, short blocks can be obtained from the long ones \eqref{eq:chiralantichirallongblocks} at the unitarity bounds by using
\beq\label{unitarityonbloks}
\mathcal{G}_{\mathcal{A}_{[a,a],r=0,\ell}^{\Delta=\ell+2+2a}}\,=\,
\mathcal{G}_{\hat{\mathcal{C}}_{[a,a],\ell}}+\,
\mathcal{G}_{\hat{\mathcal{C}}_{[a+1,a+1],\ell-1}}\,,
\eeq
where we identify $\hat{\mathcal{C}}_{[a,a],-1}\equiv \hat{\mathcal{B}}_{[a+1,a+1]}$. This is consistent with the multiplet decomposition at the unitarity bound: $\mathcal{A}_{[a,a],r=0,\ell}^{\Delta=\ell+2+2a}\rightarrow  
\hat{\mathcal{C}}_{[a,a],\ell} \oplus \hat{\mathcal{C}}_{[a+1,a+1],\ell-1} \oplus \text{``extra''}$, where $\text{``extra''}$ does not 
contribute to the block.
%

\subsubsection{Superconformal blocks for the chiral channel.}
\label{sec:Superconformal blocks for the chiral channel}

We denote the superconformal blocks contributing to this channel as  $\widetilde{\mathcal{G}}_{\chi}(z,\bar{z},w)$,
where $\chi$ labels the representations being exchanged from the list \eqref{chiralchiralselectionsrules}.
As in the case of the \nonchiral channel, we start with an expansion of the superblocks in conformal times $\SU(3)$
blocks and impose the superconformal Ward identities, \eqref{eq:WardIdentities}.
Specifically we take
\beq\label{blockassum2}
\widetilde{\mathcal{G}}_{\chi}(z,\bar{z},w)\,=\,
\sum_{\alpha\,\in\,\widetilde{S}_\chi}\,\tilde{c}_{\alpha}(\chi)\,
g_{\Delta_{\alpha},\ell_{\alpha}}(z,\bar{z})\,\widetilde{h}_{[2(R-n_{\alpha}),n_{\alpha}]}(w)\,.
\eeq 
 It appears, perhaps not too surprisingly, that the $\SU(3)$ R-symmetry blocks $\widetilde{h}_{[2m,n]}(w)$ in this channel  coincide with  $\SU(2)$ blocks, where here and in the following  we take $m=2(R-n)$. 
 They  take the form\footnote{
One can recognize the appearance of Legendre polynomials as 
$(-1)^m{}_2F_1(-m,m+1,1,w^{-1})=P_m(\tfrac{2}{w}-1)$.} 
\beq\label{Rblocks1}
\widetilde{h}_{[2m,n]}(w)\,=\,h_m^{\SU(2)}(w)=(-1)^m\,\binom{2m}{m}^{-1}{}_2F_1(-m,m+1,1,w^{-1})\,,
\eeq
where the normalization is chosen so that $\widetilde{h}_{[2m,n]}(w)\sim w^m$ for $w\sim 0$, and we omit the label $n$ since it is related to $m$.
The set $\widetilde{S}_{\chi}$ is determined by looking at the content of the representation $\chi$ using characters.
Using this information, all the coefficients $\tilde{c}_{\alpha}(\chi)$ are then fixed by the requirement that \eqref{blockassum2}
satisfies the superconformal WI \eqref{eq:WardIdentities}.

With a little inspection on the solutions,
 one recognizes that the superblocks in this channel are the 
 $\mathcal{N}=2$ superconformal blocks that contribute
  to the four-point function of $ \hat{\mathcal{B}}_{\mathcal{N}=2}$ supermultiplets \cite{Dolan:2001tt,Dolan:2004mu,Nirschl:2004pa}.
The identification is given by
\beq\label{N=3andN=2}
\widetilde{\mathcal{G}}^{\mathcal{N}=3}_{\chi}(z,\bar{z},w)\,=\,
\mathcal{G}^{\mathcal{N}=2}_{p(\chi)}(z,\bar{z},w)\,,
\eeq
where $p$ maps the $\mathcal{N}=3$ representations being exchanged in the \chiral channel, 
see \eqref{chiralchiralselectionsrules}, to an $\mathcal{N}=2$ representations as follows
\beq\label{pmap}
p
\begin{pmatrix}
\hat{\mathcal{B}}_{[2R,0]}\\
\hat{\mathcal{C}}_{[2(R-1),0],(\frac{\ell+1}{2},\frac{\ell}{2})}\\
\overline{\mathcal{B}}^{r=4R}_{[2(R-a),a],0}\\
\overline{\mathcal{C}}^{r=4R-1}_{[2(R-a),a-1],(\frac{\ell+1}{2},\frac{\ell}{2})}\\
\mathcal{A}^{\Delta,r=4R-2}_{[2(R-a),a-2],(\frac{\ell}{2},\frac{\ell}{2})}
\end{pmatrix}
\,=\,
\begin{pmatrix}
\hat{\mathcal{B}}_{R}\\
\hat{\mathcal{C}}_{R-1,\ell+1}\\
\mathcal{A}^{2R}_{R-a,0}\\
\mathcal{A}^{2R+\ell+1}_{R-a,\ell+1}\\
\mathcal{A}^{\Delta+1}_{R-a,\ell}
\end{pmatrix}\,.
\eeq
The equality \eqref{N=3andN=2} is not accidental,  we will comment on its origin in the remark below.

The resulting superblocks in the parametrization \eqref{solutionWI} 
are given in table \ref{tab:blockssummary2}. 
Once again note that the meromorphic function $\widetilde{\fc}(z)$ has a decomposition in $2d$ blocks, in this case $\slf(2)$ blocks
\beq
\label{eq:sl22dblocksdefinition}
\sl_h(z)=z^h\,_{2}F_1(h,h,2h,z)\,.
\eeq
These are the blocks relevant for the decomposition of the chiral algebra correlators in the chiral channel, since each $2d$ $\NN=2$ multiplet contributes with a single $\slf(2)$ primary to the OPE of two $2d$ $\NN=2$ chiral operators.
%

\begin{table}
\centering
\renewcommand{\arraystretch}{1.6}
\begin{tabular}{%
| l
                |>{\centering }m{3.2cm}
             |>{\centering\arraybackslash}m{6.35cm}|
}
\hline
$\chi$
& 
$\widetilde{\fc}_{\chi}$
&
$\widetilde{\Hc}_{\chi}$
\\\hline
$\hat{\mathcal{B}}_{[2R,0]}$
&
$\sl_R$
&
$\sum_{k=0}^{R-2}\,g_{R+k+2,R-k-2}\,h^{\SU(2)}_{k}$
\\\hline
$\hat{\mathcal{C}}_{[2R-2,0],(\frac{\ell+1}{2},\frac{\ell}{2})}$
&
$\sl_{R+\ell+2}$
&
$\sum_{k=0}^{R-2}\,g_{R+\ell+k+4,R+\ell-k}\,h^{\SU(2)}_{k}$
\\\hline
$\overline{\mathcal{B}}^{r=4R}_{[2(R-a),a],0}$
&
$0$
&
$g_{2R+2,0}\,\,h^{\SU(2)}_{R-a}$
\\\hline
$\overline{\mathcal{C}}^{r=4R-1}_{[2(R-a),a-1],(\frac{\ell+1}{2},\frac{\ell}{2})}$
&
$0$
&
$g_{2R+\ell+3,\ell+1}\,\,h^{\SU(2)}_{R-a}$
\\\hline
$\mathcal{A}^{\Delta,r=4R-2}_{[2(R-a),a-2],(\frac{\ell}{2},\frac{\ell}{2})}$
&
$0$
&
$g_{\Delta+3,\ell}\,\,h^{\SU(2)}_{R-a}$
\\\hline
\end{tabular}
\renewcommand{\arraystretch}{1.0}
\caption{
Superconformal blocks contributing to \eqref{chiralchiralselectionsrules}
in the parametrization \eqref{solutionWI}, with $f$ and $H$ replaced by $\widetilde{f}$ and $\widetilde{H}$ to indicate they correspond to the chiral channel blocks. 
These expressions are consistent with the decompositions of superblocks at unitarity bounds,
see \eqref{unitboundchiralchiral}.
We recall that the explicit expression of the $\slf(2)$ and R-symmetry blocks is given in  \eqref{eq:sl22dblocksdefinition} and  \eqref{Rblocks1} respectively. 
}
\label{tab:blockssummary2}
\end{table}
The unitarity bound relevant for the \chiral channel is 
\beq\label{unitboundchiralchiral}
\mathcal{A}^{\Delta,r=4R-2}_{[2(R-a),a-2],(\frac{\ell}{2},\frac{\ell}{2})}\,
\stackrel{\Delta=\ell+2R-1}{\xrightarrow{\hspace*{1.2cm}}}\,
\overline{\mathcal{C}}^{r=4R-2}_{[2(R-a),a-2],(\frac{\ell}{2},\frac{\ell}{2})}
\oplus
\underline{
\overline{\mathcal{C}}^{r=4R-1}_{[2(R-a),a-1],(\frac{\ell}{2},\frac{\ell-1}{2})}}\,,
\eeq
where 
$\overline{\mathcal{C}}^{r=4R-1}_{[2(R-a),a-1],(0,-\frac{1}{2})}=
\overline{\mathcal{B}}^{r=4R-1}_{[2(R-a),a],0}$. 
Only  the underlined term contributes to the  superblocks $\widetilde{\mathcal{G}}$, as can be seen in table~\ref{tab:blockssummary2}.

\begin{remark}
\label{remark: Hesloppaper}
\normalfont
In \cite{Doobary:2015gia}, the authors derived 
superconformal blocks for  scalar four-point functions on a super Grassmannian space $Gr(m|n,2m|2n)$.
It is an interesting problem to generalize the analysis  of \cite{Doobary:2015gia} to the more general case of $Gr(m|n,M|N)$.
The example we just studied corresponds to $Gr(2|1,4|3)$. The example of chiral superfields (in the traditional sense) for $\mathcal{N}$-extended supersymmetry corresponds to the  super Grassmannian  $Gr(2|0,4|\mathcal{N})$ and the corresponding superblocks were given in  \cite{Fitzpatrick:2014oza}.
The simplicity of the results \eqref{eq:chiralantichirallongblocks} and \eqref{N=3andN=2} and the one presented in 
\cite{Fitzpatrick:2014oza}
 suggests a simple unified picture.
\end{remark}

\section{Crossing equations}
\label{sec:crossing}

Equipped with the superconformal blocks relevant for the four-point function of half-BPS operators we are finally ready to write the crossing equations. Most of this section treats the case of arbitrary external dimension $R$, and in the final section we focus on the cases of a dimension two and three operator.
Crossing symmetry can be written in terms of the functions $f_R(x)$ and $H_R(\x_1,\x_2,y)$ used to parametrize the solution of the Ward identities \eqref{solutionWI}.
As expected, the chiral algebra correlator $f_R(x)$ satisfies a crossing equation on its own, that we solve analytically. This amounts to input about the exchange of Schur operators, that we feed into the full set of crossing equations for $H_R(\x_1,\x_2,y)$.
These give rise to a system of (three) six  equations of the two bosonic cross-ratios, for ($R=2$) $R=3$, which are the subject of the numerical analysis of section \ref{sec:numerics}.

\subsubsection*{First equation}
Consider the four-point function \eqref{eq:BBbBBbcorr}, where we take pairwise identical operators. Imposing that it is invariant upon the exchange of points $1 \leftrightarrow 3$
implies the crossing equation
\beq
\label{crossingCA}
G_R(\x_1^{},\x_2^{},y)=\left(\frac{\x_1\x_2}{y}\right)^R\,G_R(\x_1^{-1},\x_2^{-1},y^{-1})\,.
\eeq
This is due to the fact that the matrix $\mathcal{Z}$, given in \eqref{Zmatrix}, transforms to its inverse, up to a similarity transform, if points 1 and 3 are exchanged. In terms of the solution of the WI \eqref{solutionWI}, equation \eqref{crossingCA} implies that 
the single variable function $f_R(x)$ satisfies a crossing equation on its own:
\beq
\label{crossingSINGLEvar}
f_R(x)\,=\,x^R\,f_R(x^{-1})\,.
\eeq
Note that the above is a specialization of \eqref{crossingCA} to $(\x_1,\x_2,y)=(x,y,y)$, and thus corresponds to the crossing equation for the two-dimensional correlator of a dimension $\tfrac{R}{2}$ operator.
The function $f_R(x)$ is easily argued to be a polynomial of degree $R$ in $x$ as we shall show in section~\ref{sec:determinationoff}. 
Imposing \eqref{crossingSINGLEvar}, together with the normalization $f_R(0)=1$,
 implies this function is fixed in terms of $\tfrac{R}{2}$
 (respectively $\tfrac{R-1}{2}$)  independent parameters for $R$ even (respectively  odd). 

The remaining constraints from crossing symmetry \eqref{crossingCA} translate into the following equation
\beq\label{Hcrossing1}
\frac{(\x_1\x_2)^{R+1}}{y^{R-2}}
\,H_R(\x_1^{-1},\x_2^{-1},y^{-1})-H_R(\x_1^{},\x_2^{},y)=\frac{y^{2-R}}{\x_1^{-1}-\x_2^{-1}}
\left(
\x_2\,A_R(\x_2,y)\,f_R(\x_1)-
\x_1\leftrightarrow \x_2\right) \,,
\eeq
where we have made use of \eqref{crossingSINGLEvar}, and defined
\beq\label{ARdef}
A_R(x,y)\,:=\,\frac{x^{R-1}-y^{R-1}}{x^{}-y}\,,
\eeq
which is a polynomial in its arguments.
Recall that $H_R(\x_1^{},\x_2^{},y)$ is a polynomial of degree $R-2$ in $y^{-1}$ and thus, with the exception of $R=2$, \eqref{Hcrossing1} encodes a system of crossing equations.
%

\subsubsection*{Second equation}
In the channel where one takes the OPE of the two chiral operators it is convenient to relabel the points in \eqref{eq:BBbBBbcorr}
to obtain
\beq
\begin{split}
\label{eq:BBbBBbcorr2}
\langle \hat{\BB}_{[R,0]}(1) \hat{\BB}_{[R,0]}(2)  \hat{\BB}_{[0,R]}(\bar 3) \hat{\BB}_{[0,R]}(\bar 4) \rangle 
 \,&=\,
 \left[(1\bar{3})\,(2\bar{4})\,
\left(\frac{\tilde{w}}{\tilde{z}\tilde{\bar{z}}}\right)
  \right]^{R}
 \widetilde{G}_R(\tilde{z},\tilde{\bar{z}},\tilde{w})\\
 &=\, (2\bar{3})^R\,(1\bar{4})^R\,G_R(\widehat{z}_1,\widehat{z}_2,\widehat{y}) \,,
 \end{split}
\eeq
where we have defined
\beq\label{Zhatmatrix}
\widehat{\mathcal{Z}}\,:=\,
X_{2\bar{3}}^{}\,
X_{1\bar{3}}^{-1}\,
X_{1\bar{4}}^{}\,
X_{2\bar{4}}^{-1}
\sim\,
\text{diag}(\widehat{z}_1,\widehat{z}_2,\widehat{y})
\,,
\eeq
and
$(\tilde{z},\tilde{\bar{z}},\tilde{w}):=(1-\widehat{z}_1,1-\widehat{z}_2,1-\widehat{y}_1)$.
If the superspace coordinates
are $\SV=\overline{\SV}=0$, the cross-ratios above are related to the ones entering 
\eqref{eq:BBbBBbcorr} as 
$(\tilde{z},\tilde{\bar{z}},\tilde{w})=(z,\bar{z},w)$ 
and $(\widehat{z}_1,\widehat{z}_2,\widehat{y})=((1-\x_1)^{-1},(1-\x_2)^{-1},(1-y)^{-1})$.
The first equality in \eqref{eq:BBbBBbcorr2} is to be understood as defining the function 
$\widetilde{G}_R$.\footnote{
The strange prefactor is the natural supersymmetric completion of  $(x_{12}^{2}x_{34}^2)^{-R}$.}
The second one is a rewriting of \eqref{eq:BBbBBbcorr}, relating the chiral channel to the non-chiral channel.
The function $\widetilde{G}_R$ satisfies the same superconformal Ward identities as $G_R$.
We thus parametrize it as in \eqref{solutionWI}, with the functions $f_R$ and $H_{R}$ replaced by $\widetilde{f}_R$ and $\widetilde{H}_R$,
and the variables $\x_{1},\x_2,y$ replaced by $\tilde{z},\tilde{\bar{z}},\tilde{w}$.
 An immediate consequence of \eqref{eq:BBbBBbcorr2} is the relation
  \beq
  \label{eq:crossing2}
\widetilde{G}_R(z,\bar{z},w) \,=
\left(\frac{(\tfrac{z}{1-z})(\tfrac{\bar{z}}{1-\bar{z}})}{(\tfrac{w}{1-w})}
\right)^R\,
  G_R(1-z,1-\bar{z},1-w)\,.
  \eeq
Note that $\tilde{z},\tilde{\bar{z}},\tilde{w}$ are the usual cross-ratios for the correlator \eqref{eq:BBbBBbcorr2}, and so we rename them as $(\tilde{z},\tilde{\bar{z}},\tilde{w})\rightarrow (z,\bar{z},w)$.
The relation \eqref{eq:crossing2}
 implies a relation for the single variable function $\widetilde{f}_R$: 
\beq
\label{fandftilderel}
\widetilde{f}_R(z) \,=
\left(\frac{z}{1-z}\right)^{R}\,
  f_R(1-z)\,,
\eeq
which again follows from the fact that the single variable functions are capturing a two-dimensional correlator. For the $\tilde{H}$ function we get from \eqref{eq:crossing2} that
\beq
\label{secondequationHs}
(-1)^R\,\frac{(\x_1\x_2)^{R+1}\!\!\!\!}{y^{{R-2}}}
\, H_R(1-z,1-\bar{z},1-w)-\widetilde{H}_R(z,\bar{z},w)\,=\,
\frac{\x_1^{R-1}A_R(\x_1,y)\,\widetilde{f}_{R}(\bar{z})-(z \leftrightarrow \bar{z})}{z^{-1}-\bar{z}^{-1}}\,,
\eeq
where $A_R(x,y)$ was defined in \eqref{ARdef} and we remind that in \eqref{z1z2yfromzzw} we set
$(\x_1,\x_2,y)=(\tfrac{z}{z-1},
\tfrac{\bar{z}}{\bar{z}-1},\tfrac{w}{w-1})$.
As in the first crossing equation \eqref{Hcrossing1}, the dependence on $w$ disappears from 
\eqref{secondequationHs} for $R=2$.

\subsubsection*{Third equation} 
Since we consider the case of identical $\hat{\BB}_{[R,0]}$ operators, Bose symmetry under the exchange $1\leftrightarrow 2$ in \eqref{eq:BBbBBbcorr2} requires
\beq
\label{eq:crossing3}
\widetilde{G}_R(z,\bar{z},w) \,=
 (-1)^R \widetilde{G}_R(\tfrac{z}{z-1},\tfrac{\bar{z}}{\bar{z}-1},\tfrac{w}{w-1})\,.
  \eeq
Plugging in the solution of the WI and using \eqref{idforWI}, the above equation translates into 
 \beq
 \label{thirdHandf}
 \widetilde{f}_R(z)=
 (-1)^R\widetilde{f}_R(\tfrac{z}{z-1})\,,
 \qquad
 \widetilde{H}_{R}(z,\bar{z},w)=
 (-1)^R\widetilde{H}_R(\tfrac{z}{z-1},\tfrac{\bar{z}}{\bar{z}-1},\tfrac{w}{w-1})\,,
 \eeq
with the first equation again following from Bose symmetry in the chiral algebra.
 


\subsection{From the chiral algebra to numerics}
\label{subsubsec:chiraltonumerics}

In the following subsections we turn the crossing equations
 \eqref{Hcrossing1}, \eqref{secondequationHs}, \eqref{thirdHandf}
 into a system ready for the numerical analysis, by fixing all the chiral algebra data.
To do so we proceed as follows:
\begin{itemize}
\item The first step, undertaken in subsection \ref{sec:determinationoff}, is to analytically solve the chiral algebra crossing equations for  $f_{R}(x)$ and $\tilde{f}_{R}(x)$.
\item Decomposing these functions in the blocks of tables \ref{tab:blockssummary} and \ref{tab:blockssummary2} allows us to fix an infinite number of Schur operator OPE coefficients. We recall these operators are the ones in the OPEs \eqref{selectionchiantichi} and \eqref{chiralchiralselectionsrules} that contribute to the chiral algebra, see \eqref{eq:CCSchur} and the following.
\item From the block decomposition of $H_R$, $\widetilde{H}_R$, also given in tables \ref{tab:blockssummary} and \ref{tab:blockssummary2}, we see that they receive contributions from some of the multiplets contributing to  $f_{R}(x)$ and $\tilde{f}_{R}(x)$. 
\item Therefore we split the expansion into a sum over the exchange of Schur operators
 \beq
 \label{eq:exchangedSchur}
\hat{ \mathcal{B}}_{[a,a]}\,,\quad \hat{\mathcal{C}}_{[a,a],\ell}\,,\quad \hat{\mathcal{B}}_{[2R,0]}\,,\quad \hat{\mathcal{C}}_{[2R-2,0](\tfrac{\ell+1}{2},\tfrac{\ell}{2})}\, ,
  \eeq
  and a sum of the remaining operators.\footnote{
 This split does not coincide in general with the separation between long and short operators, as can be seen in the \chiral channel.} 
 \item  The final step is to sum the contribution of Schur operators to the  $H_R$ and $\widetilde{H}_R$ functions,
 denoted  as $H_{R,\text{short}}$ and $\widetilde{H}_{R,\text{short}}$ in the following. (We sometimes omit the index $R$ and write just $H_{\text{short}}$ and $\widetilde{H}_{\text{short}}$.) We deal with these functions in section~\ref{subsec:ContributionofSchur} and our final results are  given in \eqref{firttermto_toesum} and \eqref{eq:expressionforwidetildeHRshort}.
\end{itemize}
In general, knowledge of the function $f_R(x)$ alone is not sufficient to determine $H_{R,\text{short}}$  unambiguously, in contrast with the \chiral channel function $\widetilde{H}_{R,\text{short}}$, which is fixed in terms of $\tilde{f}_R(x)$.
 This is because different $\mathcal{N}=3$ supermultiplets give the same contribution,  in the sense of $2d$ blocks,  to the functions $f_R(x)$ and $\tilde{f}_R(x)$.
 As we will see in section~\ref{subsec:ContributionofSchur}, \emph{assuming the absence of supermultiplets containing conserved currents of spin greater than two}, the function $\widetilde{H}_{R, \text{short}}$
  and the component of  $H_{R,\text{short}}$ 
  in the R-symmetry singlet channel can be extracted \emph{unambiguously} from the knowledge of $f_R(x)$.

\subsubsection*{Summary of the result}
Following the procedure that we just outlined  one arrives at the following system of crossing equations:
\begin{align}\label{fullbootstrap_equations}
\sum_{\chi\,\in\,\hat{\mathcal{B}}_{[R,0]}\times \hat{\mathcal{B}}_{[0,R]}|_{\mathcal{A}}}
\!\!\!\!\!\!
 |\lambda_{\chi}|^2\,
\begin{bmatrix}
\mathcal{F}_{\chi}\\
+\mathcal{F}^{\mathsf{b}}_{-, \chi}\\
-\mathcal{F}^{\mathsf{b}}_{+,\chi}
\end{bmatrix}
\,\,\,\,\,\,+\!\!
\sideset{}{'}
\sum_{\chi\,\in\,\hat{\mathcal{B}}_{[R,0]}\times \hat{\mathcal{B}}_{[R,0]}
|_{\mathcal{A},\overline{\mathcal{C}},\overline{\mathcal{B}}}}
\!\!\!\!\!\!
 |\widetilde{\lambda}_{\chi}|^2
\begin{bmatrix}
0\\
\widetilde{\mathcal{F}}_{-,\chi}\\
\widetilde{\mathcal{F}}_{+,\chi}
\end{bmatrix}=
\begin{bmatrix}
\mathcal{F}^{(0)}_{\text{short}}\\
\mathcal{F}^{(-)}_{\text{short}}\\
\mathcal{F}^{(+)}_{\text{short}}
\end{bmatrix}\,.
\end{align}
We now have to make several remarks in order to explain our notation. 
\begin{itemize}
\item[a)] We have defined the functions
\begin{align}
\label{Fchi1}
\mathcal{F}_{\chi}\,&=\,
\frac{\left[(1-z)(1-\zb)\right]^{R+1}}{(1-w)^{R-2}}\,\Hc_{\chi}(z,\zb,w)
-(-1)^R
\Big[(z,\zb,w)\leftrightarrow (1-z,1-\zb,1-w)\Big]\,,\\
\label{Fbraided}
\mathcal{F}^{\mathsf{b}}_{\pm,\chi}\,&=\,
\frac{\left[(1-z)(1-\zb)\right]^{R+1}}{(1-w)^{R-2}}\,\Hc_{\chi}(\tfrac{z}{z-1},\tfrac{\zb}{\zb-1},\tfrac{w}{w-1})\pm
\Big[(z,\zb,w)\leftrightarrow (1-z,1-\zb,1-w)\Big]\,,\\
\label{Ftilde}
\widetilde{\mathcal{F}}_{\pm,\chi}\,&=\,
\frac{\left[(1-z)(1-\zb)\right]^{R+1}}{(1-w)^{R-2}}\,\widetilde{\Hc}_{\chi}(z,\zb,w)
\pm
\Big[(z,\zb,w)\leftrightarrow (1-z,1-\zb,1-w)\Big]\,.
\end{align}
The explicit form of the functions $\Hc_{\chi}$, $\widetilde{\Hc}_{\chi}$ is given in tables  \ref{tab:blockssummary} and 
 \ref{tab:blockssummary2} for each representation $\chi$. Note that the above functions still have a dependence on the R-symmetry cross-ratio, and thus each equation in \eqref{fullbootstrap_equations} will give several equations, once this dependence is expanded out.
\item[b)] The functions $\mathcal{F}^{(0,\pm)}_{\text{short}}$ receive contributions from two sources. The first one comes from the right hand side of  
 \eqref{Hcrossing1}, \eqref{secondequationHs} and contains the function $f_{R}$ explicitly.  The second one corresponds to the contribution of Schur operators to the left hand side of  \eqref{Hcrossing1}, \eqref{secondequationHs}.
Specifically, we have
\beq
\label{eq:explicitformFshort}
\mathcal{F}^{(0,\pm)}_{\text{short}}\,=\,
\mathcal{F}^{(0,\pm)}_{\text{short}}[f]-
\mathcal{F}^{(0,\pm)}_{\text{short}}[
H_{\text{short}},\widetilde{H}_{\text{short}}]\,,
\eeq
with the explicit form of $\mathcal{F}^{(0,\pm)}_{\text{short}}[f]$ and $\mathcal{F}^{(0,\pm)}_{\text{short}}[
H_{\text{short}},\widetilde{H}_{\text{short}}]$ given in appendix \ref{app:forFshort}.
\item[c)] The precise range of summation in \eqref{fullbootstrap_equations} is specified by the selection rules
\eqref{selectionchiantichi} and  \eqref{chiralchiralselectionsrules}, where we only take the operators that are
 \textit{not} of  Schur type, \ie, $\mathcal{A}$ in the \nonchiral channel and $\mathcal{A}, \overline{\mathcal{B}}, \overline{\mathcal{C}}$ in the \chiral one. The prime in the second sum $\sum'$ indicates that the parity of the spin label $\ell$ of the exchanged operator is fixed in terms of its  R-symmetry representation. Specifically, only even spins appear for irreps in the $\mathrm{sym}([R,0]\otimes[R,0])$, while for irreps in $\mathrm{antisym}([R,0]\otimes[R,0])$ only odd spins appear. This follows from \eqref{thirdHandf} and the braiding relations \eqref{braidingsuperblocks} of individual blocks.
\item[d)] The second and third equations in  \eqref{fullbootstrap_equations} are obtained respectively from the antisymmetrization and symmetrization 
of the superconformal block expansion of \eqref{secondequationHs} with respect to the exchange
 $(z,\zb,w)\leftrightarrow (1-z,1-\zb,1-w)$.
 An important remark, relevant for the numerical implementation, is that the arguments of four dimensional superconformal and 
 R-symmetry blocks entering \eqref{Fbraided}, namely $(\tfrac{z}{z-1},\tfrac{\zb}{\zb-1},\tfrac{w}{w-1})$ and their inverses,
  can be traded for $(z,\bar{z},w)$ and $(1-z,1-\bar{z},1-w)$ using the braiding properties of the conformal blocks \eqref{braidingsuperblocks}.
This fact justifies the use of the suffix $\mathsf{b}$  to denote ``braided'' in \eqref{Fbraided}.
\item[e)] Finally, as customary,   the identification $|\lambda_{\chi}|^2\,=\,\sum_{\mathcal{O}| \chi(\mathcal{O})=\chi}|\lambda_{\mathcal{O}}|^2$ is understood. By $\chi(\mathcal{O})$ we mean the representation $\chi$ in which the operator $\mathcal{O}$ transforms.
 \end{itemize}

\subsection{Fixing the chiral algebra contributions}
\label{subsec:ContributionofSchur}

We have defined above the functions $H_{R,\text{short}}$ and $\widetilde{H}_{R,\text{short}}$  as the contribution 
from the exchange of  Schur  operators to the $H_R$ and $\widetilde{H}_R$ functions, entering 
 \eqref{Hcrossing1} and \eqref{secondequationHs}. We will now discuss to which extent these contributions can be extracted from the knowledge of $f_R(x)$, or more generally,
 from the knowledge of the chiral algebra.

\subsubsection{Determination of the function \texorpdfstring{$f_R(x)$}{f}}
\label{sec:determinationoff}

The cohomological reduction of the correlator 
\eqref{eq:BBbBBbcorr},  which in superspace corresponds to a specialization of the superspace coordinates 
in  \eqref{Xandstuff} to $X=\text{diag}(z,y,y)$, $\SV=(\theta,0,0)^T$ and $\overline{\SV}=(\bar{\theta},0,0)$,
gives the holomorphic correlator
\beq
\label{eq:BBbBBbcorrreduced}
 \langle \WW(z_{-,1},\theta_1) \WB (z_{+,2},\bar{\theta}_2)  \WW(z_{-,3},\theta_3) \WB (z_{+,4},\bar{\theta}_4)\rangle 
 = \frac{f_R(x)}{(z_{1\bar{2}}z_{3\bar{4}})^R}\,,
 \qquad x=\frac{z_{1\bar{2}}z_{3\bar{4}}}{z_{3\bar{2}}z_{1\bar{4}}}\,,
\eeq
where $z_{\pm}=z\pm \tfrac{1}{2}\theta\bar{\theta}$ and $z_{1\bar2}=z_{1,-}-z_{2,+}+\theta_1\bar\theta_2$.
For the following discussion we set the fermionic coordinates $\theta=\bar{\theta}=0$.
We can view the correlator above as a meromorphic function of $z_1$,  whose poles correspond to singular terms in the OPE of $\mathcal{W}(1)$ with the remaining operators. The \chiral OPE is non-singular, so there is no pole when $z_1\sim z_3$ (corresponding to $x\sim 1$).
The singularity  for $z_1\sim z_2$ (corresponding to $x\sim 0$), on the other hand, is already taken care of by 
the prefactor in the right hand side of \eqref{eq:BBbBBbcorrreduced}.
Finally, for  $z_1\sim z_4$ (corresponding to $x\sim \infty$) we have $f_R(x)\sim x^R$.
There is no other singularity and so $f_R(x)$ is a polynomial of degree $R$ in $x$, that we normalize as $f_R(0)= 1$,
subject to the crossing relation \eqref{crossingSINGLEvar}. It is thus fixed in terms of 
$\left \lfloor{\tfrac{R}{2}}\right \rfloor$ constants.
If follows from the exchange of the two-dimensional stress tensor that the small $x$ expansion of the correlator takes the form
 $f_R(x)=1+\tfrac{R^2}{4c_{4d}}x+\dots$, where $c_{4d}\equiv c$ is the central charge of the four-dimensional theory, thus fixing one of the $\left \lfloor{\tfrac{R}{2}}\right \rfloor$ constants.\footnote{If the subscript is omitted, it is understood that $c$ is the four-dimensional central charge.}
For $R=1$ the crossing relation \eqref{crossingSINGLEvar} implies that $f_{1}(x)=1+x$, forcing the central charge to take the value $c_{4d}=\frac{1}{4}$, which corresponds to  $\mathcal{N}=4$ SYM with gauge group $\U(1)$.

\subsubsection*{Non-chiral channel}
Consider the expansion of the function $f_R(x)$ in holomorphic $\mathcal{N}=2$ (global $\ospf(2|2)$) blocks as
\beq
\label{fexpansion}
f_{R}(x)=1+\sum_{h=1}^{\infty}\,b_{h}^{(R)}\,\Ssl_h(x)\,.
\eeq
Using the result given in table \ref{tab:blockssummary}, and the selection rules 
  \eqref{selectionchiantichi},
it is clear that in general one cannot reconstruct the four-dimensional 
OPE coefficients corresponding to Schur operators \eqref{eq:exchangedSchur} from the knowledge of the expansion \eqref{fexpansion}.
This is best illustrated by looking at the following examples
\begin{align}
\label{lambdafromb}
R&=1\,:\,\quad
 b_1^{(1)}=|\lambda^{\hat{\mathcal{B}}}_1|^2\,,
\quad
b_{h>1}^{(1)}=(-1)^{h+1}|\lambda^{\hat{\mathcal{C}}}_{0,h-2}|^{2}\,,\nonumber\\
R&=2\,:\,
\quad
 b_1^{(2)}=|\lambda^{\hat{\mathcal{B}}}_1|^2\,,
\quad
b_{2}^{(2)}=|\lambda^{\hat{\mathcal{B}}}_{2}|^{2}\,,\,
\quad
b_{h>2}^{(2)}=(-1)^{h}
|\lambda^{\hat{\mathcal{C}}}_{1,h-3}|^{2}\,,
\\
R&=3\,:\,
\quad
 b_1^{(3)}=|\lambda^{\hat{\mathcal{B}}}_1|^2\,,
\quad
b_{2}^{(3)}=|\lambda^{\hat{\mathcal{B}}}_{2}|^{2}\,,
\quad
b_{3}^{(3)}=
|\lambda^{\hat{\mathcal{B}}}_{3}|^{2}\,-
|\lambda^{\hat{\mathcal{C}}}_{1,0}|^{2}\,,
\quad
b_{h>3 }^{(3)}=
(-1)^h(|\lambda^{\hat{\mathcal{C}}}_{1,h-3}|^{2}
-|\lambda^{\hat{\mathcal{C}}}_{2,h-4}|^{2})\,,\nonumber
\end{align}
and so on.
Above we used the compact notation $\lambda^{\hat{\mathcal{B}}}_a\equiv \lambda^{(R)}_{\hat{\mathcal{B}}_{[a,a]}}$
and $\lambda^{\hat{\mathcal{C}}}_{a,\ell}\equiv\lambda^{(R)}_{\hat{\mathcal{C}}_{[a,a],\ell}}$.
Of course, $\lambda$'s from different rows (\ie, for external operators with different values of $R$) in \eqref{lambdafromb} are not the same, even though this is
not captured by the notation.
The general pattern is quite simple and one finds 
\beq
 b^{(R)}_{1\leq h\leq R  }=|\lambda^{\hat{\mathcal{B}}}_{h}|^{2}
-\sum_{a=1}^{h-2}(-1)^{h-a}\,|\lambda^{\hat{\mathcal{C}}}_{a,h-a-2}|^{2}\,,
\qquad
b^{(R)}_{h>R }=-\sum_{a=1}^{R-1}(-1)^{h-a}\,|\lambda^{\hat{\mathcal{C}}}_{a,h-a-2}|^{2}\,.
\eeq

Note that compared to the results that can be obtained from table \ref{tab:blockssummary} and the selection rules 
  \eqref{selectionchiantichi}, we \textit{omitted by hand} the supermultiplets $\hat{\mathcal{C}}_{[0,0],\ell}$ for external fields with $R\geq 2$, because they are the supermultiplets  that contain higher spin conserved currents. 
They are included only in the  free field case $R=1$.
For $R\geq 2$, while allowed  
by the selection rules   \eqref{selectionchiantichi},
 we want to demand that they are absent in order to focus on interacting theories.
We remark further that the OPE coefficient 
\beq
b_1^{(R)}\,=\,|\lambda^{(R)}_{\hat{\mathcal{B}}_{[1,1]}}|^2\,=\,\frac{R^2}{4c_{4d}}\,,
\eeq
corresponding to the exchange of the stress-tensor
supermultiplet $\hat{\mathcal{B}}_{[1,1]}$ can be extracted unambiguously.

It follows from the above considerations
 that also $|\lambda^{\hat{\mathcal{B}}}_{2}|^{2}$ can be extracted without ambiguity.
However, in general, the four-dimensional OPE coefficients cannot be extracted uniquely from the expansion \eqref{fexpansion}.
As  discussed in section~\ref{sec:fixingB33} and section~\ref{sec:boundingOPEcoeff} using the knowledge of the chiral algebra and some extra assumptions one can 
find, in the case $R=3$, only two allowed values  for  
$\lambda^{(3)}_{\hat{\mathcal{B}}_{[3,3]}}$ and $\lambda^{(3)}_{\hat{\mathcal{C}}_{[1,1],0}}$.

Let us now investigate the structure of $H_{R,\text{short}}$. By definition, we have
\beq
\label{eq:definitionofHRshort}
H_{R,\text{short}}\,\colonequals\,\sum_{a=2}^R\,
|\lambda_{\hat{\mathcal{B}}_{[a,a]}}|^2\,\Hc_{\hat{\mathcal{B}}_{[a,a]}}+
\sum_{a=1}^{R-1}\sum_{\ell=0}^{\infty}
|\lambda_{\hat{\mathcal{C}}_{[a,a],\ell}}|^2\,\Hc_{\hat{\mathcal{C}}_{[a,a],\ell}}\,,
\eeq
which we can express in terms of the blocks $h_{[a,a]}(w)$ and $\mathcal{G}^{\mathcal{N}=1}_{\Delta,\ell}(z,\zb)$
 given in  \eqref{Rblocks}, \eqref{N=1blocks} as
\beq
\label{firttermto_toesum}
\begin{split}
&H_{R,\text{short}}(\x_1,\x_2,y)\,=\,\sum_{h=2}^{\infty}\,
(-1)^h \,b_h^{(R)}\,\mathcal{G}^{\mathcal{N}=1}_{h+2,h-2}(z,\bar z )\\&-
\begin{cases}
-h_{[0,0]}(w) \sum_{n=0}^{\infty} |\lambda^{(R)}_{\hat{\mathcal{C}}_{[0,0],n}}|^2\mathcal{G}^{\mathcal{N}=1}_{n+4,n}\,,\,\,\,\, & R=2 \\
h_{[1,1]}(w)\sum_{n=0}^{\infty} \, 
|\lambda^{(R)}_{\hat{\mathcal{C}}_{[2,2],n-1}}|^2\,
\mathcal{G}^{\mathcal{N}=1}_{n+6,n}(z,\zb)
-h_{[0,0]}(w) \sum_{n=0}^{\infty} |\lambda^{(R)}_{\hat{\mathcal{C}}_{[0,0],n}}|^2\mathcal{G}^{\mathcal{N}=1}_{n+4,n}\,,\,\,\,\,
& R=3\\
\sum_{t=0}^{R-2}\,h_{[t,t]}(w)\,C_t^{(R)}(z,\zb)
\,,\,\,\,\, &\text{general $R$}
\end{cases}\,.
\end{split}
\eeq
In \eqref{eq:definitionofHRshort} the first summation starts from $h=2$, since $\Hc_{\hat{\mathcal{B}}_{[1,1]}}=0$.
In writing this equation we allowed for higher-spin currents to have a non-vanishing OPE coefficient, such that it becomes clear how they would contribute to the crossing equations. 
Looking at table \ref{tab:blockssummary} we see that if higher-spin currents are present they contribute exactly the same way as the R-symmetry singlet long multiplet at the unitarity bound $\Delta= \ell + 2$.
After setting them zero $|\lambda^{(R)}_{\hat{\mathcal{C}}_{[0,0],n}}|^2=0$, only the part of $H_{R,\text{short}}$ in the R-symmetry singlet channel is completely fixed in terms of the function $f_R(x)$. The explicit expression for the function $C_t^{(R)}(z,\zb)$ is easily worked out, but will not be relevant here.
We finally remark that the summation of the first term in \eqref{firttermto_toesum} can be done explicitly for any $R$. See appendix \ref{app:summationforFshort} for details.

\vspace{0.5cm}
\noindent
\emph{Example:} For $R=2$, we find
\beq
\label{eq:decompositionR2offR}
f_2(x)=1+c^{-1}x +x^2=1+\sum_{h=1}^{\infty} b_h^{(2)}\Ssl_h(x)\,,
\eeq
where from \eqref{lambdafromb} we take,
$|\lambda_{\hat{\mathcal{B}}_{[1,1]}}|^2=b_1^{(2)}=c^{-1}, |\lambda_{\hat{\mathcal{B}}_{[2,2]}}|^2=b_2^{(2)}=1-\tfrac{1}{3c}$ and
\beq
|\lambda_{\hat{\mathcal{C}}_{[1,1],\ell}}|^2= (-1)^{\ell+1}b_{\ell+3}^{(2)}=\frac{ (\ell+2) (4)_{\ell+1}}{2^{2\ell+2}\left(\frac{5}{2}\right)_{\ell+1}}-\frac{\Gamma (\ell+4)}{2^{2 \ell+5}  \left(\frac{1}{2}\right)_{\ell+3}}c^{-1}\,.
 \eeq 
 Note that if higher-spin currents are present the above identification of OPE coefficients with $b_h$ cannot be made for $h>1$.

\subsubsection*{Chiral channel}
Now we expand the function $\widetilde{f}_{R}$, related to $f_{R}$ by \eqref{fandftilderel},
in $\mathcal{N}=2$ holomorphic blocks, which in this channel coincide with ordinary $\slf(2)$ blocks, see \eqref{eq:sl22dblocksdefinition}. 
Specifically
\beq\label{fexpansion2}
\widetilde{f}_{R}(z)=
\sum_{\substack{h=R\\ h+R  \,\,\,\text{even}}}^\infty
\,\,\widetilde{b}_{h}^{(R)}\,    \sl_{h}(z)\,,
\eeq
where we note that the sum starts from $h=R$, which is due to the fact that the relevant OPE is non-singular. 
Moreover, the index $h$ has the same parity as $R$ as follows from the braiding relations of individual blocks \eqref{braidingSL2} 
together with \eqref{thirdHandf}. 
By looking at the selection rules given in \eqref{chiralchiralselectionsrules},
and after a quick look at table  \ref{tab:blockssummary2}, one concludes that 
\beq
\widetilde{b}_{R}^{(R)}\,=\,\big|\widetilde{\lambda}_{\hat{\mathcal{B}}_{[2R,0]}}\big|^2\,,
\qquad
\widetilde{b}_{R+1}^{(R)}=0\,,\qquad 
\widetilde{b}_{\ell+R+2}^{(R)}\,=\,\big|\widetilde{\lambda}_{\hat{\mathcal{C}}_{[2R-2,0],((\ell+1)/2,\ell/2)}}\big|^2\,,
\eeq
where $\ell\geq 0$.
Note that in this channel we can reconstruct the four-dimensional OPE coefficients of Schur operator
completely in terms of the OPE coefficients of the cohomologically  reduced problem.
We can thus uniquely determine the contribution of these operators to $\widetilde{H}_R$:
\begin{align}
\label{eq:expressionforwidetildeHRshort}
\widetilde{H}_{R,\text{short}}\,=&\,
\widetilde{b}_{R}^{(R)}
\,\widetilde{\Hc}_{\hat{\mathcal{B}}_{[2R,0]}}+\,\,
\sum_{n=0}^{\infty}
\widetilde{b}_{R+2n+2}^{(R)}
\,\widetilde{\Hc}_{\hat{\mathcal{C}}_{[2(R-1),0],(n+\frac{1}{2},n)}}\,.
\end{align}
The summation of this expression is straightforward and similar to the one done in appendix \ref{app:summationforFshort}.
The final result is given by
\beq
\widetilde{H}_{R,\text{short}}(z,\zb,w)\,=\,
\frac{\widetilde{f}_R(\zb)\,\gamma_R(z,w)-(z\leftrightarrow \zb)}
{z^{-1}-\zb^{-1}}\,,
\eeq
where we have defined the kinematic factor
\beq
\gamma_R(z,w)=\sum_{a=0}^{R-2} k_{2a+2}(z)h_a^{\SU(2)}(w)\,,
\eeq 
with  $k_{\beta}(z)$ and $h_a^{\SU(2)}(w)$ given in \eqref{app:kdef} and \eqref{Rblocks1} respectively.

We have now obtained explicit expressions for $H_{R,\text{short}}$ and $\widetilde{H}_{R,\text{short}}$, and can compute the functions $\mathcal{F}_{\text{short}}^{(0,\pm)}$ entering the crossing equations \eqref{fullbootstrap_equations}. See appendix \ref{app:forFshort}
for more details.


\subsection{Explicit form of the bootstrap equations for \texorpdfstring{$R=2,3$}{R=2,3}}
\label{subsec:explicitboostrapequationsforR23}

We will now show the explicit form of \eqref{fullbootstrap_equations}  in the cases $R=2,3$. In order to do so, it is convenient to define the combinations of conformal blocks (compare to \eqref{Fchi1}, \eqref{Fbraided}, \eqref{Ftilde})
\beq
\label{Req2FfromFc}
\begin{split}
\mathsf{F}_{\pm,\Delta, \ell} &
:=  \left[(1-z) (1-\zb)\right]^{R+1}  (z \zb)^{-\tfrac{1}{2}}\g^{1,1}_{\Delta+3,\ell}(z,\zb)\pm \left[(z,\zb)\leftrightarrow (1-z,1-\zb) \right]\,,\\ %
\mathsf{F}^{\mathsf{b}}_{\pm, \Delta, \ell} 
&:= (-1)^\ell
 \left[(1-z) (1-\zb)\right]^{R+2} (z \zb)^{-\tfrac{1}{2}}\g^{-1,1}_{\Delta+3,\ell}(z,\zb)\pm\left[ (z,\zb)\leftrightarrow (1-z,1-\zb)\right]\,,\\
\widetilde{\mathsf{F}}_{\pm, \Delta, \ell}&:=
\left[ (1-z) (1-\zb)\right]^{R+1} \g^{0,0}_{\Delta+3,\ell}(z,\zb) \pm \left[  (z,\zb)\leftrightarrow (1-z,1-\zb)\right]\,.
\end{split}
\eeq
As before, we suppressed the index $R$ from the notation, its value should be clear from the context.
\begin{table}[h]
\centering
\renewcommand{\arraystretch}{1.6}
\begin{tabular}{%
| l
                |>{\centering }m{3.2cm}
             |>{\centering\arraybackslash}m{6.35cm}|
}
\hline
\text{Multiplet } $\chi$
& 
$\fc_\chi(\x)$
&
$\Hc_\chi(z,\bar z )$
\\\hline
$\text{Identity}$
&$\Ssl_{0}(x)=1$
& 0
\\\hline
$\hat{\BB}_{[1,1]}$
&$\Ssl_{1}(x)	$
& 0
\\\hline
$\hat{\CC}_{[0,0],\ell}$
&$(-1)^{\ell+1}\Ssl_{\ell+2}(x)$
&$0$
\\\hline
$\hat{\BB}_{[2,2]}$
&$\Ssl_{2}(x)$
& $\mathcal{G}^{\calN=1}_{4,0}=(z \zb)^{-\tfrac{1}{2}}\g^{1,1}_{5,0}(z,\zb)$
\\\hline
$\hat{\CC}_{[1,1],\ell}$
&$(-1)^{\ell+1} \Ssl_{\ell+3}(x)$
&$ \mathcal{G}^{\calN=1}_{\ell+5,\ell+1}=(z \zb)^{-\tfrac{1}{2}}\g^{1,1}_{\ell+6,\ell+1}(z,\zb)$ 
\\\hline
$\AA_{[0,0],\ell}^{\Delta > \ell+2}$
&0	
& $\mathcal{G}^{\calN=1}_{\Delta+2,\ell}=(z \zb)^{-\tfrac{1}{2}}\g^{1,1}_{\Delta+3,\ell}(z,\zb)$
\\\hline
\end{tabular}
\renewcommand{\arraystretch}{1.0}
\caption{Contributions of the various $\Nm=3$ multiplets appearing in the \nonchiral OPE \eqref{selectionchiantichi}, for $R=2$, to the functions $\fc_\chi(x_1)$ and $\Hc_\chi(z,\bar z)$. The multiplets $\hat{\CC}_{[0,0],\ell}$ contain conserved currents of spin larger than two, and must be set to zero for an interacting theory \cite{Maldacena:2011jn,Alba:2013yda}. We recall that $\hat{\BB}_{[1,1]}	$ is the stress-tensor multiplet. When the long multiplet $\AA_{[0,0],\ell}^{\Delta > \ell+2}$ hits the unitarity bound $\Delta=\ell+ 2$ it decomposes in a $\hat{\CC}_{[0,0],\ell}$ and a $\hat{\CC}_{[1,1],\ell-1}$, where $\hat{\CC}_{[1,1],-1}=\hat{\BB}_{[2,2]}$. 
Note that while long multiplets arbitrarily close to the unitarity bound mimic higher-spin conserved currents, they do not mimic the stress tensor.
}
\label{tab:chiralachiralaeq2blocks}
\end{table}
\begin{table}[h]
\centering
\renewcommand{\arraystretch}{1.6}
\begin{tabular}{%
| l
                |>{\centering }m{3.2cm}
             |>{\centering\arraybackslash}m{6.35cm}|
}
\hline
\text{Multiplet} $\chi$
& 
$\tilde{\fc}_{\chi}(z) $
&
$\widetilde{\Hc}_\chi(z,\zb)$
\\\hline
$\hat{\BB}_{[4,0]}$
&$\sl_{2}(z)	$
& $\g_{4,0}(z,\zb) $
\\\hline
$\hat{\CC}_{[2,0], (\tfrac{\ell+1}{2}, \tfrac{\ell}{2})}$
&$\sl_{4+\ell}(z)$
&$\g_{\ell+6,\ell+2}(z,\zb) $
\\\hline
$\bar{\BB}_{[0,2], r=8, (0,0)}$
& 0
& $\g_{6,0}(z,\zb) $
\\\hline
$\bar{\CC}_{[0,1], r=7, (\tfrac{\ell+1}{2},\tfrac{\ell}{2})}$
&0
&$\g_{\ell+7,\ell+1}(z,\zb)$
\\\hline
$\mathcal{A}_{[0,0], r=6, \ell}^{\Delta > 3+\ell}$
& $0$
& $\g_{\Delta+2,\ell}(z,\zb)$
\\\hline
\end{tabular}
\renewcommand{\arraystretch}{1.0}
\caption{Contributions of the various $\NN=3$ multiplets appearing in the \chiral  OPE \eqref{chiralchiralselectionsrules}, for $R=2$, to the functions $\tilde{\fc}_\chi(z)$ and $\widetilde{\Hc}_\chi(z,\bar z)$.
Note that at the unitarity bound (see \eqref{unitboundchiralchiral}) of the long multiplet we find two types (for $\ell=0$ and $\ell \neq0$) of short multiplets which do not contribute to the chiral algebra, namely $\bar{\BB}_{[0,2], r=8, (0,0)}$ and $\bar{\CC}_{[0,1], r=7, (\tfrac{\ell+1}{2},\tfrac{\ell}{2})}$. When considering identical $\hat{\BB}_{[2,0]}$ operators Bose symmetry requires $\ell$ to be even for $\mathcal{A}_{[0,0], r=6, \ell}^{\Delta > 3+\ell}$ and odd for $\bar{\CC}_{[0,1], r=7, (\tfrac{\ell+1}{2},\tfrac{\ell}{2})}$.
}
\label{tab:chiralchiralaeq2blocks}
\end{table}
\subsubsection*{The case $R=2$}
The bootstrap equations  \eqref{fullbootstrap_equations}  in this case are independent of the R-symmetry variables $w$. Using the $R=2$ specializations of the tables~\ref{tab:blockssummary} and \ref{tab:blockssummary2}, namely table~\ref{tab:chiralachiralaeq2blocks} and \ref{tab:chiralchiralaeq2blocks}, we obtain
\begin{align}
\label{eq:Req2crossing}
\sum_{\Delta >  \ell + 2} |\lambda_{\Delta, \ell}|^2
\begin{bmatrix}
\,\mathsf{F}_{-,\Delta, \ell}\\
+\mathsf{F}^{\mathsf{b}}_{-, \Delta, \ell}\\
-\mathsf{F}^{\mathsf{b}}_{+, \Delta, \ell}
\end{bmatrix}
+
\sum_{\substack{\Delta \geqslant \ell + 3\\ \ell \text{ even }}} |\tilde{\lambda}_{\Delta, \ell}|^2
\begin{bmatrix}
0\\
\widetilde{\mathsf{F}}_{-, \Delta, \ell}\\
\widetilde{\mathsf{F}}_{+, \Delta, \ell}
\end{bmatrix}=
\begin{bmatrix}
\mathsf{F}^{(0)}_{\text{short}}\\
\mathsf{F}^{(-)}_{\text{short}}\\
\mathsf{F}^{(+)}_{\text{short}}
\end{bmatrix}\,.
\end{align}
Here $\lambda_{\Delta, \ell}$ and $\tilde{\lambda}_{\Delta, \ell}$ denote the OPE coefficients of the longs multiplets with dimension $\Delta$ and spin $\ell$ appearing in the \nonchiral and \chiral channels respectively, 
and $\mathsf{F}_{\text{short}}^{(0,\pm)}:=\mathcal{F}_{\text{short}}^{(0,\pm)}$, see appendix \ref{app:forFshort} .

\subsubsection*{The case $R=3$} 
For $R=3$ we unpack the superconformal blocks given in tables tables~\ref{tab:blockssummary} and \ref{tab:blockssummary2} in tables~\ref{tab:chiralachiralaeq3blocks} and \ref{tab:chiralchiralaeq3blocks}.
To write down the crossing equations   \eqref{fullbootstrap_equations}  in components we need to fix a basis in the space of R-symmetry polynomials. There is a natural choice which follows by noticing that 
\beq
\label{eq:Req3crossing}
\begin{split}
(1-w)\,\mathcal{F}_{\mathcal{A}^{\Delta,r=0}_{[0,0],\ell}}\,&=\,
+\tfrac{1}{2} (1-y^{-1})\,\mathsf{F}_{+,\Delta, \ell}+\tfrac{1}{2} (1+y^{-1})\,\mathsf{F}_{-,\Delta, \ell}\,,\\
(1-w)\,\mathcal{F}_{\mathcal{A}^{\Delta,r=0}_{[1,1],\ell}}\,&=\,
-\tfrac{2}{3} (1-y^{-1})\,\mathsf{F}_{+,\Delta, \ell}+\tfrac{1}{3} (1+y^{-1})\,\mathsf{F}_{-,\Delta, \ell}\,,\\
\mathcal{F}^{\mathsf{b}}_{\pm,\mathcal{A}^{\Delta,r=0}_{[0,0],\ell}}\,&=\,
+\tfrac{1}{2}\left(\tfrac{1}{1-w}+\tfrac{1}{w}\right)\,\mathsf{F}^{\mathsf{b}}_{\pm,\Delta, \ell}+
\tfrac{1}{2} \left(\tfrac{1}{1-w}-\tfrac{1}{w}\right)\,\mathsf{F}^{\mathsf{b}}_{\mp,\Delta, \ell}\,,\\
\mathcal{F}^{\mathsf{b}}_{\pm,\mathcal{A}^{\Delta,r=0}_{[1,1],\ell}}\,&=\,
+\tfrac{5}{6}\left(\tfrac{1}{1-w}+\tfrac{1}{w}\right)\,\mathsf{F}^{\mathsf{b}}_{\pm,\Delta, \ell}
-\tfrac{1}{6} \left(\tfrac{1}{1-w}-\tfrac{1}{w}\right)\,\mathsf{F}^{\mathsf{b}}_{\mp,\Delta, \ell}\,,\\
\widetilde{\mathcal{F}}_{\pm,\mathcal{A}^{\Delta,r=10}_{[0,1],\ell}}\,&=\,
+\tfrac{1}{2}\left(\tfrac{1}{1-w}+\tfrac{1}{w}\right)\,\widetilde{\mathsf{F}}_{\pm,\Delta, \ell}+
\tfrac{1}{2} \left(\tfrac{1}{1-w}-\tfrac{1}{w}\right)\,\widetilde{\mathsf{F}}_{\mp,\Delta, \ell}\,,\\
\widetilde{\mathcal{F}}_{\pm,\mathcal{A}^{\Delta,r=10}_{[2,0],\ell}}\,&=\,
+\tfrac{3}{4}\left(\tfrac{1}{1-w}+\tfrac{1}{w}\right)\,\widetilde{\mathsf{F}}_{\pm,\Delta, \ell}-
\tfrac{1}{4} \left(\tfrac{1}{1-w}-\tfrac{1}{w}\right)\,\widetilde{\mathsf{F}}_{\mp,\Delta, \ell}\,.
\end{split}
\eeq
The equations for $\overline{\mathcal{B}}$ and $\overline{\mathcal{C}}$ in the chiral channel
 follow from the last two entries in \eqref{eq:Req3crossing} at the unitarity bound, as can be seen from table \ref{tab:chiralchiralaeq3blocks}.
Let us go back to the bootstrap equations  \eqref{fullbootstrap_equations} specialized to the case $R=3$.
Using the relations \eqref{eq:Req3crossing},
 the equations  \eqref{fullbootstrap_equations}  are easily recognized to be equivalent to 
\begin{align}
\label{eq:Req2crossingR3}
\begin{split}
&\sum_{\Delta >  \ell + 2} |\lambda_{0,\Delta, \ell}|^2
\begin{bmatrix}
+\tfrac{1}{2}\,\mathsf{F}_{-,\Delta, \ell}\\
+\tfrac{1}{2}\,\mathsf{F}_{+,\Delta, \ell}\\
+\tfrac{1}{2}\,\mathsf{F}^{\mathsf{b}}_{-, \Delta, \ell}\\
+\tfrac{1}{2}\,\mathsf{F}^{\mathsf{b}}_{+, \Delta, \ell}\\
-\tfrac{1}{2}\,\mathsf{F}^{\mathsf{b}}_{+, \Delta, \ell}\\
-\tfrac{1}{2}\,\mathsf{F}^{\mathsf{b}}_{-, \Delta, \ell}
\end{bmatrix}
+
\sum_{\Delta >  \ell + 4} |\lambda_{1,\Delta, \ell}|^2
\begin{bmatrix}
+\tfrac{1}{3}\,\mathsf{F}_{-,\Delta, \ell}\\
-\tfrac{2}{3}\,\mathsf{F}_{+,\Delta, \ell}\\
+\tfrac{5}{6}\,\mathsf{F}^{\mathsf{b}}_{-, \Delta, \ell}\\
-\tfrac{1}{6}\,\mathsf{F}^{\mathsf{b}}_{+, \Delta, \ell}\\
-\tfrac{5}{6}\,\mathsf{F}^{\mathsf{b}}_{+, \Delta, \ell}\\
+\tfrac{1}{6}\,\mathsf{F}^{\mathsf{b}}_{-, \Delta, \ell}
\end{bmatrix}
\\
+&\sum_{\substack{\Delta \geqslant \ell + 5\\ \ell \text{ even }}} |\tilde{\lambda}_{0,\Delta, \ell}|^2
\begin{bmatrix}
0\\
0\\
+\tfrac{1}{2}\,\widetilde{\mathsf{F}}_{-, \Delta, \ell}\\
+\tfrac{1}{2}\,\widetilde{\mathsf{F}}_{+, \Delta, \ell}\\
+\tfrac{1}{2}\,\widetilde{\mathsf{F}}_{+, \Delta, \ell}\\
+\tfrac{1}{2}\,\widetilde{\mathsf{F}}_{-, \Delta, \ell}
\end{bmatrix}
+
\sum_{\substack{\Delta \geqslant \ell + 5\\ \ell \text{ odd }}} |\tilde{\lambda}_{1,\Delta, \ell}|^2
\begin{bmatrix}
0\\
0\\
+\tfrac{3}{4}\,\widetilde{\mathsf{F}}_{-, \Delta, \ell}\\
-\tfrac{1}{4}\,\widetilde{\mathsf{F}}_{+, \Delta, \ell}\\
+\tfrac{3}{4}\,\widetilde{\mathsf{F}}_{+, \Delta, \ell}\\
-\tfrac{1}{4}\,\widetilde{\mathsf{F}}_{-, \Delta, \ell}
\end{bmatrix}=
\vec{\mathsf{F}}_{\text{short}}\,.
\end{split}
\end{align}
More explicitly, we  extract the coefficients of $(1-w)^{-1}(1\pm y^{-1})$ of the first line of  \eqref{fullbootstrap_equations} 
and the coefficients of $\left(\tfrac{1}{1-w}\pm\tfrac{1}{w}\right)$ of the second and third line of  \eqref{fullbootstrap_equations}.
The expression for $\vec{\mathsf{F}}_{\text{short}}$ follows from the expansion of  $\mathcal{F}_{\text{short}}^{(0,\pm)}$ 
given in  appendix \ref{app:forFshort}  in this basis.
In the above equation $\lambda_{a,\Delta, \ell}$ and $\tilde{\lambda}_{a,\Delta, \ell}$ denote the OPE coefficients of the longs multiplets $\AA_{[a,a],\ell}^{\Delta}$  and $\AA_{[2(1-a),a],\ell}^{\Delta, r=10}$ respectively.
As a consistency check, we verified that the bootstrap equations above are satisfied with positive coefficients for the cases of free $\U(1)$ $\mathcal{N}=4$ SYM (considered a special $\mathcal{N}=3$ theory) and  for the generalized free theory discussed in appendix \ref{app: generalized free theory}.

\begin{table}[h]
\centering
\renewcommand{\arraystretch}{1.6}
\begin{tabular}{%
| l
                |>{\centering }m{3.3cm}
             |>{\centering\arraybackslash}m{6cm}|
}
\hline
\text{Multiplet} $\chi$
& 
$\fc_\chi(x) $
&
$\Hc_\chi(z,\bar z,w)$
\\\hline 
$\text{Identity}$
&$\Ssl_{0}(z)=1$
& 0
\\\hline
$\hat{\BB}_{[1,1]}$
&$\Ssl_{1}(z)$
& 0 
\\\hline
$\hat{\BB}_{[2,2]}$
&$\Ssl_{2}(z)$
&$\mathcal{G}^{\calN=1}_{4,0}(z,\bar z)$
\\\hline
$\hat{\BB}_{[3,3]}$
&$\Ssl_{3}(z)$
&$-\mathcal{G}^{\calN=1}_{5,1}(z,\bar z)-\mathcal{G}^{\calN=1}_{4,0}(z,\bar z)h_{[1,1]}(w)$
\\\hline
$\hat{\CC}_{[0,0],\ell}$
&$(-1)^{\ell+1} \Ssl_{\ell+2}(z)$
& 0
\\\hline
$\hat{\CC}_{[1,1],\ell}$
&$(-1)^{\ell+1} \Ssl_{\ell+3}(z)$
&$\mathcal{G}^{\calN=1}_{\ell+5,\ell+1}(z,\bar z)$
\\\hline
$\hat{\CC}_{[2,2],\ell}$
&$(-1)^{\ell+1} \Ssl_{\ell+4}(z)$
&$ -\mathcal{G}^{\calN=1}_{\ell+6,\ell+2}(z,\bar z)-\mathcal{G}^{\calN=1}_{\ell+7,\ell+1}(z,\bar z)h_{[1,1]}(w)$
\\\hline
$\AA_{[0,0],\ell}^{\Delta> \ell+2}$
&0
& $\mathcal{G}^{\calN=1}_{\Delta+2,\ell}(z,\bar z)$
\\\hline
$\AA_{[1,1],\ell}^{\Delta> \ell+4}$
&0
&$-\mathcal{G}^{\calN=1}_{\Delta+2,\ell}(z,\bar z)h_{[1,1]}(w)$
\\\hline
\end{tabular}
\renewcommand{\arraystretch}{1.0}
\caption{Contributions of the various $\NN=3$ multiplets appearing in the \nonchiral OPE \eqref{selectionchiantichi}, for $R=3$, to the functions $\fc_\chi(x)$ and $\Hc_\chi(z,\bar z,w)$. Note that we can make the identification  $\hat{\CC}_{[k,k],\ell=-1}= \hat{\BB}_{[k+1,k+1]}$, and in the text we take the latter to be a special case of the former. 
}
\label{tab:chiralachiralaeq3blocks}
\end{table}

\begin{table}[h]
\centering
\renewcommand{\arraystretch}{1.6}
\begin{tabular}{%
| l
                |>{\centering }m{3.3cm}
             |>{\centering\arraybackslash}m{6cm}|
}
\hline
\text{Multiplet} $\chi$
& 
$\tilde{\fc}_{\chi}(z)$
&
$\widetilde{\Hc}_\chi(z,\zb,w) $
\\\hline 
$\hat{\BB}_{[6,0]}$
&$\sl_{3}(z)$
&$\g_{5,1}(z,\bar z)+g_{6,0}(z,\bar z)h^{\SU(2)}_1(w)$
\\\hline
$\hat{\CC}_{[4,0],(\tfrac{\ell+1}{2},\tfrac{\ell}{2})}$
&$\sl_{\ell+5}(z)$
&$\g_{\ell+7,\ell+3}(z,\bar z)+g_{8+\ell,2+\ell}(z,\bar z)h^{\SU(2)}_1(w)$
\\\hline
$\mbar{\BB}_{[2,2],0}^{r=12}$
&0
&$\g_{8,0}(z,\bar z) h^{\SU(2)}_1(w)$
\\\hline
$\mbar{\BB}_{[0,3],0}^{r=12}$
&0
&$\g_{8,0}(z,\bar z)$
\\\hline
$\mbar{\CC}^{r=11}_{[2,1],(\tfrac{\ell+1}{2},\tfrac{\ell}{2})}$
&0
&$\g_{\ell+9,\ell+1}(z,\zb)h^{\SU(2)}_1(w)$
\\\hline
$\mbar{\CC}^{r=11}_{[0,2],(\tfrac{\ell+1}{2},\tfrac{\ell}{2})}$
&0
&$\g_{\ell+9,\ell+1}(z,\zb)$
\\\hline
$\AA_{[2,0],\ell}^{\Delta > \ell+5, r=10}$
&0
&$\g_{\Delta+3,\ell}(z,\zb)h^{\SU(2)}_1(w)$
\\\hline
$\AA_{[0,1],\ell}^{\Delta> \ell+5, r=10}$
&0
&$\g_{\Delta+3,\ell}(z,\zb)$
\\\hline
\end{tabular}
\renewcommand{\arraystretch}{1.0}
\caption{Contributions of the various $\calN=3$ multiplets appearing in the \chiral  OPE \eqref{chiralchiralselectionsrules}, for $R=3$, to the functions $\tilde{\fc}_\chi(z)$ and $\widetilde{\Hc}_\chi(z,\bar z,w)$. Since we are interested in the correlation functions of identical operators, Bose symmetry under the exchange of the two identical operators forbids the multiplet $\mbar\BB_{[0,3],0}^{r=12}$ from appearing and restricts the $\ell$ to be even for $\AA_{[2,0],\ell}^{r=10}$, $\mbar{\CC}_{[0,2],(\tfrac{\ell+1}{2},\tfrac{\ell}{2})}^{r=11}$  and $\hat{\CC}_{[4,0],(\tfrac{\ell+1}{2},\tfrac{\ell}{2})}$, and odd for $\AA_{[0,1],\ell}^{r=10}$ and $ \mbar{\CC}_{[2,1],(\tfrac{\ell+1}{2},\tfrac{\ell}{2})}^{r=11}$. 
}
\label{tab:chiralchiralaeq3blocks}
\end{table}

\section{Numerical results}
\label{sec:numerics}

We are finally ready to apply the numerical bootstrap machinery to our crossing equations.
Our goal is to chart out the allowed parameter space of $\NN = 3$ theories, but also to ``zoom in'' to particular solutions of the crossing equations that correspond to individual $\NN=3$ SCFTs. 

After a short review of numerical methods we start by considering the multiplet containing a Coulomb branch operator of dimension two, which we recall also contains extra supercharges. This is a warm-up example that will allow us to check the consistency of our setup. In the remainder of the section we then focus on a Coulomb branch operator of dimension three for various values of the central charge. In general it is hard to exclude solutions that have enhanced $\Nm=4$ symmetry, and also to impose that the Coulomb branch operator is a generator.\footnote{One could imagine setting up a mixed correlator system where the multiplets containing the extra supercharges, or the candidate generators for which our operator could be a composite are exchanged.} In order to avoid $\Nm=4$ solutions, at the end of this section we input knowledge of the specific chiral algebra \cite{Nishinaka:2016hbw} that is conjectured  to correspond to the simplest known $\NN=3$ SCFT .

\subsection{Numerical methods}

The crossing equations written in \eqref{fullbootstrap_equations} are too complicated to study exactly, beyond focusing on special limits, or protected subsectors, as done in section \ref{sec:chiral algebra}.
Therefore we proceed to analyze these equations using the numerical techniques pioneered in \cite{Rattazzi:2008pe} (see \eg~\cite{Rychkov:2016iqz, Simmons-Duffin:2016gjk} for reviews).

Very schematically, we have a system of crossing equations (three \eqref{eq:Req2crossing} and six \eqref{eq:Req3crossing} for the $\hat{\BB}_{[2,0]}$ and $\hat{\BB}_{[3,0]}$ respectively) of the form
\be 
\sum_{\OO_i} |\lambda_{\OO_i}|^2 \vec{V}_{\OO_i}(z,\zb) = \vec{V}_{\text{fixed}}(z,\zb)\,,
\ee
where $\vec{V}_{\text{fixed}}(z,\zb)$ collects the part of $\vec{\mathsf{F}}_{\text{short}}$ that is completely fixed from the chiral algebra, with the remainder of $\vec{\mathsf{F}}_{\text{short}}$  moved to the left-hand side.
We use the SDPB solver of \cite{Simmons-Duffin:2015qma}, and rule out assumptions on the spectrum $\{ \OO_i\}$ of local operators and their OPE coefficients $|\lambda_{\OO_i}|^2$ (CFT data), by considering linear functionals
\be 
\vec{\Phi} = \sum_{n,m=0}^{n+m \leqslant \Lambda} \vec{\Phi}_{m,n} \partial_z^m \partial_{\zb}^n \vert_{z=\zb=\tfrac{1}{2}}\,,
\label{eq:functional}
\ee
acting on the crossing equations. 
In the crossing equation \eqref{eq:Req2crossing} and \eqref{eq:Req3crossing} we will be taking derivatives $\partial_z^m \partial_{\zb}^n$ of  $\mathsf{F}_\pm, \mathsf{F}^{\mathsf{b}}_\pm,\widetilde{\mathsf{F}}_\pm $ and from their symmetry properties under $z \to 1-z$, $\zb \to 1-\zb$ we see that only even (odd) derivatives of $ \mathsf{F}_+ , \mathsf{F}^{\mathsf{b}}_+, \widetilde{\mathsf{F}}_+$ ($\mathsf{F}^{\mathsf{b}}_-, \mathsf{F}^{\mathsf{b}}_-, \widetilde{\mathsf{F}}_-$) survive.\footnote{As usual, the equations are antisymmetric in $z \leftrightarrow \zb$ and so we only need derivatives with $m < n$.}

The numerical bounds will be obtained for different values of the cutoff $\Lambda$, which effectively means we are considering a truncation of the Taylor series expansion of the crossing equations around $z=\zb=\tfrac{1}{2}$. We rule out assumptions on the CFT data by proving that they are inconsistent with the truncated system of crossing equations at order $\Lambda$ . Therefore, for each cutoff we find valid bounds, that will improve as we send $\Lambda \to \infty$. 
We refer the reader to the by now extensive literature on these numerical techniques, \eg~\cite{Simmons-Duffin:2015qma,Poland:2011ey}, for all the other technical details and approximations needed for the numerical bootstrap.

\subsection{The case \texorpdfstring{$R=2$}{R=2}}

As a warm-up, let us consider external operators $\hat{\mathcal{B}}_{[2,0]}$, $\hat{\mathcal{B}}_{[0,2]}$, which contain the extra supercharges allowing for an enhancement to $\NN=4$. For this case we will only bound the minimal allowed central charge $c$.
We recall that the OPE selection rules in this case are given by
\begin{align}
\hat{\mathcal{B}}_{[2,0]}\times \hat{\mathcal{B}}_{[0,2]} &= \mathcal{I}+ \hat{\mathcal{B}}_{[1,1]}
+\hat{\mathcal{B}}_{[2,2]} +\sum_{\ell=0}^\infty
\left[\hat{\mathcal{C}}_{[0,0],\ell}+\hat{\mathcal{C}}_{[1,1],\ell}+
\mathcal{A}^{\Delta}_{[0,0],r=0,\ell}\right]\,, \\
\hat{\mathcal{B}}_{[2,0]}\times \hat{\mathcal{B}}_{[2,0]} &= 
\hat{\mathcal{B}}_{[4,0]}+\mbar{\mathcal{B}}_{[0,2],r=8,0}+\sum_{\ell=0}^{\infty}\left[
\hat{\mathcal{C}}_{[2,0],(\frac{\ell+1}{2},\frac{\ell}{2})}+
\mbar{\mathcal{C}}^{\,r=7}_{[0,1],(\frac{\ell+1}{2},\frac{\ell}{2})}+
\mathcal{A}^{\Delta,r=6}_{[0,0],(\frac{\ell}{2},\frac{\ell}{2})}\right]
\,,
\end{align}
with each multiplet contributing with a superblock as given in tables \ref{tab:chiralachiralaeq2blocks} and \ref{tab:chiralchiralaeq2blocks}, with a positive OPE coefficient squared, and the crossing equations are given in \eqref{eq:Req2crossing}.
To obtain central charge bounds, we allow for all operators consistent with unitarity that have not been fixed by the chiral algebra.
In the \chiral channel this amounts to allowing all long operators consistent with unitarity, together with the short multiplets which sit at the long unitarity bound (which are not Schur operators).
In the \nonchiral channel the OPE coefficient of $\hat{\BB}_{[1,1]}$ is fixed unambiguously from the chiral algebra in terms of the central charge. For the remaining Schur operators the chiral algebra is not constraining enough and we are left with some ambiguities.
As shown in equation \eqref{firttermto_toesum} we can fix universally the OPE coefficients of $\hat{\CC}_{[1,1],\ell}$ and $\hat{\BB}_{[2,2]}$ in terms of those of the $\hat{\CC}_{[0,0],\ell}$ multiplets. These last multiplets contain higher-spin currents and should be absent thereby resolving the ambiguity. Nevertheless, as is also clear from \eqref{firttermto_toesum} and table \ref{tab:chiralachiralaeq2blocks}, the contribution of the $\hat{\CC}_{[0,0],\ell}$ multiplets is identical to that of long multiplets at the unitarity bound, and thus, by allowing for long multiplets to have a dimension arbitrarily close to the unitarity bound, we allow for these currents to appear with arbitrary coefficient. Therefore, we do not truly exclude free theories in the bootstrap, and we should expect to recover the solution corresponding to  $\U(1)$ $\mathcal{N}=4$ SYM theory.
\begin{figure}[htbp!]
             \begin{center}       
              \includegraphics[scale=0.4]{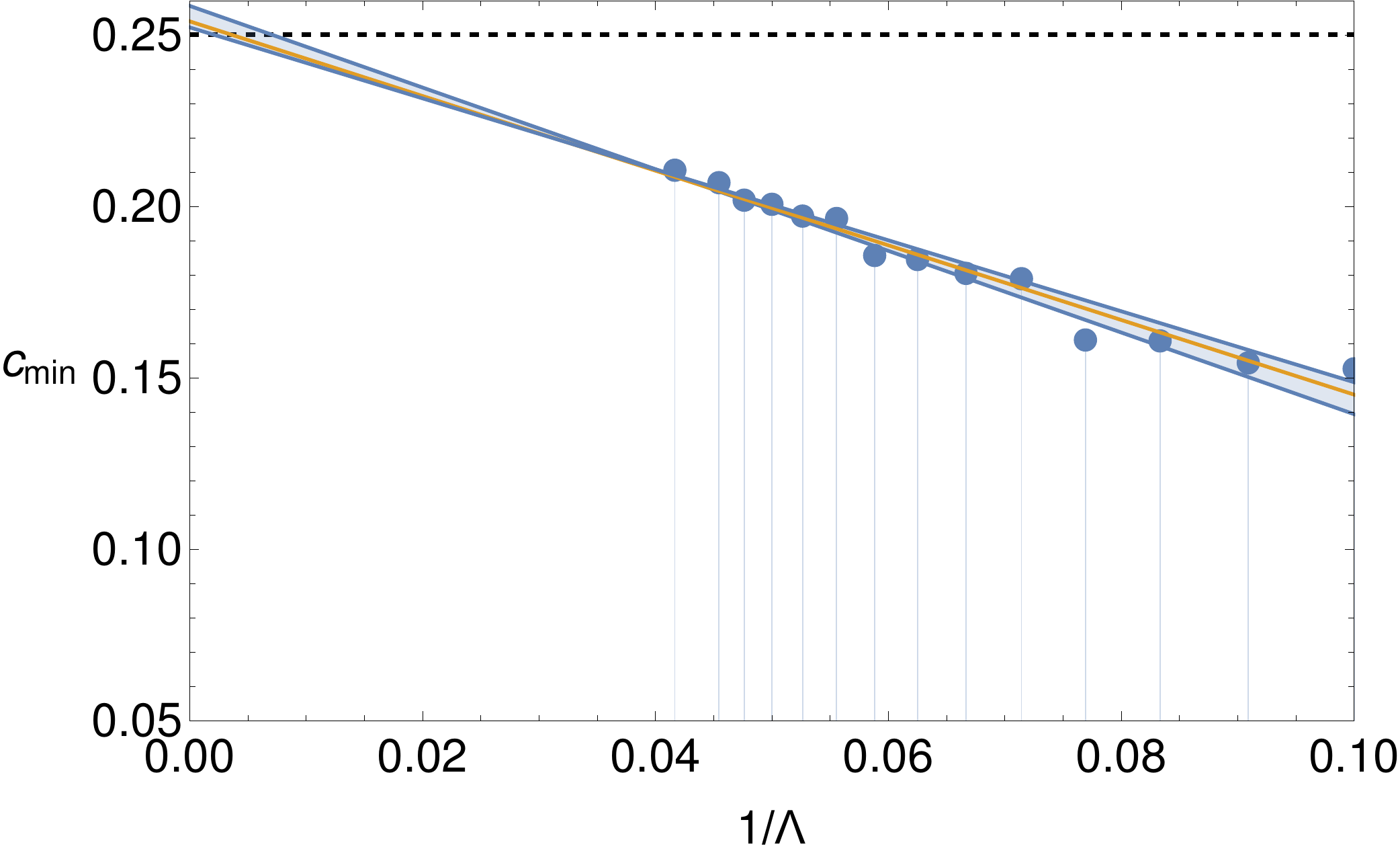}
              \caption{Numerically minimum allowed central charge for the  $\hat{\BB}_{[2,0]}$, $\hat{\BB}_{[0,2]}$ four-point function as a function of the inverse of the number of derivatives $\Lambda$. The dashed horizontal line marks the central charge of the $\U(1)$ $\mathcal{N}=4$ SYM theory. The middle orange line shows a linear fit to all the data points, while the top and bottom blue lines show fits to different subsets of the points.}
              \label{Fig:keq2_cbound}
            \end{center}
\end{figure}

The numerical lower $c$ bound is shown in figure~\ref{Fig:keq2_cbound} as a function of $\Lambda^{-1}$, where $\Lambda$ is the cutoff on the number of derivatives taken of the crossing equation, as defined in \eqref{eq:functional}. The solid yellow and blue lines correspond to various linear fits to subsets of points, and attempt to give a rough estimate of the $\Lambda=\infty$ bound. It seems plausible that the bound is converging to $c=\tfrac{3}{12}$ which corresponds to the central charge of $\U(1)$ $\NN=4$ SYM.
Recall that for this value of the central charge the coefficient $b_2^{(2)}=1-\tfrac{1}{3c}$ in \eqref{eq:decompositionR2offR} is negative, which means that it cannot be interpreted as arising only from a $\hat{\BB}_{[2,2]}$ multiplet, and that the conserved current multiplet $\hat{\CC}_{[0,0],0}$ has to be present. But this is exactly what our crossing equations are allowing for, as when we solve for the OPE coefficient of $\hat{\BB}_{[2,2]}$ in terms of $b_2$ and let the OPE coefficient of $\hat{\CC}_{[0,0],0}$ be arbitrary we find it contributes just as a long at the unitarity bound. Naturally, if one wanted to obtain dimension bounds on the long operators for $c=\tfrac{3}{12}$ we would have to allow for the multiplet  $\hat{\CC}_{[0,0],0}$ to be present by adding their explicit contribution, but if no gap is imposed, then allowing for long multiplets of arbitrary dimension automatically allows for these currents.

\subsection{The case \texorpdfstring{$R=3$}{R=3}}
\label{subsectionR3Bootstrap}

We now turn our attention to the correlation function of $\hat{\BB}_{[3,0]}$, $\hat{\BB}_{[0,3]}$ multiplets, whose crossing equations are given in equation \eqref{eq:Req2crossingR3}.
We recall that in the \chiral channel the OPE coefficients of all of the Schur multiplets $\hat{\CC}_{[4,0],(\tfrac{\ell+1}{2},\tfrac{\ell}{2})}$ and $\hat{\BB}_{[6,0]}$ were fixed universally from the chiral algebra correlation function.
Therefore, the undetermined CFT data in this channel amounts to
\begin{itemize}
\item Scaling dimensions and OPE coefficients of long multiplets $\AA_{[2,0],10,\ell}^{\Delta > \ell+5}$ and $\AA_{[0,1],10,\ell}^{\Delta> \ell+5}$,
\item OPE coefficients of short multiplets $\mbar{\BB}_{[2,2],0}^{r=12}$, $\mbar{\CC}^{r=11}_{[2,1],(\tfrac{\ell+1}{2},\tfrac{\ell}{2})}$ and $\mbar{\CC}^{r=11}_{[0,2],(\tfrac{\ell+1}{2},\tfrac{\ell}{2})}$,
\end{itemize}
where the last multiplets contribute the same way as the longs at the unitarity bound as seen in \eqref{unitboundchiralchiral} and table~\ref{tab:chiralchiralaeq3blocks}.

In the \nonchiral channel, various Schur multiplets were indistinguishable at the level of the chiral algebra, as manifest in table~\ref{tab:chiralachiralaeq3blocks}. Using the chiral algebra correlator we solved for the OPE coefficients of $\hat{\CC}_{[1,1],\ell}$ and $\hat{\BB}_{[2,2]}$ in terms of the remaining ones in \eqref{firttermto_toesum}, such that we were left with the following unfixed CFT data
\begin{itemize}
\item Scaling dimensions and OPE coefficients of long multiplets $\AA_{[0,0],\ell}^{\Delta> \ell+2}$ and  $\AA_{[1,1],\ell}^{\Delta> \ell+4}$,
\item OPE coefficients of the Schur multiplets  $\hat{\CC}_{[2,2],\ell}$,	$ \hat{\BB}_{[3,3]}$, and $ \hat{\CC}_{[0,0],\ell}$.
\end{itemize}
The Schur multiplets in the last line end up contributing to the crossing equations in the same way as the long multiplets in the line above at the unitarity bound (see \eqref{firttermto_toesum} and table \ref{tab:chiralachiralaeq3blocks}), following from the long decomposition at the unitarity bound \eqref{unitarityonbloks}. This implies that, unless we impose a gap in the spectrum of the corresponding long multiplets, we can never truly fix the OPE coefficients of these Schur operators. As usual, the multiplets  $ \hat{\CC}_{[0,0],\ell}$ should be set to zero for interacting theories. However this is not enough to resolve all the ambiguities, and we must resort to numerics in order to study the OPE coefficient of the remaining operators. In the last part of this section we will see how these ambiguities turn out to be useful to exclude $\NN=4$ solutions to the crossing equations by inputting the OPE coefficient of $ \hat{\BB}_{[3,3]}$ computed from the chiral algebra of an $\NN=3$ SCFT.

\subsubsection{Central charge bounds}

Let us start by placing a lower bound on $c$, allowing again for the presence of all operators consistent with unitarity.
We recall once again that long multiplets $\AA_{[0,0],\ell}^\Delta$ of arbitrary dimension allow for conserved currents of spin larger than two, and thus not excluding free theories from the analysis.
Naturally then, the $\U(1)$ $\NN=4$ SYM theory is also a solution to the crossing equations we study. Therefore, the strongest bound one could possibly hope to find corresponds to the central charge of $\U(1)$ $\NN=4$ SYM. This value is smaller than the smallest central charge of all known, nontrivial, $\NN=3$ theories, which is $c=\tfrac{15}{12}$.\footnote{By nontrivial we mean it cannot be obtained by $\NN=4$ SYM by a discrete gauging which does not change the correlation functions.} 

\begin{figure}[htb]
             \begin{center}       
              \includegraphics[scale=0.45]{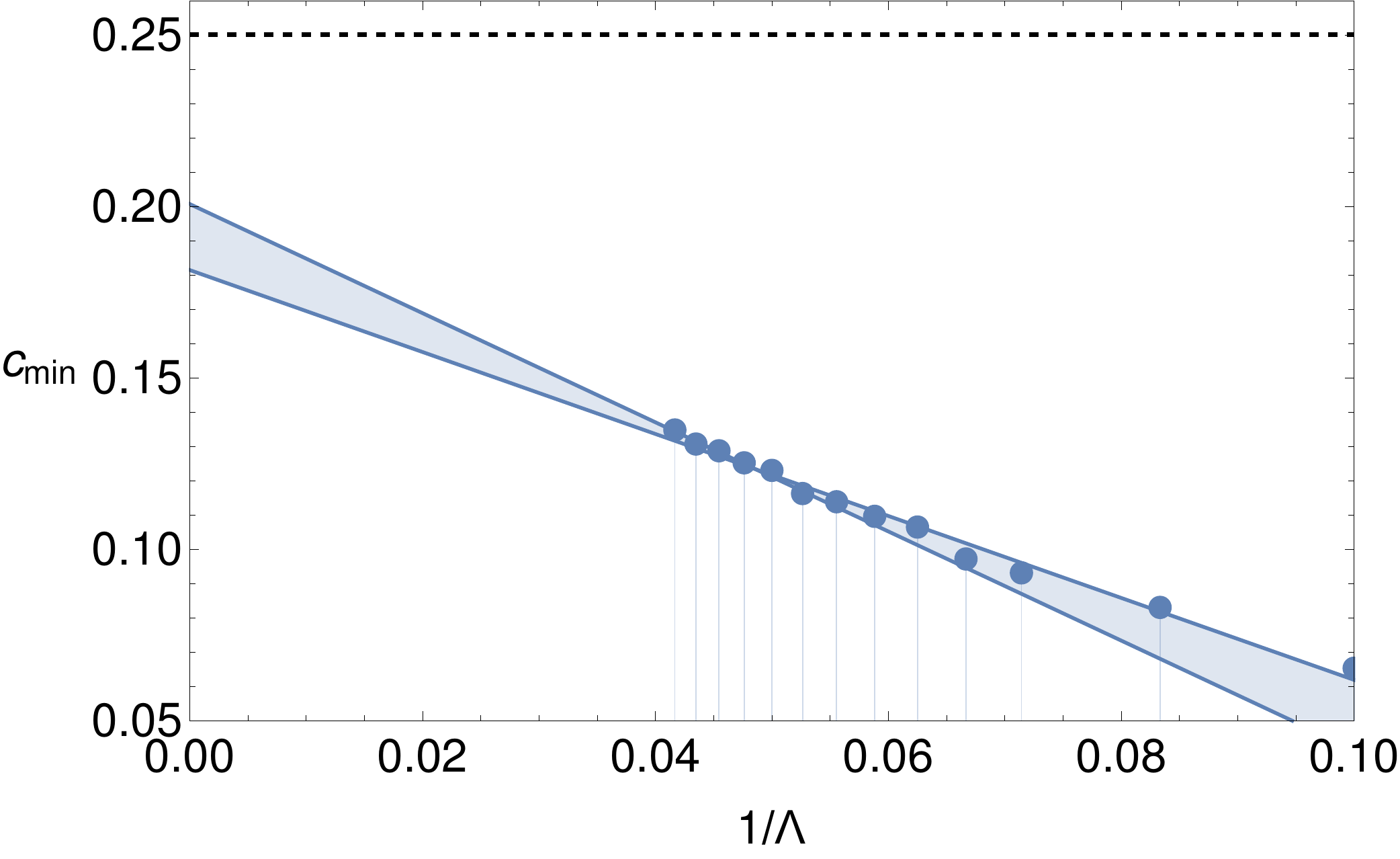}
              \caption{Minimum allowed central charge from the correlation function of $\hat{\BB}_{[3,0]}$ and its conjugate, as a function of the inverse of the number of derivatives $\Lambda$. The dashed horizontal line marks the central charge of the $\U(1)$ $\mathcal{N}=4$ SYM theory.
               The two blue lines show linear fits to different subsets of points, in order to give very rough idea of where the bound is converging to with $\Lambda \to \infty$.}
              \label{Fig:keq3_cbound}
            \end{center}
\end{figure}

In figure \ref{Fig:keq3_cbound} we show the minimal allowed central charge as a function of $\Lambda^{-1}$, the inverse of the number of derivatives. Extrapolation for infinitely many derivatives this time does not seem to converge to the value of the $\U(1)$ $\mathcal{N}=4$, which is $c=\tfrac{1}{4}=0.25$.\footnote{Similar results were also observed in the case of chiral correlators in $\NN=2$ theories \cite{Beem:2014zpa,Lemos:2015awa}.}
Since the value of the minimal allowed central charge is smaller than that of the free $\NN=4$ theory one might suspect the solution to this set of crossing equations that saturates the central charge bound does not correspond to a physical SCFT, and could imagine a mixed correlator system, \eg, adding the stress tensor multiplet, would improve on this.

\subsubsection{Bounding OPE coefficients}
\label{sec:boundingOPEcoeff}

Apart from the central charge, there are other OPE coefficients of physical interest, which were not fixed analytically and can be bounded numerically. Let us emphasize that the $\NN=3$ stress-tensor multiplet $\hat{\mathcal{B}}_{[1,1]}$ cannot recombine to form a long multiplet, unlike the $\NN=2$ stress-tensor multiplet. This has the important consequence that, when we add the stress tensor multiplet with a particular coefficient, we are truly \emph{fixing the central charge} to a particular value. In comparison,  in $\NN=2$ theories this was only accomplished when one imposed a gap in a particular channel, preventing  those long multiplets to hit the unitarity bound and mimic the stress tensor.
Therefore, we will bound the OPE coefficients as a function of the central charge for the range $ \tfrac{1}{4} \leqslant c \leqslant \infty$. The lower end of the interval corresponds to the central charge of $\U(1)$ $\NN=4$ SYM, although interacting theories should have higher central charges.
In particular there is an analytic lower bound for interacting $\NN \geqslant 2$ SCFTs of $c \geqslant \frac{11}{30}\approx 0.37$ \cite{Liendo:2015ofa}. Furthermore it can be shown, by considering the  $\NN=3$ stress tensor four-point function in the chiral algebra, that any interacting $\NN \geqslant 3$ SCFT must obey $c \geqslant \frac{13}{24}\approx 0.54$ \cite{Cornagliotto:2016}. These two bounds will be depicted as vertical dashed lines in all the numerical results.
In the limit $c \to \infty$ the stress tensor decouples and we expect that the numerical bounds converge to the values of generalized free field theory (see appendix \ref{app: generalized free theory}).

\subsubsection*{The Schur operator $\hat{\BB}_{[3,3]}$}
\begin{figure}[htb]
             \begin{center}       
              \includegraphics[scale=0.5]{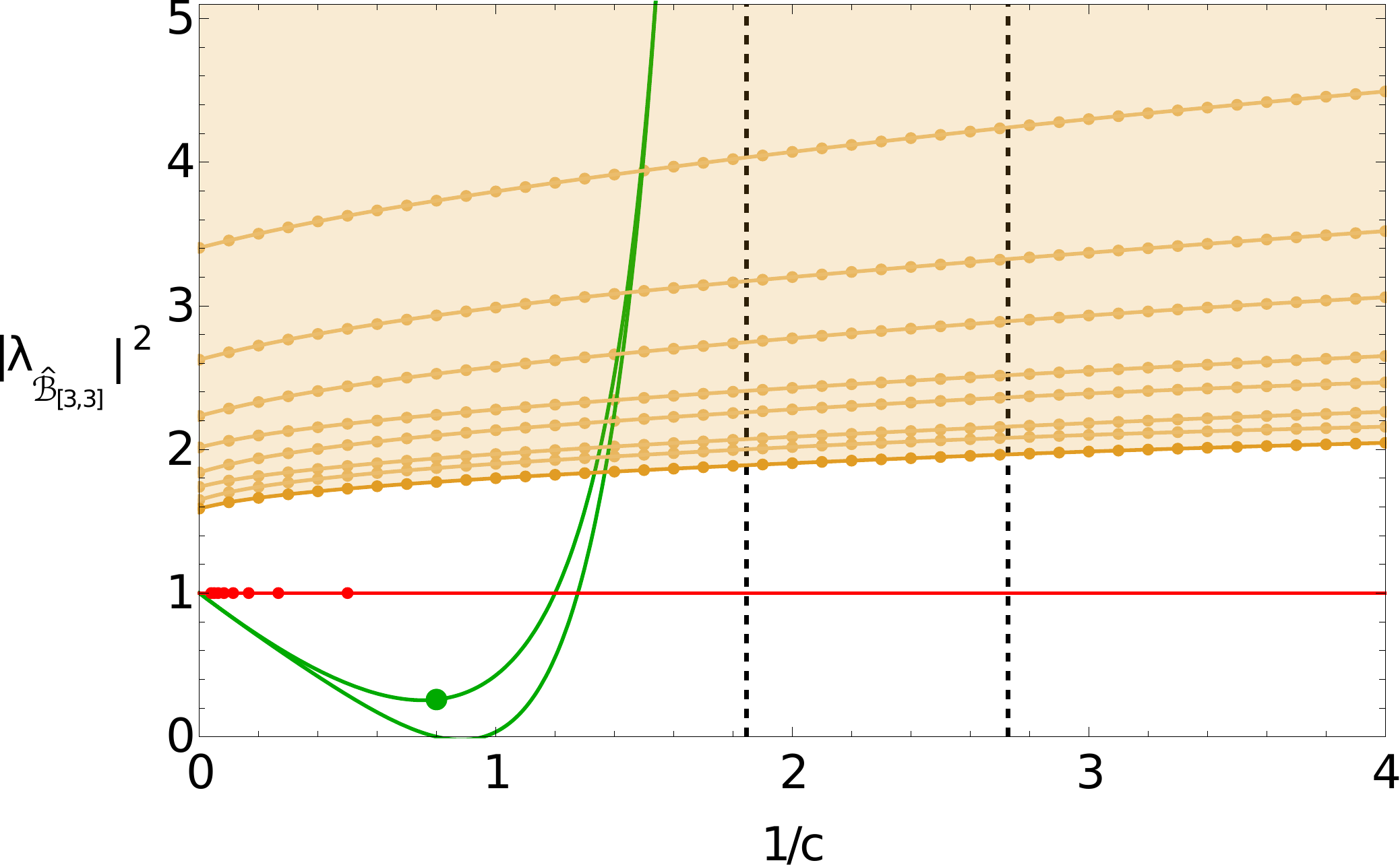}
              \caption{Upper bound on the OPE coefficient squared of $\hat{\BB}_{[3,3]}$ versus the inverse central charge $1/c$. The shaded region is excluded and the number of derivatives is increased from 10 to 24 in steps of two. The two green curves show the possible value of the  OPE coefficient computed by the chiral algebra in section \ref{sec:fixingB33}, while the green dot shows the expected value for the $\NN=3$ theory of  $1/c=0.8$, extracted from the chiral algebra of \cite{Nishinaka:2016hbw}. The red line and dots corresponds to the solution of $\NN=4$ SYM theories. The two dashed lines correspond to the minimum central charges for an interacting $\NN=2$ \cite{Liendo:2015ofa} and $\NN=3$ SCFTs \cite{Cornagliotto:2016} ($c^{-1}=\tfrac{30}{11}\approx 2.73 $ and $c^{-1}=\tfrac{24}{13}\approx 1.84$ respectively). }
              \label{Fig:keq3_B33bound}
            \end{center}
\end{figure}

A particularly interesting operator to consider is the  $\hat{\BB}_{[3,3]}$ appearing in the \nonchiral channel. Despite being captured by the two-dimensional chiral algebra, is not possible to fix its OPE coefficient universally from the chiral algebra four-point function, due to the ambiguities described in \ref{subsec:ContributionofSchur}.
Making assumptions about what particular chiral algebra corresponds to a given $4d$ theory, one can try to resolve this ambiguity, as done in section \ref{sec:fixingB33}, which gave two seemingly consistent possibilities.
However, we will first take an agnostic viewpoint, and ask what numerical constraints crossing symmetry and unitarity place on the squared OPE coefficient of this operator ($|\lambda_{\hat{\BB}_{[3,3]}}|^2$).
These are shown in figure~\ref{Fig:keq3_B33bound} as a function of the inverse of the central charge.

Since this operator is protected, we can compare the value of the bound to the well known $\NN=4$ solutions. We extracted the OPE coefficient of this multiplet from the four-point function of half-BPS operators in the $\left[ 0,3,0\right]$ representation of $\SU(4)_R$ given in \cite{Doobary:2015gia}, after projecting the $\NN=4$ multiplets to the particular $\NN=3$ multiplet we are considering.
It turns out to have a constant value of one, irrespective of the central charge of the theory. We depicted this as a red line in figure \ref{Fig:keq3_B33bound} and, to give an idea of where the physical $\NN=4$ theories sit, we also added red dots in the positions corresponding to the central charge of $\NN=4$ SYM with gauge group $\SU(n)$ ($c=\tfrac{n^2-1}{4}$) for $n\in\{3,4,\ldots\}$.

The value expected from the block decomposition of both the $\U(1)$ $\NN=4$ SYM  ($c= \tfrac{1}{4}$) and the generalized free field theory ($c=\infty$, given in appendix~\ref{app: generalized free theory}) is also one, and is marked by red dots as well. A rough extrapolation of our results for infinite central charge and for $c=\tfrac{1}{4}$ suggests the numerical bounds could converge to the values expected for these theories.

Finally, we compare the numerical bounds with the results that can be extracted from a particular chiral algebra.
Let us first consider the chiral algebra of \cite{Nishinaka:2016hbw} that is conjectured to correspond to the simplest known $\NN=3$ SCFT with $c=\tfrac{15}{12}$ (in their notation this corresponds to $\ell=3$, where of course this $\ell$ has no relation to the spin).
As discussed in section~\ref{sec:fixingB33} we can construct candidate operators, in the chiral algebra, to correspond to a  $\hat{\BB}_{[3,3]}$. In this case there is only one candidate, and if one assumes it to be in fact a $\hat{\BB}_{[3,3]}$ we find 
\be 
\big|\lambda_{\hat{\BB}_{[3,3]}}\big|^2 = \frac{22}{85}\,,
\label{eq:B33forc1512}
\ee
which is shown as a green dot in figure~\ref{Fig:keq3_B33bound}.\footnote{The other possibility, that the $\hat{\BB}_{[3,3]}$ multiplet is absent in the chiral algebra, does not appear plausible from a Higgs branch perspective.}
Note that this value lies well inside the numerical bounds, and in particular it is also smaller than the continuation to arbitrary $c$ of the value corresponding to $\SU(N)$ $\NN=4$ SYM. Since the $\NN=4$ SYM correlation function of \cite{Doobary:2015gia} which we decomposed in blocks is a solution of the crossing equations for any value of $c$, the best numerical bound one can hope to obtain is $\big|\lambda_{\hat{\BB}_{[3,3]}}\big|^2 \leq 1$. In fact, our numerical results appear consistent with the upper bound converging to one for $c=\tfrac{15}{12}$. 
Therefore, to be able to reach the known nontrivial $\NN=3$ SCFT with $c=\tfrac{15}{12}$ we must go inside these bounds, and fix the OPE coefficient of  $\hat{\BB}_{[3,3]}$ to a value that is incompatible with the $\NN=4$ solution to the crossing equation; we will do this at the end of this section.

We now turn to the chiral algebra constructed in \ref{subsec:keq3chiralalg}, with the goal of understanding the higher rank versions of the aforementioned theory. Recall that we assumed the chiral algebra of the higher rank theories to be generated solely by the Higgs branch generators, the stress tensor, and an additional dimension three operator. Under this assumption, we were able to construct a closed subalgebra of all of these chiral algebras, which is associative for generic values of $c$. In that setting we can attempt to compute $|\lambda_{\hat{\BB}_{[3,3]}}|^2$, and there were two options consistent with the large central charge behavior of the generalized free field theory and unitarity, given in \eqref{eq:B33inchiralalg} and \eqref{eq:B33inchiralalg3}, which are plotted as green curves in figure~\ref{Fig:keq3_B33bound}. Equation \eqref{eq:B33inchiralalg3} is the one that does not go through the expected value for $c=\tfrac{15}{12}$, but that we kept for arbitrary values of $c$.
If our assumptions are correct, then we see that the value of $|\lambda_{\hat{\BB}_{[3,3]}}|^2$ lies well inside the numerical bounds, and is weaker than that of $\NN=4$ SYM. This is not necessarily a downside, as one of our goals must be to determine ways to exclude the $\NN=4$ solutions to our crossing equations, and this provides such a way. \textit{By imposing the value of the OPE coefficient corresponding to \eqref{eq:B33inchiralalg} or \eqref{eq:B33inchiralalg3} we are sure to exclude $\NN=4$ from our analysis.} We will come back to this point at the end of this section.

Note that both \eqref{eq:B33inchiralalg} and \eqref{eq:B33inchiralalg3} diverge at $c=\tfrac{13}{24}$, which corresponds to the analytic central charge bound obtained in \citep{Cornagliotto:2016}, following from the fact that the norm of one of the candidate $\hat{\BB}_{[3,3]}$ operators is going to zero. We note that the chiral algebra in \ref{subsec:keq3chiralalg} was constructed with a generic central charge in mind and care was not given to possible null states arising at specific values of $c$. It is not clear that the solution we have is consistent for $c=\tfrac{13}{24}$, as null states are expected to decouple.

It is also worth noting the interesting interplay between analytical and numerical results. The analytical OPE coefficient is only consistent with the (current) numerical bounds for $1/c \lesssim  1.33 - 1.36$ depending on which curve one takes.  This provides a lower bound $c\gtrsim 0.74- 0.75$ on the central charge of any $\NN=3$ SCFT with a dimension three Coulomb branch operator ($\hat{\BB}_{[3,0]}$) of which the chiral algebra presented in \ref{subsec:keq3chiralalg} is a closed subalgebra, improving over the analytical bound $c \geqslant \frac{13}{24}\approx 0.54$ of \cite{Cornagliotto:2016}. On the other hand, this bound is lower than the one obtained using the sum-rule of \cite{Argyres:2007tq,Shapere:2008zf} for  a rank one theory with a generator of dimension three, namely $ c \geqslant \tfrac{15}{12}$. Although there are known cases where this  sum rule does not hold (see \cite{Aharony:2016kai,Argyres:2016yzz}), they correspond to theories obtained by gauging discrete symmetries, so this bound could be valid for theories which are not of this type.\footnote{See \cite{Argyres:2016yzz} for a proposed correction of this formula to hold also in the case of discretely gauged theories.}

\subsubsection*{The multiplets $\bar{\BB}_{[2,2]}$ and $\bar{\CC}_{[0,2],(\tfrac{1}{2},0)}$}
\begin{figure}[htbp!]
             \begin{center}       
              \includegraphics[scale=0.35]{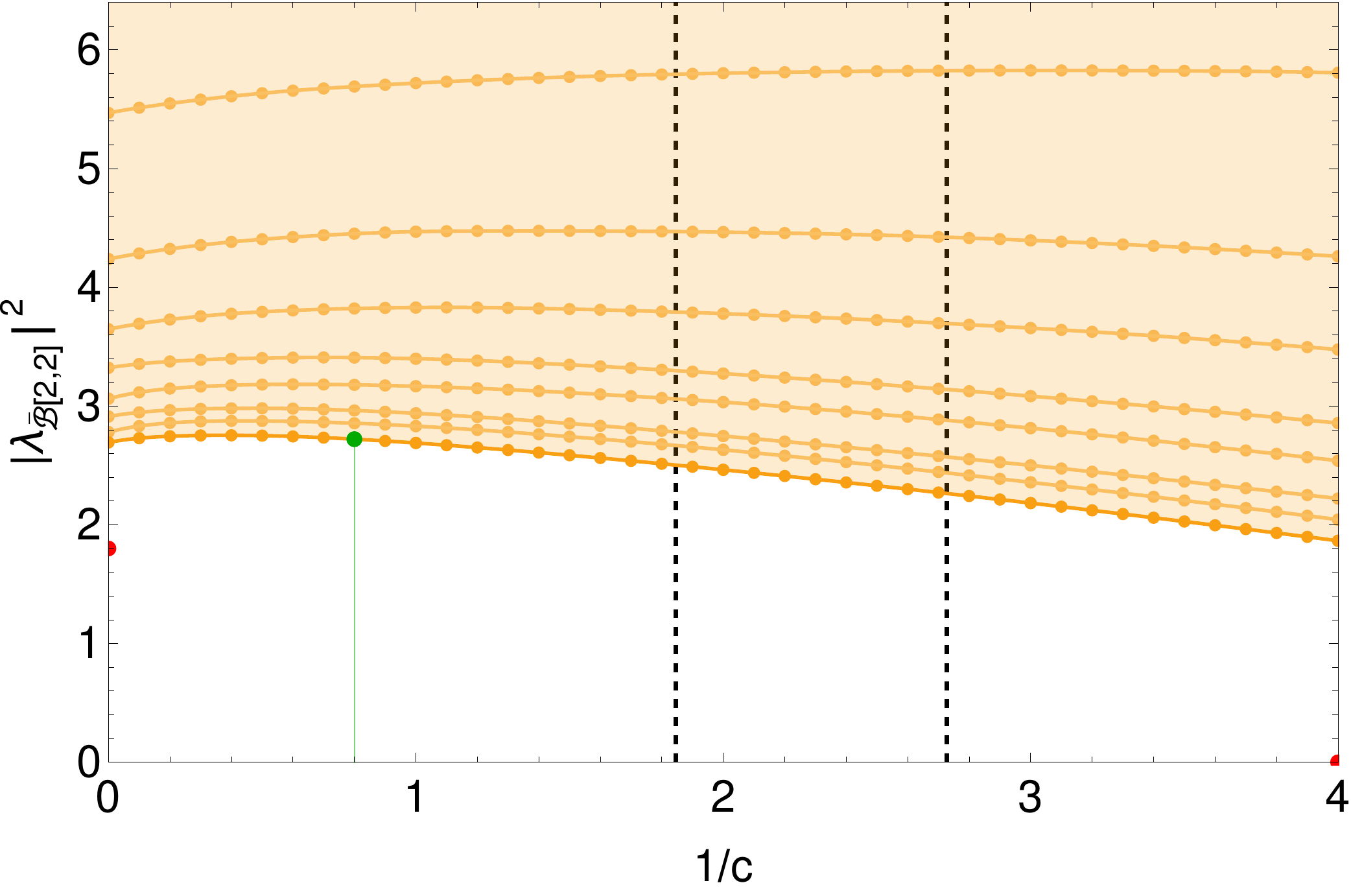}\qquad  \includegraphics[scale=0.35]{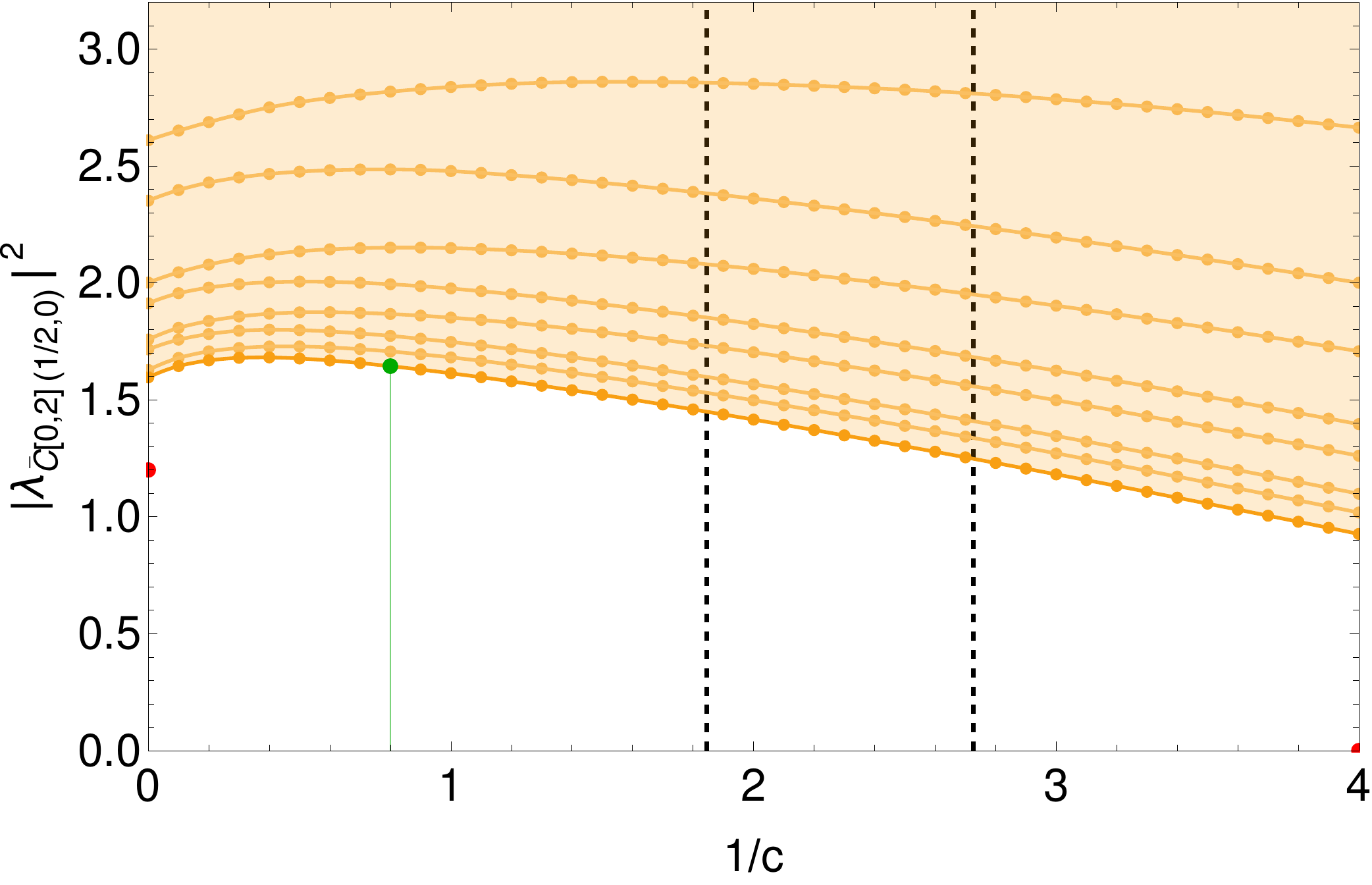}
                        \caption{Upper bound on the OPE coefficient squared of $\bar{\BB}_{[2,2]}$ ($|\tilde{\lambda}_{\bar{\BB}_{[2,2]}}|^2$, depicted on the left) and of $\bar{\CC}_{[0,2],(\tfrac{1}{2},0)}$ ($|\tilde{\lambda}_{\bar{\CC}_{[0,2],(\tfrac{1}{2},0)}}|^2$, shown on the right) versus the inverse central charge $1/c$. The first vertical dashed line marks $c=\tfrac{13}{24}$ and the second $c=\tfrac{11}{30}$ (the minimal central charges for $\NN=3$ and $\NN=2$ interacting theories respectively \cite{Liendo:2015ofa,Cornagliotto:2016}). The number of derivatives $\Lambda$ is increased from 10 to 24 in steps of two. The red dots mark the value of this OPE coefficient for generalized free field theory and $\U(1)$ $\NN=4$ SYM, while the green line marks the central charge $c=\tfrac{15}{12}$ of the simplest known $\NN=3$ SCFT, with the green dot providing an upper bound for the OPE coefficients of this theory.}
             \label{Fig:keq3_B22C02bound}
            \end{center}
\end{figure}

Next we turn our attention to the short multiplets in the \chiral OPE that sit at the unitarity bound of the long multiplets, and are not captured by the chiral algebra. As representatives, we show the upper bounds on the OPE coefficients squared of the multiplets $\bar{\BB}_{[2,2]}$ and $\bar{\CC}_{[0,2],(\tfrac{1}{2},0)}$ in figure \ref{Fig:keq3_B22C02bound}. Again we focus on the region of central charges larger than that of $\U(1)$ $\NN=4$ SYM. 

We show in figure \ref{Fig:keq3_B22C02bound} the value of these OPE coefficients in the case of the generalized free field theory, and of the $\U(1)$ $\NN=4$ SYM as the two red dots at $c^{-1}=0$ and $c^{-1}=4$ respectively. The convergence of our numerical results is rather slow and one cannot conclude if they will converge for these central charges to the known solutions, although they are not incompatible with this possibility.
The green lines in the plots mark the central charge of the ``minimal'' $\NN=3$ SCFT ($c=\tfrac{15}{12}$) with the green dot providing a valid upper bound for the OPE coefficients of this theory.

Finally, to better understand what is failing in the crossing symmetry equations if one tries to go below the minimal numerically allowed central charge ($c_{\text{min}}$ in figure \ref{Fig:keq3_cbound}), it is instructive to look at the OPE coefficient bounds near those central charges. One finds (not shown), that while the bounds on the squared OPE coefficients of both $\hat{\BB}_{[3,3]}$ and $\bar{\BB}_{[2,2]}$ have a very sharp drop near $c_{\text{min}}$, the upper bound on the squared $\bar{\CC}_{[0,2],(\tfrac{1}{2},0)}$ OPE coefficient has as smooth drop and becomes negative right after $c_{\text{min}}$, which is inconsistent with unitarity. This suggests it is this last multiplet that is responsible for the lower bound on the central charge, and that the solution at $c_{\text{min}}$ would have the other two short operators present. Note that both $\bar{\BB}_{[2,2]}$ and $\bar{\CC}_{[0,2],(\tfrac{1}{2},0)}$ have zero OPE coefficient for the $\U(1)$ $\NN=4$ SYM theory.

\subsubsection{Dimension bounds}

Next we turn to the dimensions of the  lowest lying scalar long operators in the various channels. In doing so we must worry about the short multiplets whose OPE coefficients we bootstrapped in the previous subsection, as they all sit at the unitarity bound of the different long multiplets we study (see the tables \ref{tab:chiralachiralaeq3blocks} and \ref{tab:chiralchiralaeq3blocks}). 
By allowing for long multiplets with arbitrary dimension, these short multiplets can appear with any coefficient. Even if we were to explicitly add by hand the short multiplets with a given OPE coefficient, the long multiplet at the bound would mimic those shorts, and in practice we would only be imposing the OPE coefficient of the short multiplets to be greater or equal to a given value.
However, once we impose a gap in the spectrum of the long operator, then we can truly fix the OPE coefficient of the corresponding short multiplet.

In the \nonchiral channel, we focus on the dimension of the first scalar long of each type
\beq
\AA_{[0,0],0}^{\Delta>2}\quad \text{ and }\quad  \AA_{[1,1],0}^{\Delta> 4}\,,
\eeq
while in the \chiral channel we focus on the first scalar long multiplet\footnote{Table \ref{tab:chiralchiralaeq3blocks} contains also long multiplets $\AA_{[0,1],10,\ell}^{\Delta > 5+\ell}$ but for those the spin $\ell$ must be odd by Bose symmetry.}
\beq
\AA_{[2,0],10,0}^{\Delta > 5}\,.
\eeq

\subsubsection*{Non-chiral channel}
\begin{figure}[htbp!]
             \begin{center}       
              \includegraphics[scale=0.35]{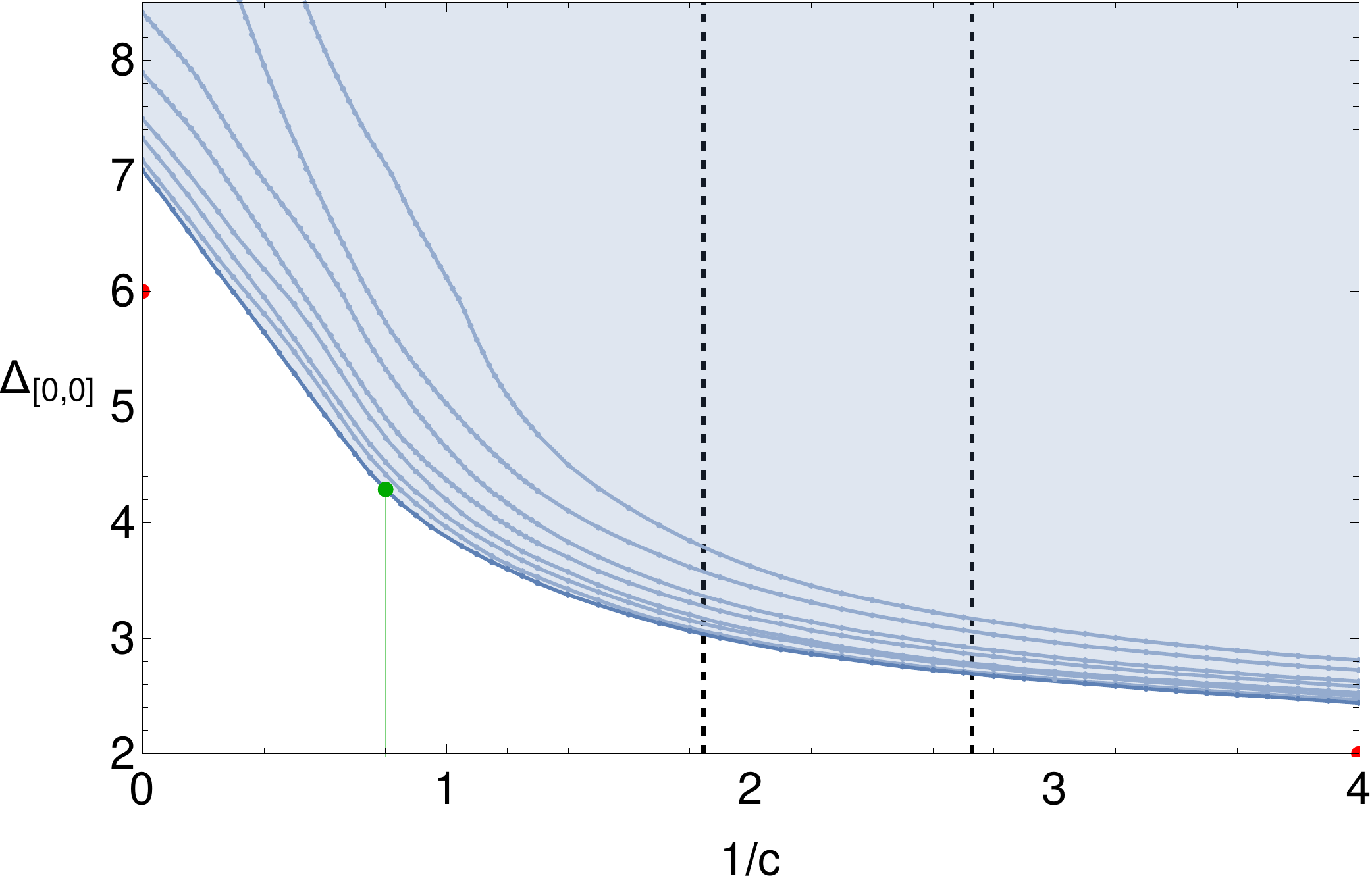}\qquad  \includegraphics[scale=0.35]{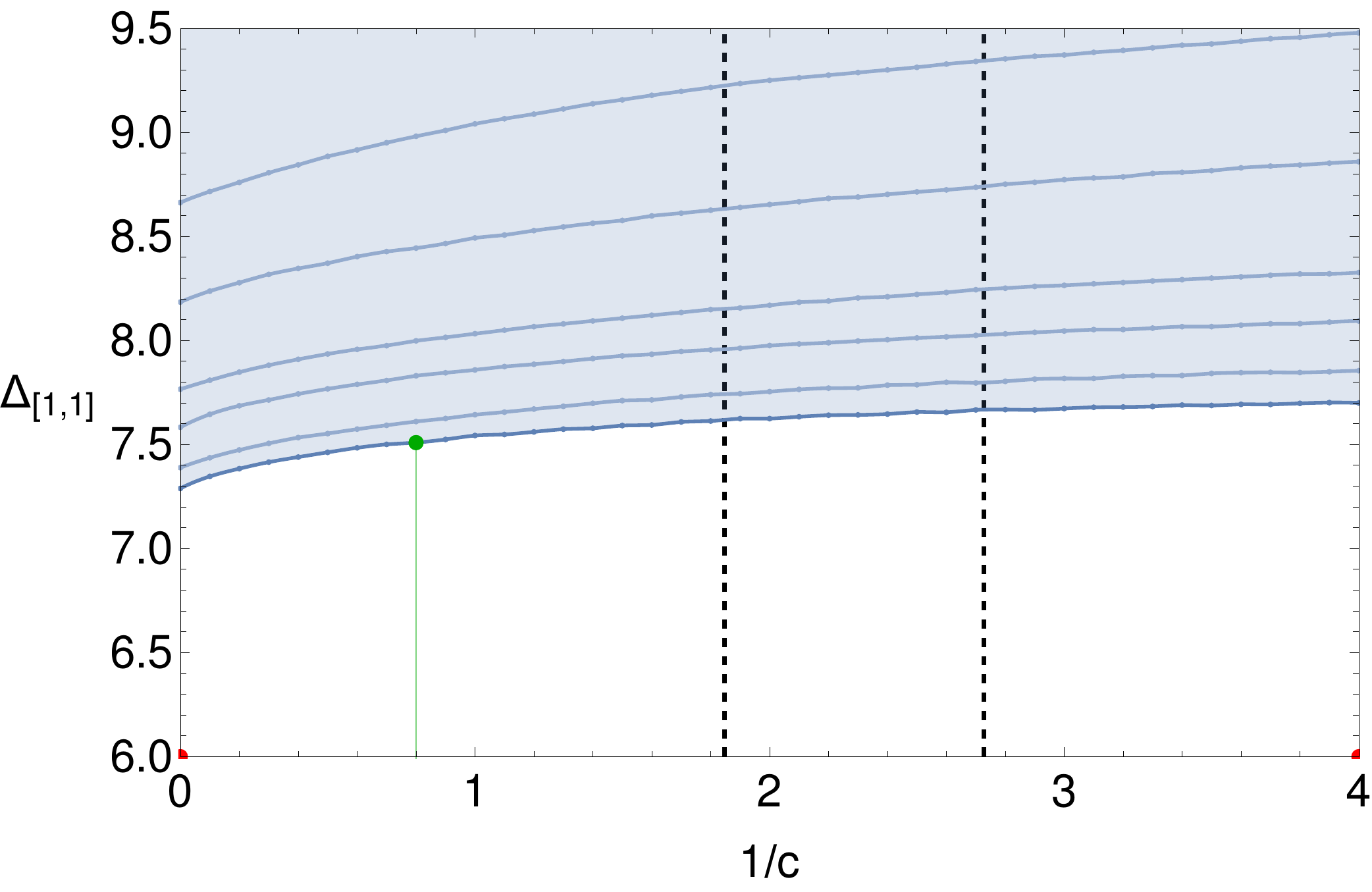}
                        \caption{Upper bound on the dimensions of long multiplets $\AA_{[0,0],0}^{\Delta>2}$ (left) and $\AA_{[1,1],0}^{\Delta> 4}$ (right) for different values of the inverse of the central charge $c$.
                        The maximum number of derivatives is $\Lambda=24$, and the weaker bounds correspond to decreasing the number of derivatives by two. The red dots mark the dimension of the first long operators for generalized free field theory and $\U(1)$ $\NN=4$ SYM, while the green line marks the central charge $c=\tfrac{15}{12}$ of the simplest known $\NN=3$ SCFT, with the green dot providing an upper bound for this theory.
                        The two dashed lines correspond to the minimum central charges for an interacting $\NN=2$ \cite{Liendo:2015ofa} and $\NN=3$ SCFTs \cite{Cornagliotto:2016}.}
             \label{Fig:dimbounds_long00and11}
            \end{center}
\end{figure}

The upper bounds on the dimensions $\Delta_{[0,0]}$ and $\Delta_{[1,1]}$, of the first long multiplets $\AA_{[0,0],0}^{\Delta>2}$ and $\AA_{[1,1],0}^{\Delta> 4}$ respectively, as functions of the inverse central charge are depicted in figure \ref{Fig:dimbounds_long00and11}. Once again red dots mark the dimension of the lowest dimensional operator in the generalized free field theory and the $\U(1)$ $\NN=4$ SYM solutions.
In both cases the green vertical line ending on a dot marks the central charge of the simplest known nontrivial $\NN=3$ SCFT, and provides an upper bound for the dimension of these two operators in this theory. We will improve on the latter bound at the end of this section.

At the unitarity bound, the long multiplet of type $\AA_{[0,0],0}^{\Delta>2}$ mimics a higher spin conserved current multiplet ($\hat{\CC}_{[0,0] \ell=0}$), expected to be absent in an interacting theory, and therefore when obtaining the bound on the left side of \ref{Fig:dimbounds_long00and11} we do not allow for such a multiplet to be present. This explains why the upper bound is presumably converging to the unitarity bound $\Delta=2$ for $c^{-1}=4$, since such currents should be present in the $\U(1)$ $\NN=4$ solution, as indicated by the red dot. For larger central charges the upper bound is far away from unitarity, and thus theories saturating the upper bound do not contain the $\hat{\CC}_{[0,0] \ell=0}$ multiplet, although they could have the higher spin versions of this multiplet which also contain higher-spin conserved currents.

On the other hand, the multiplet that sits at the unitarity bound of $\AA_{[1,1],0}^{\Delta> 4}$ is  the $\hat{\BB}_{[3,3]}$ discussed in the previous subsection, and in obtaining the bounds for $\Delta_{[1,1]}$ we allowed the short multiplet to be present with arbitrary OPE coefficient. We can obtain a stronger bound for specific $\NN=3$ SCFTs by fixing the short OPE coefficient according to section \ref{sec:fixingB33}, as we shall do later in figure~\ref{Fig:OPEvsDimensionB33} for the case of $c=\tfrac{15}{12}\Rightarrow c^{-1}=0.8$.

\subsubsection*{Chiral channel}
\begin{figure}[htbp!]
             \begin{center}       
              \includegraphics[scale=0.35]{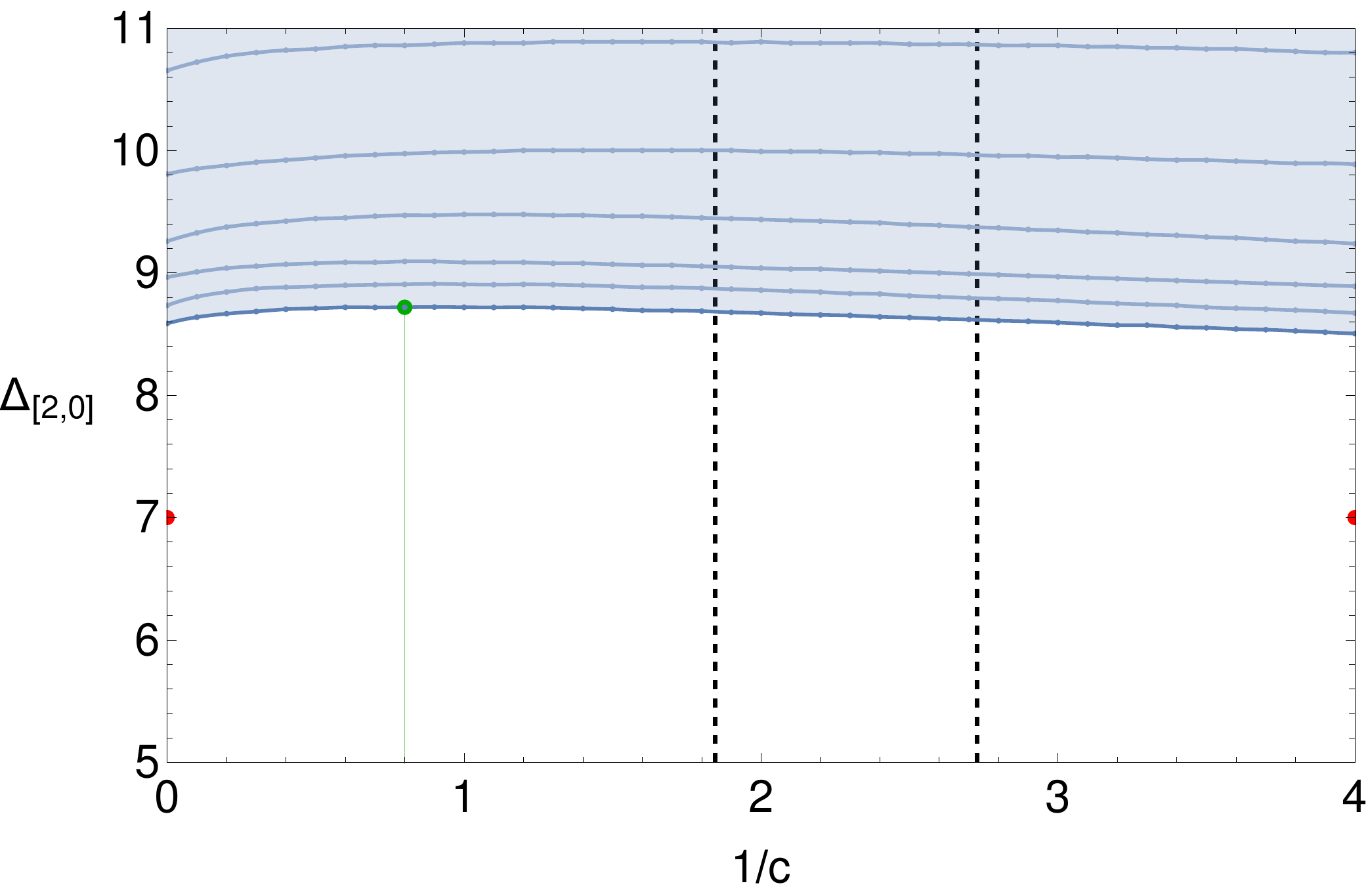}\qquad  \includegraphics[scale=0.35]{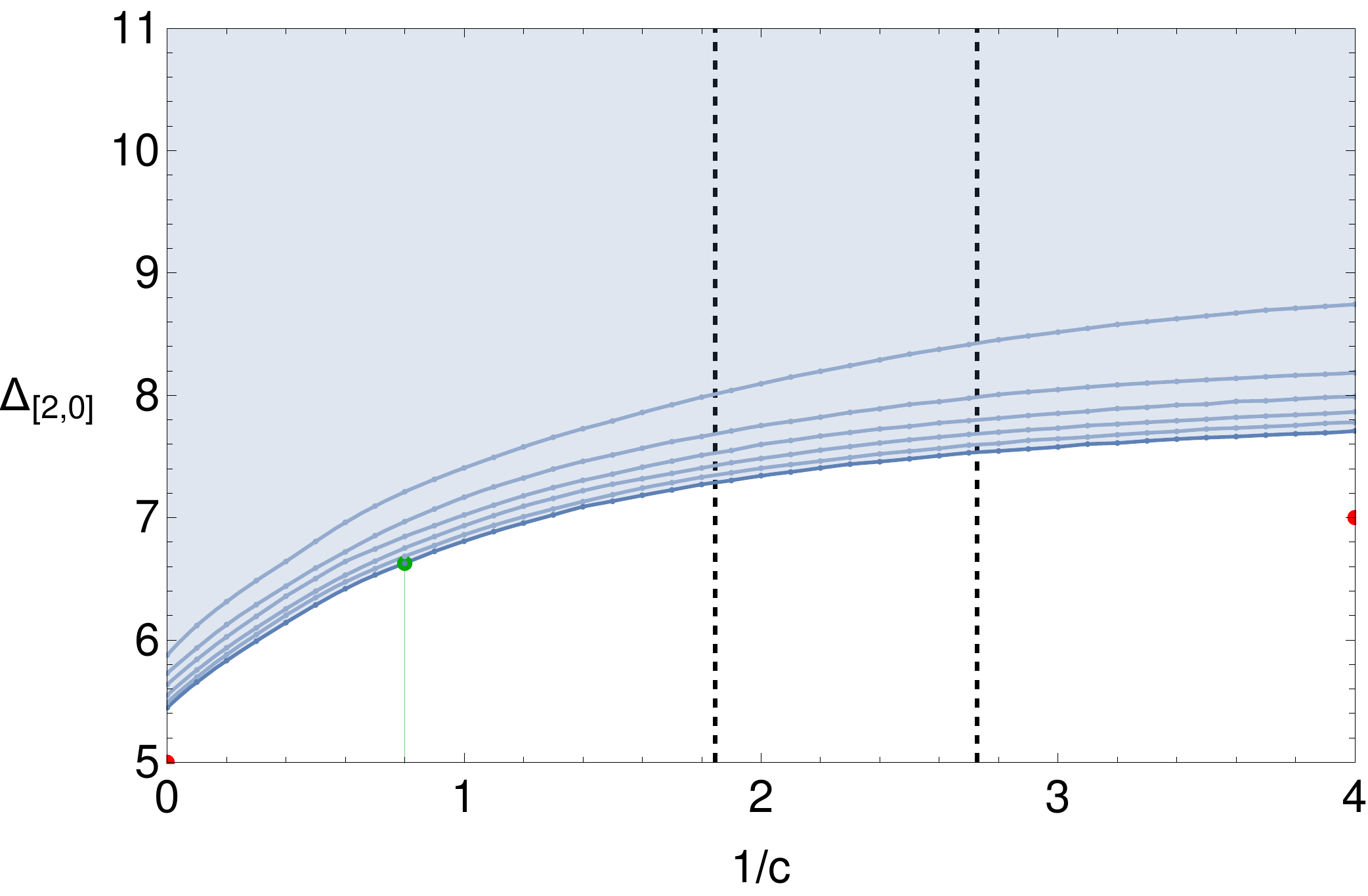}
                        \caption{Upper bound on the dimensions of the long multiplet $\AA_{[2,0],10,0}^{\Delta > 5}$, allowing for 
                     (left) and disallowing for (right) the short multiplet $\bar{\BB}_{[2,2]}$.                      
                       The strongest bound corresponds to 24 derivatives, and they are reduced in steps of two.
                       The red dots mark the dimension of the first long operators for generalized free field theory and $\U(1)$ $\NN=4$ SYM, in the right plot the red dot of generalized free field theory is at the unitarity bound, meaning that the short multiplet is present in this solution. The green line marks the central charge $c=\tfrac{15}{12}$ of the simplest known $\NN=3$ SCFT, with the green dot providing an upper bound for this theory.
                       The two dashed lines correspond to the minimum central charges for an interacting $\NN=2$ \cite{Liendo:2015ofa} and $\NN=3$ SCFTs \cite{Cornagliotto:2016}.}
             \label{Fig:dimbounds_longcc}
            \end{center}
\end{figure}

Turing to the scalar long operator appearing in the \chiral OPE, we obtain an upper bound for the first $\AA_{[2,0],10,0}^{\Delta > 5}$ multiplet. In imposing a gap in this channel we must decide on whether the short multiplet $\bar{\BB}_{[2,2]}$ is present or not. Recall that, unlike the shorts at the unitarity bound of long operators appearing in the \nonchiral channel, this short is not captured by the chiral algebra and thus we have no reason to expect it to be present or absent.
Therefore, we show a bound on the dimension $\Delta_{[2,0]}$ of this long multiplet both allowing for (left plot in \ref{Fig:dimbounds_longcc}) and disallowing for (right plot in \ref{Fig:dimbounds_longcc}) the presence of $\bar{\BB}_{[2,2]}$. Once again the red dots depict the value of these dimensions expected for the $\U(1)$ $\NN=4$ SYM and generalized free field theories. 
We observe that for $c\rightarrow \infty$, the right hand side of figure \ref{Fig:dimbounds_longcc} comes close to the unitarity bound $\Delta=5$. In fact, a simple extrapolation seems to suggest that for $\Lambda \to \infty$ the bound will converge to around $5$.
This is consistent with the fact that this multiplet is \textit{present} in the generalized free theory solution (see appendix~\ref{app: generalized free theory}), \ie~the bounds force the long multiplet to ``become short'' for $c\rightarrow \infty$. (Said multiplet is absent in the $\U(1)$ $\NN=4$ SYM solution.)
For values of $c$ around the value relevant for the ``minimal''  $\NN=3$ SCFT, marked as green lines in the plots, there seems to be a solution of the crossing equations with this multiplet absent. 

\subsubsection*{Carving out solutions inside the bounds}
As the final point of this section we come back to the issue of distinguishing $\NN=4$ solutions to the crossing equations from pure $\NN=3$ ones.
One possibility is to extract the spectrum of the extremal solution \cite{ElShowk:2012hu} saturating each of the above bounds and check if it is consistent or inconsistent with $\NN=4$ supersymmetry. However, we would like to do better, and to be able to exclude the $\NN=4$ solution altogether.
Our explorations in the first part of this section provide such a way, namely by fixing the OPE coefficient of $\hat{\BB}_{[3,3]}$ to the value expected to correspond to the $\NN=3$ theories of interest (see section \ref{sec:fixingB33}). This value is smaller than the one of $\SU(N)$ $\NN=4$ SYM and in its derivation in the chiral algebra we did not allow for the currents enhancing the supersymmetry to $\NN=4$.
As usual, because a long multiplet at the unitarity bound ($\AA_{[1,1],0}^{\Delta= 4}$) mimics the contribution of this short multiplet to the crossing equations, we cannot really fix its OPE coefficient unless we impose a gap.
\begin{figure}[h!]
             \begin{center}       
              \includegraphics[scale=0.5]{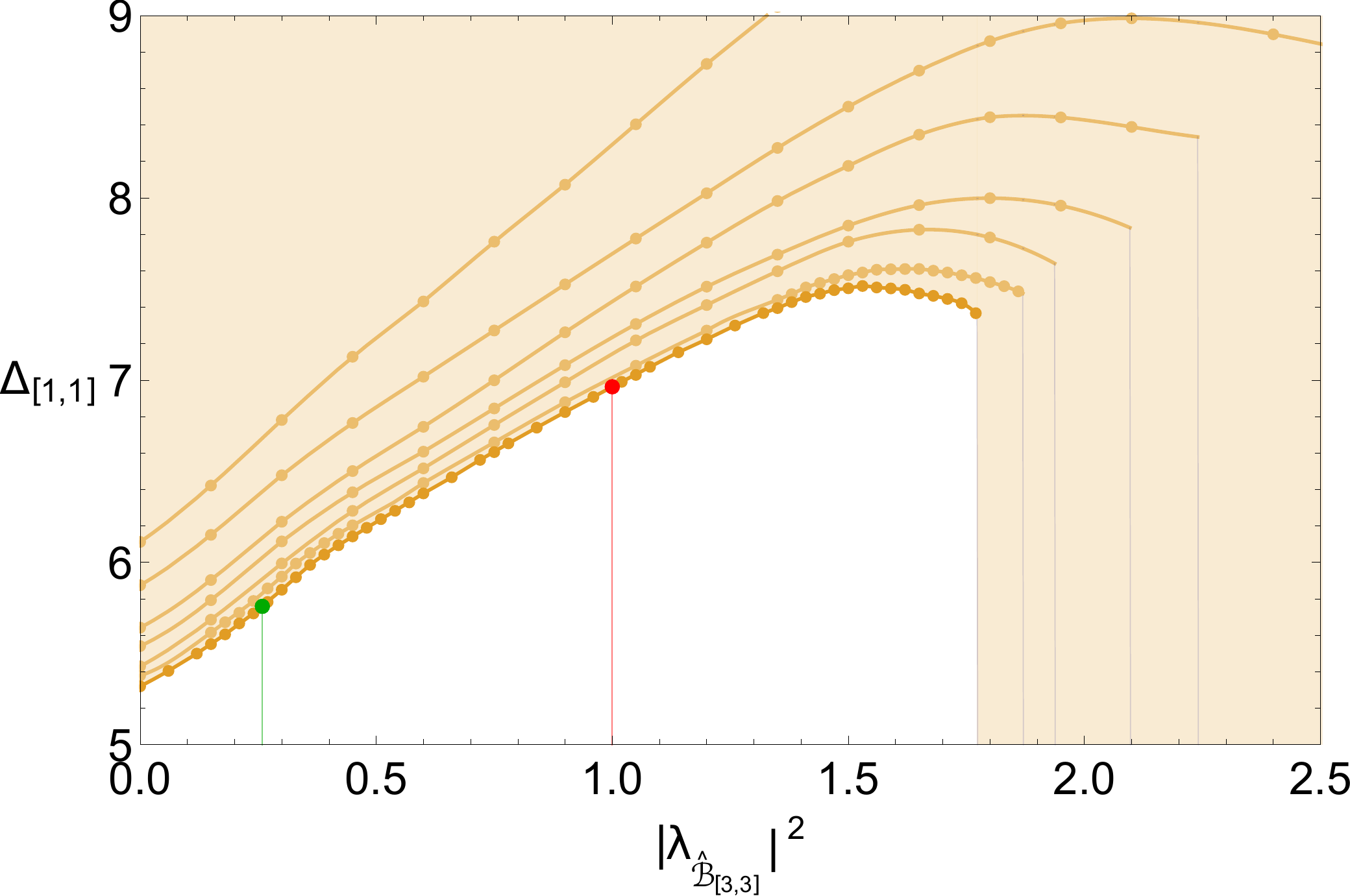}
              \caption{
              Upper bound on the dimension of the long $\AA_{[1,1],0}^{\Delta> 4}$ as a function of the OPE coefficient squared of $\hat{\BB}_{[3,3]}$ for $c^{-1}=0.8$. For each cutoff $\Lambda$ the bounds end abruptly at the value corresponding to the maximum value 
$|\lambda_{\hat{\BB}_{[3,3]}}|^2$ can have, as read off from figure \ref{Fig:keq3_B33bound} at $c^{-1}=0.8$.  
The green line marks the expected value for the OPE coefficient for the  $c^{-1}=0.8$ $\NN=3$ SCFT \eqref{eq:B33forc1512} with the green dot providing an upper bound for this OPE coefficient, while the red line marks the value for $\NN=4$ SYM.
The strongest bound corresponds to 24 derivatives, and they are reduced in steps of two.
              }
              \label{Fig:OPEvsDimensionB33}
            \end{center}
\end{figure}

This is what is done when bounding the lowest dimensional $\AA_{[1,1],0}^{\Delta> 4}$, and so we repeat the analysis leading to the right side of figure \ref{Fig:dimbounds_long00and11}, but now fixing the OPE coefficient of $\hat{\BB}_{[3,3]}$.
The result is shown in figure \ref{Fig:OPEvsDimensionB33}, where we plot the upper bound on the dimension as a function of the OPE coefficient for fixed $c=\tfrac{15}{12}$. The red line marks the value of the OPE coefficient for the $\NN=4$ solution with this particular value of $c$. While the green line marks the value of the OPE coefficient expected for the $\NN=3$ SCFT we are interested in \eqref{eq:B33forc1512}, and provides an upper bound for the dimension $\Delta_{[1,1]}$ in this theory, which improves significantly on the one obtained from figure \ref{Fig:keq3_B22C02bound}. This shows that, at least in figure \ref{Fig:keq3_B22C02bound}, the theory saturating the bound does not correspond to the $\NN=3$ SCFT we were after, and thus, to zoom in to this specific theory we must carve further inside the bounds as done here. This however does not guarantee the theory now sits at the bound.\footnote{We seem to observe a small bump for $|\lambda_{\hat{\BB}_{[3,3]}}|^2\approx 0.33$, and preliminary functional analysis suggest this is correlated to the fact that to the left of the bump a conserved current $\hat{\CC}_{[0,0] \ell=0}$ is allowed, and to the right disallowed. This does not necessarily imply that the conserved currents are present for the $c=\tfrac{15}{12}$ extremal solution, but could mean that to get closer to an interacting $\NN=3$ SCFT we should simultaneously impose a gap in the $\AA_{[0,0],0}^{\Delta>2}$ long channel.}

Similarly, we can repeat this analysis for the central charges of the higher rank theories and we find that, at fixed $|\lambda_{\hat{\BB}_{[3,3]}}|^2$, if the central charge is (increased) decreased the bound seems to get (stronger) weaker (not shown). Due to the dependence of \eqref{eq:B33inchiralalg} and \eqref{eq:B33inchiralalg3} on $c$ the upper bound on $\Delta_{[1,1]}$ does not change significantly.

\section{Conclusions}
\label{sec:conclusions}

In this paper we have initiated the $\NN=3$ superconformal bootstrap program with two goals in mind.
First, to constrain the space of four-dimensional $\NN=3$ SCFTs,
and second, to focus on specific examples of $\NN=3$ theories with the hope of obtaining information about their spectrum.  In order to zoom in on the known $\NN=3$ SCFTs we relied on a combination of numerical bootstrap results and analytical results from two-dimensional chiral algebras, with particular emphasis on the ``minimal'' $\NN=3$ SCFT, and its higher-rank versions.
We approached these theories from the point of view of the Coulomb branch, focusing mostly on a half-BPS operator of dimension three, which is the only Coulomb branch generator of the ``minimal'' $\NN=3$ SCFT, and which is also present in its higher-rank versions.

A basic requirement for any bootstrap study is the conformal block expansion of the four-point function.
In section \ref{sec:superblocksmain}  we showed that for $\NN=3$ half-BPS operators there are no nilpotent invariants, allowing us to concentrate on superconformal primaries without any loss of information.
Demanding the absence of singularities when turning on the fermionic coordinates places strong restrictions on the form of the four-point function, giving rise to the Ward identities. In the case at hand, these identities were not enough to completely fix the superblock (unlike the cases of $\NN=2$ and $\NN=4$ half-BPS superblocks \cite{Nirschl:2004pa,Dolan:2004mu}). For superblocks associated to short multiplets, we used information coming from the $2d$ chiral algebra, while for long blocks we leveraged knowledge of $\NN=1$ blocks.
In the end, we packaged our solution in an elegant way in terms of a \emph{single} $\NN=1$ conformal block with shifted arguments.

The existence of a protected subsector captured by the $2d$ chiral algebra allowed us to solve the crossing equations exactly within the subsector. Which in turn fixed the OPE coefficients of certain short operators universally, \ie, without needing to specify a particular four-dimensional theory. However, some operators appear indistinguishable at the level of the chiral algebra, leading to ambiguities in fixing the corresponding OPE coefficients. Some of these ambiguities can be resolved by knowledge of the specific chiral algebra associated to the $\NN=3$ theory in question, but this is not always the case.

An important question is the defining characteristics of the chiral algebra associated to $\NN=3$ SCFTs.
To that end, we determined which $\NN=3$ superconformal multiplets are captured by the $2d$ chiral algebra, and some of their general properties, which could allow distinguishing between the aforementioned operators. 
Taking advantage of the chiral algebra conjectured to correspond to the ``minimal'' SCFT \cite{Nishinaka:2016hbw}, we were able to compute the OPE coefficient $|\lambda_{\hat{\BB}_{[3,3]}}|^2$.
Moreover, we proposed, under certain assumptions, a closed subsector for the higher-rank versions of this theory, and used it to compute $|\lambda_{\hat{\BB}_{[3,3]}}|^2$ in this case.

To go beyond the protected subsector, or even to distinguish between operators appearing identically in the chiral algebra, one needs numerical bootstrap techniques. These provided constraints on the spectrum of unprotected long operators, and on the OPE coefficients of various short operators. For the particular OPE coefficient $|\lambda_{\hat{\BB}_{[3,3]}}|^2$ that we were able to fix from the chiral algebra, we compared the general numerical results valid for \emph{any} $\NN=3$ SCFT, with the ones of the \emph{specific} theories captured by the chiral algebra that we constructed. This comparison provided a numerical lower bound on the central charge for theories captured by our chiral algebra.

A natural limitation of any $\NN=3$ bootstrap program, as it was also for the $\NN=2$ bootstrap, is that theories with more supersymmetry will generically be solutions of the crossing equations we consider. In order to restrict to pure $\NN=3$ SCFTs, one would have to exclude the presence of superconformal multiplets containing the currents allowing for this enhancement. However, the multiplets that are physically relevant for the study of these theories (for example the ones considered in \cite{Beem:2014zpa,Lemos:2015awa,Liendo:2015ofa} in the $\NN=2$ case) usually do not allow for the multiplets containing the extra supercurrents to be exchanged in their OPEs, and therefore we cannot set them to zero. 
To overcome this limitation we input into the numerical bootstrap information arising from the chiral algebras of pure $\NN=3$ SCFTs, namely, the OPE coefficient  $|\lambda_{\hat{\BB}_{[3,3]}}|^2$. This allowed us explore inside the numerical bounds, and zoom in on the $\NN=3$ solutions with this particular value of the OPE coefficient. By fixing the central charge to that of the ``minimal'' $\NN=3$ theory, and fixing the OPE coefficient accordingly, it is plausible that this theory sits at the bound of figure~\ref{Fig:OPEvsDimensionB33}, although currently there is no evidence this has to be the case, and we would have to provide more information (such as adding stress tensors as external operators).
Nevertheless, the ambiguity in fixing OPE coefficients turned out to be crucial in excluding the $\NN=4$ solution to the crossing equations.
For the higher rank versions, one would have to also consider the four-point functions of the additional Coulomb branch operator, which is a natural next step in the $\NN=3$ bootstrap, along with the study of stress tensor four-point functions.


\acknowledgments

We have greatly benefited from discussions with
P.~Argyres,
M.~Martone,
L.~Rastelli,
D.~Regalado, and
B.~van~Rees.
The research leading to these results has received funding from the People Programme (Marie Curie Actions) of the European Union’s Seventh Framework Programme FP7/2007-2013/ under REA Grant Agreement No 317089 (GATIS).
M.~L., P.~L., and C.~M. thank the Galileo Galilei Institute for Theoretical Physics for hospitality and the INFN for partial support during the completion of this work during the workshop ``Conformal field theories and renormalization group flows in dimensions $d>2$''. 
P.~L. acknowledges the hospitality of Universidad de Santiago de Chile during the last stages of this work.
The authors gratefully acknowledge the computing time granted on the DESY Theory and BIRD clusters and on the supercomputer Mogon at Johannes Gutenberg University Mainz (hpc.uni-mainz.de).


\appendix
\section{Unitary representations of the \texorpdfstring{$\NN=3$}{N=3} superconformal algebra}
\label{app:shortening}

We summarize the unitary representations of the four-dimensional $\NN=3$ superconformal algebra, which fall in the classification of \cite{Dobrev:1985qv,Minwalla:1997ka,Kinney:2005ej} and which were recently discussed with emphasis on $\NN=3$ theories in \cite{Aharony:2015oyb,Cordova:2016xhm,Cordova:2016emh}. We list the possible representations in table~\ref{tab:RepresentationsN3}. The first column lists the name we give to the representation, inspired by the conventions of \cite{Dolan:2002zh}, while the second one uses the notation of \cite{Cordova:2016xhm}. The third column list the quantum numbers of the superconformal primary, denoted by $(j,\bar\jmath)_{[R_1,R_2],r}^{\Delta}$, where $(j,\bar \jmath)\in \frac{\mathbb{N}_0}{2}\times \frac{\mathbb{N}_0}{2}$  are the double of the left/right spins,\footnote{An irreducible representation of label $(j,\bar \jmath)$ has dimension $(2j+1)(\bar 2\jmath+1)$.} $\Delta\in \mathbb{R}$ is the conformal dimension, $(R_1,R_2)\in \mathbb{N}_0\times \mathbb{N}_0$ are the Dynkin labels of $\SU(3)_R$ and $r\in \mathbb{R}$ is the $\U(1)_r$ R-charge.  We follow the $\NN=3$ R-charge conventions of \cite{Cordova:2016xhm}, while for the $\NN=2$ R-charges we follow the conventions of Dolan and Osborn \cite{Dolan:2002zh}. Lastly, we make two remarks:
\begin{itemize} 
\item When dealing with symmetric-traceless representations, we shall label the spins by $j=\bar \jmath=\frac{\ell}{2}$, and by an abuse of notation we will replace the two spin labels $(j,\bar \jmath)$ by $\ell$ in these cases. For example, we have  $\mathcal{A}^{\Delta}_{[R_1,R_2],r,\ell}\equiv \mathcal{A}^{\Delta}_{[R_1,R_2],r,(\tfrac{\ell}{2},\tfrac{\ell}{2})}$.
\item If the $r$ label is zero, we will often omit it. Furthermore, in order to keep some equations compact, we will freely write it up or down, \eg~$\mathcal{A}^{\Delta}_{[R_1,R_2],r,\ell}\equiv \mathcal{A}^{\Delta,r}_{[R_1,R_2],\ell}$.
\end{itemize}

\begin{table}
\centering
\renewcommand{\arraystretch}{1.6}
\begin{tabular}{|c|c|c|c|}

\hline
Name
& 
Name in \cite{Cordova:2016xhm}
&
Superconformal primary
&
Conditions
\\
\hline
$\mathcal{A}^{\Delta}_{[R_1,R_2],r,(j,\bar \jmath)}$
& $L\bar L$
& $(j,\bar \jmath)^{\Delta}_{[R_1,R_2],r}$
& $\footnotesize{\begin{array}{l}\Delta>2+2j+\frac{2}{3}(2R_1+R_2)-\frac{r}{6}\\\Delta>2+\bar 2\jmath+\frac{2}{3}(R_1+2R_2)+\frac{r}{6}\end{array}}$
\\\hline
$\mathcal{B}_{[R_1,R_2],r,\bar \jmath}$
& $B_1\bar L$
& $(0,\bar \jmath)_{[R_1,R_2],r}^{\frac{2}{3}(2R_1+R_2)-\frac{r}{6}}$
& $-6\bar \jmath+2 (R_1-R_2)-6> r$
\\\hline
$\bar{\mathcal{B}}_{[R_1,R_2],r,j}$
& $L\bar B_1$
& $(j,0)_{[R_1,R_2],r}^{\frac{2}{3}(R_1+2R_2)+\frac{r}{6}}$
& $6j+2(R_1-R_2)+6<r$
\\\hline
$\hat{\mathcal{B}}_{[R_1,R_2]}$
& $B_1\Bar B_1$
& $(0,0)_{[R_1,R_2],2(R_1-R_2)}^{R_1+R_2}$
&
\\\hline
$\mathcal{C}_{[R_1,R_2],r,(j,\bar\jmath)}$
&$A_1\bar L$
& $(j,\bar \jmath)_{[R_1,R_2],r}^{2+2j+\frac{2}{3}(2R_1+R_2)-\frac{r}{6}}$
& $6(j-\bar \jmath)+2(R_1-R_2)>r$
\\\hline
$\mbar{\mathcal{C}}_{[R_1,R_2],r,(j,\bar\jmath)}$
&$L\bar A_1$
& $(j,\bar \jmath)_{[R_1,R_2],r}^{2+2\bar\jmath+\frac{2}{3}(R_1+2R_2)+\frac{r}{6}}$
&$ 6(j-\bar \jmath)+2(R_1-R_2)<r$
\\\hline
$\hat{\mathcal{C}}_{[R_1,R_2],(j,\bar\jmath)}$
&$A_1\bar A_1$
&$(j,\bar\jmath)_{[R_1,R_2],6(j-\bar \jmath)+2(R_1-R_2)}^{2+j+\bar\jmath+R_1+R_2}$
& 
\\\hline
$\mathcal{D}_{[R_1,R_2],\bar \jmath}$
&$B_1\bar A_1$
&$(0,\bar \jmath)^{1+\bar \jmath+R_1+R_2}_{[R_1,R_2],2 (R_1-R_2)-6-6\bar \jmath}$
&
\\\hline
$\mbar{\mathcal{D}}_{[R_1,R_2],j}$
&$A_1\bar B_1$
&$(j,0)^{1+j+R_1+R_2}_{[R_1,R_2],2 (R_1-R_2)+6+6j}$
&
\\\hline
\end{tabular}
\renewcommand{\arraystretch}{1.0}
\caption{ We list here the unitary representations of $\calN=3$ with the name that we give them in the present work accompanied by the one that they have in \cite{Cordova:2016xhm}, which was based on the type of shortening condition that they obey. The third column shows the charges of the superconformal primary in the representation, while the fourth one lists the conditions that the charges have to obey. The $A_2$, respectively $\bar A_2$ shortening cases are obtained by putting $j=0$, respectively $\bar \jmath=0$. This changes the null states drastically, but not our labels.}
\label{tab:RepresentationsN3}
\end{table}

\subsection{Decomposition in \texorpdfstring{$\NN=2$}{N=2} multiplets}

Since $\NN=3$ representations are probably less familiar to most readers than $\NN=2$ representations, we give a few examples of how $\NN=3$ multiplets decompose in $\NN=2$ multiplets. In doing so we pick an $\NN=2$ subalgebra of the $\NN=3$, and therefore the $\SU(3)_R \times \U(1)_r$ R-symmetry of the latter decomposes in
$\SU(2)_{\RN} \times \U(1)_{\rN} \times \U(1)_f$, where the first two factors are the R-symmetry of the $\NN=2$ superconformal algebra, and the last corresponds, from the $\NN=2$ point of view, to a global symmetry. Therefore when viewed as $\NN=2$ theories, all $\NN=3$ theories have a $\U(1)_f$ flavor symmetry, and we will keep this flavor grading when decomposing $\NN=3$ representations in $\NN=2$. We follow the conventions of \cite{Nishinaka:2016hbw} for the definition of the flavor charges.
We note that we follow the naming conventions of Dolan and Osborn \cite{Dolan:2002zh} for the representations of $\NN=2$, which are summarized for instance in Appendix~A of \cite{Beem:2014zpa}. While the interpretation of most of these multiplets might be obscure, the following have a natural physical interpretation\footnote{For a more detailed description see, \eg, section~2 of \cite{Beem:2014zpa}}
\begin{itemize}
\item $\hat{\CC}_{0,(0,0)}$ is the stress tensor multiplet of an $\NN=2$ SCFT, containing in addition to the stress tensor, the $\SU(2)_{\RN}$ and $\U(1)_{\rN}$ currents,
\item $\mathcal{\BB}_{R}$  are closely related to the Higgs branch of the theory, in particular the $\hat{\BB}_1$ multiplet contains conserved currents of spin one, associated to flavor currents of the theory,
\item $\mathcal{\EE}_{r,(0,0)}$ are $\NN=2$ chiral operators, and are related to the Coulomb branch of the theory,
\item $\mathcal{\DD}_{\frac{1}{2},(0,0)}$ (and conjugate) which are additional supercurrent multiplets, 
\item $\hat{\CC}_{0,(j>0,\bar{\jmath}>0)}$ contain conserved currents of spin greater than two, which signal free theories \cite{Maldacena:2011jn,Alba:2013yda}.
\end{itemize}
In addition, the multiplets dubbed ``Schur'' operators in \cite{Beem:2013sza}, that is, the ones captured by the two-dimensional chiral algebra reviewed in section~\ref{sec:chiral algebra}, also play an important role. These are $\hat{\BB}_R$, $\DD_{R (0, \bar{\jmath})}$, $\mbar \DD_{R (j, 0 )}$ and $\hat \CC_{R (j, \bar{\jmath}) }$, giving rise to two-dimensional $\slf(2)$ primaries of scaling dimension $R$, $R+\bar{\jmath} +1$,  $R+ j+1$ and $R+ j + \bar{\jmath} +2$ respectively.
The $\NN=3$ multiplets that contain such operators are listed in equations \eqref{eq:CCSchur}-\eqref{eq:DbSchurR20}, together with their decomposition in $\NN=2$, but where we omitted all $\NN=2$ multiplets not containing Schur operators. Below we present a few examples of the complete $\NN=2$ decomposition.
These decompositions are obtained by computing the characters of the $\NN=3$ multiplets of table~\ref{tab:RepresentationsN3}, following the method described in appendix~C of \cite{Beem:2014kka}, and re-writing it in terms of characters of $\NN=2$ representations, which can be obtained from the tables of \cite{Dolan:2002zh}.

The stress-tensor multiplet decomposes in the expected way, containing only Schur multiplets
\beq 
\hat{\BB}_{[1,1]} = \hat{\BB}_1  \oplus u_f^{-1} \DD_{\tfrac{1}{2},(0,0)} \oplus u_f \mbar{\DD}_{\tfrac{1}{2},(0,0)} \oplus \hat{\CC}_{0,(0,0)}\,.
\label{eq:Neq2decST}
\eeq
Also of particular importance are the half-BPS multiplets, related to the Coulomb branch of $\NN=3$ theories. Their full decomposition is given by 
\beq
\label{eq:BR0fulldec}
\hat{\BB}_{[R_1,0]} =
u_f^{-R_1}\,\hat{\mathcal{B}}_{\frac{R_1}{2}}\oplus
u_f^{-R_1+1 }\mbar{\DD}_{\frac{R_1-1}{2},(0,0)}\oplus 
\left(\bigoplus_{a=1}^{R_1-2} u_f^{-R_1+a+1} \mbar{\BB}_{\frac{R_1-a-1}{2},-a-1,(0,0)}\right)\oplus
\mbar{\EE}_{-R_1,(0,0)}\,,
\eeq
and similarly for the conjugate multiplet. An interesting question to ask is, apart from the above $\hat{\BB}_{[R_1,0]}$ and conjugate, which $\NN=3$ multiplets contain $\NN=2$ Coulomb branch operators.
An obvious place to look would be to consider $\NN=3$ chiral operators, which decompose as
\beq
\mbar{\BB}_{[0,0],r,0} =
\bigoplus_{a=0}^2 u_f^{ a-\tfrac{r}{3}} \mbar{\EE}_{-\frac{1}{2} \left(a+\frac{r}{3}\right),(\frac{1}{2} a (2-a),0)}\,,
\eeq
and their conjugates.
Note that the above decomposition contains ``exotic'' $\NN=2$ $\mbar{\EE}_{r,(j,0)}$ operators with spin $j>0$, which do not seem to occur in known $\NN=2$ SCFTs (see \cite{Buican:2014qla} for a discussion). Similarly in  \cite{Aharony:2015oyb} the question of which $\NN=3$ operators could contain operators whose vevs parametrized the Coulomb branch was addressed.
The authors of  \cite{Aharony:2015oyb} argue that the only type of such multiplets are $\hat{\BB}_{[R_1,0]}$ and conjugates, since the $\mbar{\BB}_{[0,0],r,0}$ multiplet would not be consistent with the three different $\NN=2$ subalgebras $\NN=3$ contains.

We finish this appendix with the example of the decomposition of a generic long $\NN=3$ multiplet. Considering a multiplet whose highest weight transforms in the symmetric traceless representation for simplicity, $\AA_{[R_1,R_2],r, \ell}^\Delta$, there appears to be a simple prescription for the decomposition into $\NN=2$ multiplets, which we have checked in a variety of cases.
Namely, we first decompose the $\SU(3)_R \times \U(1)_r$ representation $([R_1,R_2],r)$ of the superconformal primary of the $\NN=3$ multiplet in representations of $\SU(2)_{\RN} \times \U(1)_{\rN} \times \U(1)_f$. Let $\{(R',r',F)\}$ be the list of representations appearing in that decomposition. To each such representation we associate an $\calN=2$ multiplet $\mathcal{A}^{\Delta}_{R',r',\ell}$, graded by the corresponding $\U(1)_f$ charge $(u_f)^F$.
Finally, in the decomposition of the $\NN=3$ multiplet, each of these $\NN=2$ multiplets will be accompanied by the following list of long multiplets:
\begin{align}
\begin{split}
&(u_f)^F\left(\mathcal{A}_{R',r'-1,(\frac{\ell }{2},\frac{\ell }{2})}^{\Delta +1} u_f^{2}+u_f^{-2} \mathcal{A}_{R',r'+1,(\frac{\ell }{2},\frac{\ell }{2})}^{\Delta +1}+u_f^{-1} \mathcal{A}_{R',r'+\frac{1}{2},(\frac{\ell }{2},\frac{\ell +1}{2})}^{\Delta +\frac{1}{2}} +u_f^{-1} \mathcal{A}_{R',r'+\frac{1}{2},(\frac{\ell }{2},\frac{\ell -1}{2})}^{\Delta +\frac{1}{2}}\right.\\
&+u_f^{-1} \mathcal{A}_{R',r'+\frac{1}{2},(\frac{\ell +1}{2},\frac{\ell }{2})}^{\Delta +\frac{3}{2}}
+u_f^{-1}  \mathcal{A}_{R',r'+\frac{1}{2},(\frac{\ell -1}{2},\frac{\ell }{2})}^{\Delta +\frac{3}{2}}+\mathcal{A}_{R',r'-\frac{1}{2},(\frac{\ell }{2},\frac{\ell +1}{2})}^{\Delta +\frac{3}{2}}u_f +\mathcal{A}_{R',r'-\frac{1}{2},(\frac{\ell }{2},\frac{\ell -1}{2})}^{\Delta +\frac{3}{2}}u_f \\
&+\mathcal{A}_{R',r'-\frac{1}{2},(\frac{\ell +1}{2},\frac{\ell }{2})}^{\Delta +\frac{1}{2}}u_f+\mathcal{A}_{R',r'-\frac{1}{2},(\frac{\ell -1}{2},\frac{\ell }{2})}^{\Delta +\frac{1}{2}}u_f+\mathcal{A}_{R',r',(\frac{\ell }{2},\frac{\ell }{2})}^{\Delta +2}+\mathcal{A}_{R',r',(\frac{\ell +1}{2},\frac{\ell +1}{2})}^{\Delta +1}+\mathcal{A}_{R',r',(\frac{\ell +1}{2},\frac{\ell -1}{2})}^{\Delta +1}\\
&\left.+\mathcal{A}_{R',r',(\frac{\ell -1}{2},\frac{\ell +1}{2})}^{\Delta +1}+\mathcal{A}_{R',r',(\frac{\ell -1}{2},\frac{\ell -1}{2})}^{\Delta +1}\right)\,.
\end{split}
\end{align}

\section{OPEs of the chiral algebra}
\label{app:chiral_algebra}

In this appendix we collect the OPEs corresponding to the chiral algebra constructed in section \ref{subsec:keq3chiralalg}, with generators given by \eqref{eq:generatorschiralalg}. Here we show all the OPE coefficients already fixed to the values dictated by the Jacobi identities. These computations were performed using the Mathematica package \textsf{SOPEN2defs} of \cite{Krivonos:1995bk} and we follow their conventions.
In what follows we take a product of operators $\OO_1 \OO_2 \cdots \OO_{n-1} \OO_n$ to mean the normal ordered product
$(\OO_1 (\OO_2 (\cdots (\OO_{n-1} \OO_n))))$.

Since all generators, with the exception of the stress-tensor multiplet, are super Virasoro primaries, the OPE of a generator $\OO$ of dimension $\Delta_\OO$ and $\U(1)_f$ charge $f_\OO$ with the stress-tensor current $\JJ$ is fixed to be
\be 
\JJ(Z_1) \OO(Z_2) \sim  \frac{\Delta_{\OO} \th_{12}\bth_{12} \OO }{Z_{12}^2} + \frac{-f_{\OO} \OO - \th_{12} \DD \OO +\bth_{12} \DDb \OO + \th_{12} \bth_{12} \partial \OO}{Z_{12}}\,.
\label{eq:superPrimaryOPE}
\ee
The stress-tensor multiplet has the standard self-OPE given in \eqref{eq:STOPE}, while
the OPEs $\WW(Z_1) \WW(Z_2)$  and $\WB(Z_1) \WB(Z_2)$ are regular. The $\WW(Z_1) \WB(Z_2)$ OPE is given in a general form in \eqref{eq:WWBOPE} where the sum is taken to run over all uncharged generators, composites and/or (super)derivatives thereof. The coefficients $\lambda_{\OO_{h}}$ in \eqref{eq:WWBOPE} are completely fixed by the Jacobi identities to
\be
\label{eq:lambdainWWB}
\begin{split}
&\lambda_{\mathbf{1}}= -\frac{c_{2d}}{9}\,, \quad 
\lambda_{\JJ}=1\,, \quad
\lambda_{\JJ \JJ}=-\frac{4}{c_{2d}-1}\,,\quad
\lambda_{\DD \DDb \JJ}= \frac{c_{2d}-9}{6 (c_{2d}-1)}\,,\quad
\lambda_{\JJ'}= \frac{1}{2}\,, \\
&\lambda_{\OOh2}= -\frac{4 (5 c_{2d}+27)}{\beta  (c_{2d}-9) (c_{2d}-1)}\,,
\end{split}
\ee
where $\beta$ is related to the norm of $\OOh2$. The remaining non-trivial OPEs were found to be
\begin{align}
&\WW(Z_1) \OOh2(Z_2)  \sim  -\frac{\beta  (c_{2d}-9) (c_{2d}+15) \th _{12} \bth_{12} \WW}{2 (5 c_{2d}+27) Z_{12}^3} \nn \\
& +\frac{\beta  (c_{2d}+15)}{12 (5 c_{2d}+27)} \frac{18  \th _{12} \bth_{12} \JJ \WW-2  (c_{2d}-9) \th _{12} \DD \WW- (c_{2d}-27) \th _{12} \bth_{12} \WW'-6 (c_{2d}-9)  \WW}{ Z_{12}^2}\nn \\
&+\frac{\beta}{12 (5 c_{2d}+27)} \frac{  6(c_{2d}+63) \th _{12} \JJ\DD \WW+54  (c_{2d}-1) \th _{12} \WW\DD \JJ- (c_{2d}-9) (c_{2d}+39) \th _{12} \DD \WW'}{Z_{12}} \nn \\
&+\frac{ \beta  (c_{2d}+15)}{6 (5 c_{2d}+27)} \frac{18  \JJ\WW-  (c_{2d}-27) \WW'}{Z_{12}}\,,
\end{align}
and 
\begin{align}
&\WB(Z_1) \OOh2(Z_2)  \sim \frac{\beta  (c_{2d}-9) (c_{2d}+15)}{2 (5 c_{2d}+27)}\frac{ \th _{12} \bth_{12} \WB}{ Z_{12}^3} \nn \\
&+\frac{\beta  (c_{2d}+15)}{12 (5 c_{2d}+27)}  \frac{18 \th _{12} \bth_{12} \JJ\WB- 2 (c_{2d}-9)  \bth_{12} \DDb\WB+ (c_{2d}-27)  \th _{12} \bth_{12} \WB'- 6 (c_{2d}-9)  \WB}{Z_{12}^2} \nn \\
& -\frac{\beta}{12 (5 c_{2d}+27)}  \frac{ 6(c_{2d}+63) \bth_{12} \JJ\DDb\WB +54  (c_{2d}-1) \bth_{12} \WB\DDb \JJ (c_{2d}-9) (c_{2d}+39) \bth_{12} \DDb\WB'}{Z_{12}} \nn \\
& - \frac{ \beta  (c_{2d}+15)}{6 (5 c_{2d}+27)} \frac{18  \JJ\WB +   (c_{2d}-27)  \WB'}{Z_{12}}\,,
\end{align}
with the most complicated one being
\begin{small}
\begingroup
\allowdisplaybreaks[1]
\begin{align}
&\OOh2(Z_1) \OOh2(Z_2) \sim - \frac{(c_{2d}-9)^2 (c_{2d}-1) (c_{2d}+15)  \beta ^2 }{72(5 c_{2d}+27)^2}\frac{c_{2d} + 6 \JJ \th _{12} \bth_{12}}{Z_{12}^4}  \nn \\
&+\frac{\beta ^2 (c_{2d}-9)^2 (c_{2d}-1) (c_{2d}+15) }{12 (5 c_{2d}+27)^2 } \frac{-\bth_{12} \DDb \JJ+\th_{12} \DD \JJ-\th_{12} \bth_{12} \JJ'}{Z_{12}^3}\nn \\
& +\frac{1}{Z_{12}^2}\Bigg(\frac{\beta ^2 \DD \DDb \JJ' \th _{12} \bth_{12} (c_{2d}-9)^3}{8 (5 c_{2d}+27)^2}+\frac{c_{2d} (c_{2d}+15) \beta ^2 \DD \DDb \JJ (c_{2d}-9)^2}{36 (5 c_{2d}+27)^2}+\frac{(c_{2d}+15) \beta ^2 \JJ\JJ (c_{2d}-9)^2}{12 (5 c_{2d}+27)^2}\nn\\
&+\frac{(c_{2d}+15) (2 c_{2d}-3) \beta ^2 \DD \JJ' \th _{12} (c_{2d}-9)^2}{36 (5 c_{2d}+27)^2}+\frac{(c_{2d}+15) \beta ^2 \JJ\DD\JJ \th _{12} (c_{2d}-9)^2}{12 (5 c_{2d}+27)^2} +\frac{(c_{2d}+15) \beta ^2 \JJ\DDb\JJ \bth_{12} (c_{2d}-9)^2}{12 (5 c_{2d}+27)^2}
\nn \\
&+\frac{(c_{2d}-21) \beta ^2 \JJ\DD \DDb\JJ \th _{12} \bth_{12} (c_{2d}-9)^2}{6 (5 c_{2d}+27)^2} -\frac{\beta ^2 \DD \JJ\DDb\JJ \th _{12} \bth_{12} (c_{2d}-9)^2}{8 (5 c_{2d}+27)}-\frac{(c_{2d}+15) (2 c_{2d}-3) \beta ^2 \DDb\JJ' \bth_{12} (c_{2d}-9)^2}{36 (5 c_{2d}+27)^2}
\nn\\
&
+\frac{(c_{2d}+63) \beta ^2 \JJ\JJ\JJ \th _{12} \bth_{12} (c_{2d}-9)}{2 (5 c_{2d}+27)^2}-\frac{27 (c_{2d}-1) \beta ^2 \JJ\JJ' \th _{12} \bth_{12} (c_{2d}-9)}{2 (5 c_{2d}+27)^2}-\frac{27 (c_{2d}-1)^2 \beta ^2 \WW\WB \th _{12} \bth_{12} (c_{2d}-9)}{8 (5 c_{2d}+27)^2}
\nn\\
&-\frac{\left(c_{2d}^3-11 c_{2d}^2-105 c_{2d}+243\right) \beta ^2 \th _{12} \bth_{12} \JJ'' (c_{2d}-9)}{24 (5 c_{2d}+27)^2}-\frac{1}{6} (c_{2d}+3) \OOh2 \beta -\frac{1}{12} (c_{2d}+3) \beta  \DD\OOh2 \th _{12}
\nn\\
&+\frac{\left(c_{2d}^2-8 c_{2d}+135\right) \beta  \DD \DDb\OOh2 \th _{12} \bth_{12}}{8 (5 c_{2d}+27)}-\frac{(7 c_{2d}-135) \beta  \JJ\OOh2 \th _{12} \bth_{12}}{2 (5 c_{2d}+27)}-\frac{27 (c_{2d}-1) \beta  \th _{12} \bth_{12} \OOh2'}{4 (5 c_{2d}+27)}-\frac{1}{12} (c_{2d}+3) \beta  \DDb\OOh2 \bth_{12}\Bigg) \nn \\
&+\frac{1}{Z_{12}}\Bigg(\frac{\beta ^2  \JJ'\DD \DDb\JJ \th _{12} \bth_{12} (c_{2d}-9)^3}{6 (5 c_{2d}+27)^2}+\frac{c_{2d} (c_{2d}+15) \beta ^2 \DD \DDb\JJ' (c_{2d}-9)^2}{72 (5 c_{2d}+27)^2}+\frac{(c_{2d}+15) \beta ^2 \JJ\JJ' (c_{2d}-9)^2}{12 (5 c_{2d}+27)^2}\nn \\
&+\frac{(c_{2d}-33) \beta ^2 \JJ\DD\JJ' \th _{12} (c_{2d}-9)^2}{24 (5 c_{2d}+27)^2}+\frac{(c_{2d}-33) \beta ^2 \JJ\DDb\JJ' \bth_{12} (c_{2d}-9)^2}{24 (5 c_{2d}+27)^2}+\frac{3 (c_{2d}-1) \beta ^2 \DDb\JJ\DD \DDb\JJ \bth_{12} (c_{2d}-9)^2}{16 (5 c_{2d}+27)^2}\nn \\
&+\frac{(c_{2d}-33) \beta ^2 \JJ\DD \DDb\JJ' \th _{12} \bth_{12} (c_{2d}-9)^2}{12 (5 c_{2d}+27)^2}+\frac{\beta ^2 \DDb\JJ'\DD\JJ \th _{12} \bth_{12} (c_{2d}-9)^2}{12 (5 c_{2d}+27)}\nn \\
&+\frac{(c_{2d}-1) (2 c_{2d}+15) \beta ^2 \th _{12} \left(\DD\JJ'\right)' (c_{2d}-9)^2}{96 (5 c_{2d}+27)^2}-\frac{\beta ^2 \DD\JJ'\DDb\JJ \th _{12} \bth_{12} (c_{2d}-9)^2}{12 (5 c_{2d}+27)}\nn \\
&-\frac{3 (c_{2d}-1) \beta ^2 \DD\JJ\DD \DDb\JJ \th _{12} (c_{2d}-9)^2}{16 (5 c_{2d}+27)^2}-\frac{(c_{2d}-1) (2 c_{2d}+15) \beta ^2 \bth_{12} \left(\DDb\JJ'\right)' (c_{2d}-9)^2}{96 (5 c_{2d}+27)^2}\nn \\
&+\frac{9 (c_{2d}-1)^2 \beta ^2 \WB\DD\WW \th _{12} (c_{2d}-9)}{8 (5 c_{2d}+27)^2}+\frac{(c_{2d}-81) (c_{2d}-1) \beta ^2 \JJ'\DD\JJ \th _{12} (c_{2d}-9)}{16 (5 c_{2d}+27)^2}\nn \\
&+\frac{(c_{2d}+63) \beta ^2 \JJ\JJ\DDb\JJ \bth_{12} (c_{2d}-9)}{2 (5 c_{2d}+27)^2}+\frac{(c_{2d}-81) (c_{2d}-1) \beta ^2 \JJ'\DDb\JJ \bth_{12} (c_{2d}-9)}{16 (5 c_{2d}+27)^2}\nn \\
&+\frac{(c_{2d}+63) \beta ^2 \JJ\JJ\JJ' \th _{12} \bth_{12} (c_{2d}-9)}{(5 c_{2d}+27)^2}-\frac{4 \beta  \OOh2\JJ' \th _{12} \bth_{12} (c_{2d}-9)}{5 c_{2d}+27}\nn \\
&-\frac{(c_{2d}+63) \beta ^2 \JJ\JJ\DD\JJ \th _{12} (c_{2d}-9)}{2 (5 c_{2d}+27)^2}-\frac{9 (c_{2d}-1)^2 \beta ^2 \WW\WB' \th _{12} \bth_{12} (c_{2d}-9)}{4 (5 c_{2d}+27)^2}\nn \\
&-\frac{9 (c_{2d}-1)^2 \beta ^2 \WB\WW' \th _{12} \bth_{12} (c_{2d}-9)}{4 (5 c_{2d}+27)^2}-\frac{9 (c_{2d}-1)^2 \beta ^2 \WW\DDb\WB \bth_{12} (c_{2d}-9)}{8 (5 c_{2d}+27)^2}\nn \\
&-\frac{(c_{2d}+3) \left(c_{2d}^2-41 c_{2d}+72\right) \beta ^2 \th _{12} \bth_{12} \JJ^{(3)} (c_{2d}-9)}{72 (5 c_{2d}+27)^2}+\frac{9 (c_{2d}-1) \beta  \OOh2\DD\JJ \th _{12}}{2 (5 c_{2d}+27)}\nn \\
&+\frac{(c_{2d}+63) \beta  \JJ\DDb\OOh2 \bth_{12}}{2 (5 c_{2d}+27)}+\frac{\left(c_{2d}^2-3 c_{2d}+162\right) \beta  \DD \DDb\OOh2' \th _{12} \bth_{12}}{12 (5 c_{2d}+27)}-\frac{(c_{2d}-1) (c_{2d}+27) \beta  \DDb\OOh2' \bth_{12}}{4 (5 c_{2d}+27)}\nn \\
&-\frac{1}{2} \beta  \DD\JJ\DDb\OOh2 \th _{12} \bth_{12}+\frac{1}{2} \beta  \DD\OOh2\DDb\JJ \th _{12} \bth_{12}-\frac{1}{12} (c_{2d}+3) \beta  \OOh2'-\frac{1}{4} \beta  \th _{12} \bth_{12} \OOh2''-\frac{(c_{2d}+63) \beta  \JJ\DD\OOh2 \th _{12}}{2 (5 c_{2d}+27)}\nn \\
&-\frac{9 (c_{2d}-1) \beta  \OOh2\DDb\JJ \bth_{12}}{2 (5 c_{2d}+27)}-\frac{3 (c_{2d}-33) \beta  \JJ\OOh2' \th _{12} \bth_{12}}{2 (5 c_{2d}+27)}-\frac{(c_{2d}-1) (c_{2d}+27) \beta  \DD\OOh2' \th _{12}}{4 (5 c_{2d}+27)}\Bigg)\,.
\end{align}
\endgroup
\end{small}

\section{Conformal blocks and generalized free field theory}
\label{app:blockology}

\subsection{Conformal block conventions}

We adopt the following conventions for the four-dimensional bosonic conformal blocks, 
\begin{align}
g_{\Delta,\ell}^{\Delta_{12},\Delta_{34}}(z,\zb)&=\frac{z \zb}{z-\zb}
\left(k_{\Delta+\ell}^{\Delta_{12},\Delta_{34}}(z)\,k_{\Delta-\ell-2}^{\Delta_{12},\Delta_{34}}(\zb)
-(z\leftrightarrow \zb)\right)\,,
\\
\label{app:kdef}
k_{\beta}^{a,b}(x)&=x^{\frac{\beta}{2}}{}_2F_1(\tfrac{\beta-a}{2},\tfrac{\beta+b}{2},\beta,x)\,.
\end{align}
We also set $k_{\beta}(x):=k_{\beta}^{0,0}(x)$ and $g_{\Delta,\ell}(z,\zb):=g_{\Delta,\ell}^{0,0}(z,\zb)$.

\subsubsection*{Braiding}

Here we collect useful identities between the blocks needed for the crossing symmetry discussion in section \ref{sec:crossing}, namely their transformation under braiding, \ie, the exchange of points one and two.
The $4d$ bosonic blocks transform as 
\be 
\label{braidingsuperblocks}
\left((1-z)(1-\zb)\right)^{-\frac{\Delta_{34}}{2}} g_{\Delta,\ell}^{-\Delta_{12},\Delta_{34}}\big(\tfrac{z}{z-1},\tfrac{\zb}{\zb-1}\big)= (-1)^\ell g_{\Delta,\ell}^{\Delta_{12},\Delta_{34}}(z,\zb)\,,
\ee
and the $\SU(2)$ R-symmetry ones as
\beq
h_m^{\SU(2)}(w)\,=\,(-1)^m\,h_m^{\SU(2)}(\tfrac{w}{w-1})\,.
\eeq
In particular \eqref{braidingsuperblocks} implies that $g_{\Delta,\ell}\big(\tfrac{z}{z-1},\tfrac{\zb}{\zb-1}\big)= (-1)^\ell g_{\Delta,\ell}(z,\zb)$. Finally, the $2d$ bosonic blocks \eqref{eq:sl22dblocksdefinition} satisfy
\beq\label{braidingSL2}
\sl_h(z)=(-1)^h\,\sl_h(\tfrac{z}{z-1})\,.
\eeq

\subsection{Generalized free theory example}
\label{app: generalized free theory}

In this appendix we present a solution to the Ward identities \eqref{eq:WardIdentities} and to the crossing equations of section~\ref{sec:crossing}. It corresponds to the solution of generalized free theory, for which the four-point function factorizes as a product of two-point functions. It reads
\beq
\label{eq:Ggeneralizedfreetheory}
G^{\text{gft}}_R(\x_1,\x_2,y)=1+\left(\frac{\x_1 \x_2}{y}\right)^R\,,
\eeq
from which we can obtain, by setting $x_2=y$, the chiral algebra correlator $f^{\text{gft}}_R(x)\,=\,1+x^R$. Using the parametrization of the WI solution \eqref{solutionWI}, we extract 
\beq
H^{\text{gft}}_R(\x_1,\x_2,y)\,=\, \frac{\x_1 \x_2 \left[\x_1^R \x_2^R y^{2-R} (\x_1-\x_2)-y   \x_1^R \x_2 (\x_1-y)+y \x_1 \x_2^R (\x_2-y)\right]}{(\x_1-\x_2) (\x_1-y) (\x_2-y)}\,.
\eeq
In particular, we have for small $R$ the expressions 
\beq
\begin{split}
H^{\text{gft}}_2(\x_1,\x_2,y)\,&=\,(\x_1\x_2)^2\,,\\
H^{\text{gft}}_3(\x_1,\x_2,y)\,&=\,(\x_1 \x_2)^2 (\x_1+\x_2)+\frac{(\x_1\x_2)^3}{y}\\&=\,(\x_1\x_2)^2\left[(\x_1+\x_2+\tfrac{1}{3}\x_1\x_2)-\x_1\x_2 \,h_{[1,1]}(w)\right]\,,
\end{split}
\eeq
with $h_{[1,1]}(w)$ given in \eqref{Rblocks}. 
The block expansion of the two-dimensional correlator $f^{\text{gft}}_R(x)$ (\ie, in the \nonchiral channel) is explicitly given by
\beq
f^{\text{gft}}_R(x)\,=\,1+\sum_{h=R}^{\infty}\,b_{R,h}^{\text{gft}}\,\Ssl_{h}(x)\,,
\qquad
b_{R,h}^{\text{gft}}=(-1)^{h+1}4^{R-h}\,\frac{(2R)_{h-R}(1-h)_{R-1}}{\Gamma(R)(R+\tfrac{1}{2})_{h-R}}\,.
\eeq
Note (and compare with the discussion around \eqref{lambdafromb}) that in the generalized free theory example we have $b^{(R)}_{h < R}=0$. In particular,  there is no stress tensor being exchanged. Similarly, we can decompose the two-dimensional correlator in the \chiral channel to find
\beq
\widetilde{f}^{\text{gft}}_{R}(z)\,=\, \sum_{\substack{h=R\\ h+R\text{ even }}}^{\infty}\,\widetilde{b}_{R,h}^{\text{gft}}\,\sl_{h}(z)\,,
\qquad
\widetilde{b}_{R,h}^{\text{gft}}=-\frac{2^{1+2R-2 h} \Gamma \left(\frac{h+R}{2}\right)(1-h)_{R-1} \left(\frac{2 R-1}{2} \right)_{\frac{h-R}{2}} }{\Gamma (R)^2 \left(\frac{2 R-1}{4}\right)_{\frac{h-R}{2}} \left(\frac{2 R+1}{4}\right)_{\frac{h-R}{2}}}\,.
\eeq


\section{Short contributions to crossing}

Here collect some bulky equations used in the crossing equations \eqref{fullbootstrap_equations} and summarize the computation used in section~\ref{subsec:ContributionofSchur} for the function $H_{R,\text{short}}$.

\subsection{Explicit expressions for \texorpdfstring{$\mathcal{F}_{\text{short}}^{(0,\pm)}$}{F short}}
\label{app:forFshort}

Here we collect the expressions for  $\mathcal{F}^{(0,\pm)}_{\text{short}}$ that we need in the crossing equations \eqref{fullbootstrap_equations}. Using the definition \eqref{ARdef} for the function $A_R$, we write
\begin{equation}
\label{forFshortAPP1}
\begin{split}
\mathcal{F}^{(0)}_{\text{short}}[f]\,=\,&
-\frac{\left[(1-z)(1-\zb)\right]^{R+1}}{(1-w)^{R-2}}\frac{y^{2-R}}{\x_1^{-1}-\x_2^{-1}}
\left(
\x_2\,A_R(\x_2,y)\,f_R(\x_1)-
\x_1\leftrightarrow \x_2\right)\,,\\
\mathcal{F}^{(0)}_{\text{short}}[H_{\text{short}},\widetilde{H}_{\text{short}}]\,=\,&
\frac{\left[(1-z)(1-\zb)\right]^{R+1}}{(1-w)^{R-2}}\,H_{R,\text{short}}(z,\zb,w)\\
&-(-1)^R
\Big[(z,\zb,w)\leftrightarrow (1-z,1-\zb,1-w)\Big]\,,
\end{split}
\end{equation}
for $\mathcal{F}^{(0)}_{\text{short}}$ and 
\beq
\begin{split}
\label{forFshortAPP2}
\mathcal{F}^{(\pm)}_{\text{short}}[f]\,=\,&
-\Bigg\{\frac{\left[(z-1)(\zb-1)\right]^{R+1}}{(w-1)^{R-2}} \frac{\x_1^{R-1}A_R(\x_1,y)\,\widetilde{f}_{R}(\bar{z})-(z \leftrightarrow \bar{z})}{z^{-1}-\bar{z}^{-1}}
\\&
\pm\Big[(z,\zb,w)\leftrightarrow (1-z,1-\zb,1-w)\Big]\Bigg\}\,,
\\
\mathcal{F}^{(\pm)}_{\text{short}}[H_{\text{short}},\widetilde{H}_{\text{short}}]\,=\,&
(-1)^R\Bigg\{\frac{\left[(1-z)(1-\zb)\right]^{R+1}}{(1-w)^{R-2}}\,\widetilde{H}_{R,\text{short}}(z,\zb,w)
\\&\pm
\Big[(z,\zb,w)\leftrightarrow (1-z,1-\zb,1-w)\Big]  \Bigg\}
\\&\mp
\Bigg\{\frac{\left[(1-z)(1-\zb)\right]^{R+1}}{(1-w)^{R-2}}\,H_{R,\text{short}}(\tfrac{z}{z-1},\tfrac{\zb}{\zb-1},\tfrac{w}{w-1})\\&\pm
\Big[(z,\zb,w)\leftrightarrow (1-z,1-\zb,1-w)\Big]\Bigg\}\,,
\end{split}
\eeq
for $\mathcal{F}^{(\pm)}_{\text{short}}$.

\subsection{Summation for \texorpdfstring{$H_{\text{short}}$}{H short}}
\label{app:summationforFshort}

Given the function $f_R(x)$ in the parametrization \eqref{solutionWI} of a four point function \eqref{eq:BBbBBbcorr},  one can associate  a contribution to the function $H$, called  $H_{\text{short}}[f]$, corresponding to the exchange 
 of the short operators which survive the cohomological truncation.
The goal of this appendix is to explicitly perform the summations in the first term in \eqref{firttermto_toesum}.
This can be done for two reasons: 
\begin{enumerate}
\item The coefficients $b_{h}^{(R)}$ in the expansion  \eqref{fexpansion} can be easily determined in terms of a finite number of parameters. This follows from the fact that $f_R(x)$ is a polynomial of degree $R$ that satisfy the crossing  property $f_R(x)=x^{R}f_{R}(x^{-1})$.
\item Each block entering the first sum  in \eqref{firttermto_toesum} has the form 
\beq
\mathcal{G}^{d=4,\mathcal{N}=1}_{h+4,h}\,=\,
\frac{t_h(z)s(\bar{z})-t_h(\bar{z})s(z)}{z-\bar{z}}\,,
\qquad
s(t)=-2 (t+\log(1-t))\,,
\eeq
where $t_h(z)=t^{h+3}{}_{2}F_{1}(h+2,h+3,2h+5,t)$.
\end{enumerate}
Each monomial term in $f_{R}(x)$, except for $x^0=1$, can be expanded in superblocks as
\beq
\label{genfrefieldsf}
x^n\,=\,\sum_{h=n}^{\infty}\,\hat{b}_{n,h}\,\Ssl_h(\tfrac{x}{x-1})\,,
\qquad
\hat{b}_{n,h}=-4^{n-h}h\,\frac{(2n)_{h-n}(1-h)_{n-1}}{\Gamma(n+1)(n+\tfrac{1}{2})_{h-n}}\,.
\eeq
It follows that the part of $H_{R,\text{short}}$ in the R-symmetry singlet channel, compare to the first term in  \eqref{firttermto_toesum},
 is
\beq\label{Honmonomial}
H^{\text{singlet}}_{R,\text{short}}[x^n]\,:=\,
\sum_{h=n}^{\infty}\,\hat{b}_{n,h}\,\mathcal{G}^{d=4,\mathcal{N}=1}_{h+4,h}(z,\zb)\,=
\frac{z^{n+1}s(\bar{z})-\bar{z}^{n+1}s(z)}{z-\bar{z}}\,,
\qquad n\geq 0\,,
\eeq
and $H^{\text{singlet}}_{R,\text{short}}[1]\,=\,0$.
Above $H^{\text{singlet}}_{R,\text{short}}$ is considered as a linear map acting on polynomials in the variable $x$.
Such maps are characterized by their actions on monomials given in \eqref{Honmonomial}.
In the cases relevant for $R=2,3$ (recall that $H_{R=1}=0$), the only solution of the crossing symmetry condition
$f_R(x)=x^{R}f_{R}(x^{-1})$ are
\beq
\begin{split}
 f_2(x)=1+c^{-1}x+x^2\,,\qquad f_3(x)&=(1+x)(1+ \gamma(c)\,x+x^2)\\
 &=1+(1+\gamma(c) ) x+(1+\gamma(c) ) x^2+x^3\,,
 \end{split}
 \eeq
with $1+\gamma(c)=\tfrac{9}{4c_{4d}}$ and $c=c_{4d}$\,.
It follows from \eqref{Honmonomial} that 
\beq
\begin{split}
&H_{R,\text{short}}^{\text{singlet}}(z,\bar{z})=
\frac{\hat{t}^{(R)}(\x_1)\hat{s}(\x_2)-\hat{t}^{(R)}(\x_2)\hat{s}(\x_1)}{\x_1-\x_2}\,,\\
&\hat{t}^{(R)}(x)=x(f_R(x)-1)\,,
\quad 
\hat{s}(\tfrac{z}{z-1})\,=\, \frac{s(z)}{1-z}\,,
\end{split}
\eeq
with a now familiar identification $(\x_1,\x_2)=(\tfrac{z}{z-1},\tfrac{\zb}{\zb-1})$.

\bibliography{./auxi/biblio}
\bibliographystyle{./auxi/JHEP}

\end{document}